\documentstyle[preprint,aps,tighten,eqsecnum]{revtex}

\begin{document}

%%%%%%%%%%%%%%%%%%%%%%%%%%%%%%%%%%%%%%%%%%%%%%%%%%%%%%%%%%%%%%%%
%%
%%    MACROS  (TO BE REMOVED LATER)
%%
%%%%%%%%%%%%%%%%%%%%%%%%%%%%%%%%%%%%%%%%%%%%%%%%%%%%%%%%%%%%%%%%

\def \sumint#1{\sum_{#1}\kern-14.0pt\int}

%%  This first group is so I can be lazy :-)
\def \ns{\enspace}
\def \ts{\thinspace}
\def \nts{\negthinspace}
\def \beq{\begin{equation}}
\def \eeq{\end{equation}}
\def \beqa{\begin{eqnarray}}
\def \eeqa{\end{eqnarray}}
\def \eps{\epsilon}
\def \vareps{\varepsilon}
\def \GeV{{\rm \enspace GeV}}
\def \trace{{\rm Tr}}
\def \half{\hbox{$1\over2$}}
%%  A clever idea Harry showed me . . .
%%  This may be turned off by redefining LABEL to label.
%\def \LABEL#1{\enspace\scriptstyle{#1}\label{#1}}
%\def \LABEL#1{\label{#1}}
 
%%  Now for the paper-specific ones . . .

%%  These are defined this way so I can easily change notation
%%  later on, if desired.
%%
%%  how to write a 2-D vector
%\def \trans#1{\vec{#1}}
\def \trans#1{\bbox{#1}}
%%
%%  momentum fraction
\def \xf{x_{{}_F}}
\def \WW{Weis\"acker-Williams}
\def \Isng{{\cal I}}
\def \Ismth{\overline{\cal I}}

\def \LQCD{\Lambda_{\rm QCD}}
\def \Jt{\bbox{J}}
\def \bt{\bbox{b}}
\def \pt{\bbox{p}}
\def \qt{\bbox{q}}
\def \rt{\bbox{r}}
\def \xit{\bbox\xi}
\def \xiprimet{\xit^\prime}
\def \deltat{\bbox\Delta}
\def \deltahat{\bbox{\hat\Delta}}
\def \sigmat{\bbox\Sigma}
\def \delt{\bbox\nabla}
\def \wt{\bbox{w}}
\def \xt{\bbox{x}}
\def \yt{\bbox{y}}
\def \xprimet{\xt^\prime}
\def \At{\bbox{A}}
\def \Bl{\bbox{\Bigl[}}
\def \Br{\bbox{\Bigr]}}
\def \Bbl{\bbox{\biggl[}}
\def \Bbr{\bbox{\biggr]}}
\def \BBl{\bbox{\Biggl[}}
\def \BBr{\bbox{\Biggr]}}
\def \Qprobe{Q}
\def \Qtotal{{\cal Z}}
\def \curlyL{{\cal L}}
\def \curlyD{{\cal D}}
\def \CHI{{\cal X}}
\def \U{{\rm U}}
\def \Uinv{\U^{-1}}
\def \pexp{{\cal P}\exp}
\def \gnum{ {{dN}\over{dq^{+} d^2\qt}} }
\def \cfun{\langle A_i^a(x^{-},\xt) A_j^b(x^{\prime -},\xprimet)\rangle}
\def \TRcfun{\langle A_i^a(x^{-},\xt) A_i^a(x^{\prime -},\xprimet)\rangle}
\def \cfuny{\langle A_i^a(y,\xt) A_j^b(y',\xprimet)\rangle}
\def \TRcfuny{\langle A_i^a(y,\xt) A_i^a(y',\xprimet)\rangle}
\def \Ihat{C}
\def \sgn{\mathop{\rm sgn}\nolimits}
\def \Ei{\mathop{\rm Ei}\nolimits}
\def \transint{{\cal T}}

\def \MV{MV}

%%%%%%%%%%%%%%%%%%%%%%%%%%%%%%%%%%%%%%%%%%%%%%%%%%%%%%%%%%%%%%%%
%%%%%%%%%%%%%%%%%%%%%%%%%%%%%%%%%%%%%%%%%%%%%%%%%%%%%%%%%%%%%%%%
%%%%%%%%%%%%%%%%%%%%%%%%%%%%%%%%%%%%%%%%%%%%%%%%%%%%%%%%%%%%%%%%
%%%%%%
%%%%%%    TITLE AND ABSTRACT
%%%%%%
%%%%%%%%%%%%%%%%%%%%%%%%%%%%%%%%%%%%%%%%%%%%%%%%%%%%%%%%%%%%%%%%
%%%%%%%%%%%%%%%%%%%%%%%%%%%%%%%%%%%%%%%%%%%%%%%%%%%%%%%%%%%%%%%%
%%%%%%%%%%%%%%%%%%%%%%%%%%%%%%%%%%%%%%%%%%%%%%%%%%%%%%%%%%%%%%%%

\draft
\preprint{
  \parbox{2in}{McGill/99--16 \\
  hep-ph/9907281
}  }

\title{Color Neutrality and the Gluon Distribution\\
in a Very Large Nucleus}
\author{C.S. Lam~\cite{LAMemail} and Gregory Mahlon~\cite{GDMemail}}
\address{Department of Physics, McGill University, \\
3600 University St., Montr\'eal, QC  H3A 2T8 \\
Canada }
\date{July 15, 1999}
\maketitle
\begin{abstract}
We improve the McLerran-Venugopalan model for the gluon 
distribution functions in very large nuclei by imposing
the condition that the nucleons should be color neutral.
We find that enforcing color neutrality cures the infrared
divergences in the transverse coordinates which are
present in the McLerran-Venugopalan model.  Since we
obtain well-defined expressions for the distribution functions,
we are able to draw unambiguous conclusions about various
features of the model.  In particular, we show that the
gluon distribution functions in the absence of quantum corrections 
behave as $1/\xf$ to all orders in the coupling constant.  
Furthermore, our distribution functions exhibit saturation
at small transverse momenta.
The normalization of the distribution function we obtain
is not arbitrary but specified in terms of the
nucleon structure.
We derive a sum rule for the integral of the gluon distribution 
function over transverse momenta, and show that the non-Abelian
contributions serve only to modify the shape of the transverse
momentum distribution.  
We obtain a relatively simple expression for 
the mean value of the transverse momentum-squared.  
The connection between the McLerran-Venugopalan
model and the Dokshitzer-Gribov-Lipatov-Altarelli-Parisi equation
is discussed quantitatively.  
Finally,
we illustrate our results in terms of a simple nuclear model
due to Kovchegov.
\end{abstract}

%%%%%%%%%%%%%%%%%%%%%%%%%%%%%%%%%%%%%%%%%%%%%%%%%%%%%%%%%%%%%%%%
%%%%%%%%%%%%%%%%%%%%%%%%%%%%%%%%%%%%%%%%%%%%%%%%%%%%%%%%%%%%%%%%
%%%%%%%%%%%%%%%%%%%%%%%%%%%%%%%%%%%%%%%%%%%%%%%%%%%%%%%%%%%%%%%%
%%%%%%
%%%%%%    INTRODUCTION -- INTRODUCTION
%%%%%%
%%%%%%%%%%%%%%%%%%%%%%%%%%%%%%%%%%%%%%%%%%%%%%%%%%%%%%%%%%%%%%%%
%%%%%%%%%%%%%%%%%%%%%%%%%%%%%%%%%%%%%%%%%%%%%%%%%%%%%%%%%%%%%%%%
%%%%%%%%%%%%%%%%%%%%%%%%%%%%%%%%%%%%%%%%%%%%%%%%%%%%%%%%%%%%%%%%

\section{Introduction}

Several years ago, McLerran and Venugopalan~\cite{paper1} recognized
the existence of a regime in which the gluon structure
functions ought to be calculable within quantum chromodynamics
(QCD).  Their idea is to recognize that in large nuclei at
small enough values of the longitudinal momentum fraction $\xf$,
the density of partons per unit area is large.  Since a large
number of charges are contributing, it is expected that
classical methods should apply, {\it i.e.}\ the vector
potential is computed from the classical Yang-Mills equations,
and quantum correlation functions are approximated by ensemble
averages.  This picture was developed more fully in subsequent
work~\cite{paper2,paper3,paper4}, and is somewhat related
to the approach of 
Mueller~\cite{paper18,paper21,paper19,paper27,paper15,paper37}.

In the region of very small $\xf$, the quantum corrections
to the distribution functions calculated in the McLerran-Venugopalan
model become large~\cite{paper4,paper7}.
In particular, these corrections
are enhanced by powers of $\alpha_s\ln(1/\xf)$, implying that
the classical approximation being made at lowest order is
breaking down in this region.
This observation has lead to the development of a
renormalization group procedure~\cite{paper9} whereby these
corrections are taken into account by incorporating
gluons with large values of $\xf$ into the charge density.
This approach,
which uses the results of the McLerran-Venugopalan model
as its input, has been subsequently developed in 
Refs.~\cite{paper9,paper11,paper39,paper40,paper43,paper34,paper44}.

The focus of this paper is not on the very small $\xf$ region,
but rather on dealing with the infrared divergence in the
transverse coordinates which appears in the McLerran-Venugopalan model.
The vector potential two-point correlation functions obtained
in Ref.~\cite{paper9} are highly infrared divergent, going like
$(x^2)^{x^2}$ at large distances.  On one hand, the region where
this poor behavior manifests itself is confined to separations
$x \gg \LQCD^{-1}$.  This is firmly in the non-perturbative
regime where one does not really believe the results of a 
perturbative calculation anyway.  However, since the gluon number
density is derived from the Fourier transform of this correlation
function, the presence of such a strong divergence makes it difficult
to obtain more than qualitative results from the theory:  one
must resort to cutting off the transverse spatial integrations
at $\LQCD^{-1}$\cite{paper10}.

In this paper we will demonstrate that the infrared
divergence described above is closely related to the issue
of color neutrality.  Physically, nucleons appear to be colorless
when viewed at length scales which greatly exceed $\LQCD^{-1}$.
Consequently, we expect that two-point correlation functions
should die off rapidly beyond distance scales of about $\LQCD^{-1}$,
because of confinement.  The quarks in two widely separated
nucleons should not feel each others presence.  
Thus, the fundamental requirement that the nucleons be color neutral
may be rephrased as a restriction on the allowable 
charge-density correlation functions $\curlyD(\xt)$.
Although the McLerran-Venugopalan model employs an ensemble of charge
distributions which respects the fact that the average charge 
should vanish, the two-point correlator computed from this ensemble 
is inconsistent with color-neutral nucleons.
We claim that the color neutrality condition should be
viewed as of primary importance, and that we should
modify the ensemble of charge distributions to satisfy it.
By doing so, we incorporate one of the major effects
brought about by confinement, parameterizing it in terms
of the precise form chosen for $\curlyD(\xt)$.
Enforcing color neutrality cures the infrared divergence
in the McLerran-Venugopalan model
by introducing a new scale into the problem, namely the
minimum length scale for which color neutrality holds.
This outcome is somewhat similar to the one considered
in Refs.~\cite{paper24,paper42}; however, neither of these
papers examine the consequences of enforcing color neutrality
in great detail.

Given a model which incorporates the color neutrality of
the nucleons at some scale, we obtain infrared
finite correlation functions.   Because our results are
well-defined everywhere,  we are able to draw firm conclusions
about various properties of the gluon distribution functions,
in spite of their highly nonlinear form.
We are able to demonstrate that 
the gluon distribution functions in the McLerran-Venugopalan
model are proportional to $1/\xf$, independent of the distribution
of charge in the longitudinal coordinate.  To evade this
conclusion apparently
requires charge-density correlation functions which
have a non-trivial longitudinal structure.
Although we have modified the form of $\curlyD(\xt)$
from the one employed in Ref.~\cite{paper9}, we still 
obtain correlation functions which are consistent with
asymptotic freedom.  That is, in the ultraviolet limit, the
non-Abelian terms drop out and we obtain a gluon distribution
function which goes like $1/\qt^2$.  At small values of
the transverse momentum $\qt$, our gluon distribution
functions either approach a finite constant or diverge 
mildly (like $\ln\qt^2$), depending on the long distance
behavior of $\curlyD(\xt)$.   The density of gluons at small
$\qt$ grows more slowly with increasing nuclear size
than the number of nucleons, consistent with 
saturation~\cite{paper37,GLR,paper35,paper45}.
We also derive a transverse momentum sum rule, which shows
that the non-Abelian corrections only serve to transfer
gluons (at fixed $\xf$) from one value of transverse momentum
to another:  the total number of gluons is not altered.
This is reflected in the fact that the structure 
functions we obtain are simply proportional to the number of nucleons:
all non-linear dependence on the amount of charge present
is suppressed at large values of the momentum transfer.
As an application of the sum rule, we make contact with
the Dokshitzer-Gribov-Lipatov-Altarelli-Parisi 
equation~\cite{DGLAP1,DGLAP2}, demonstrating the plausibility
of the McLerran-Venugopalan model.
In addition, we are able to compute a fairly simple expression for the 
mean transverse momentum-squared of the gluons which is accurate
to order unity:  that is, the logarithms that arise in our
calculation are all evaluated at a scale which is completely
calculable given a nucleon model.  We find that the leading
$\Qprobe^2$ behavior of this quantity is linear in the number
of nucleons, whereas the logarithmic corrections are
quadratic in the charge per unit transverse area.

The remainder of this paper is organized as follows.
In Sec.~\ref{Review}, we review the 
McLerran-Venugopalan model
as described in Refs.~\cite{paper1,paper2,paper3,paper4,paper9}.
We allow for a generalized dependence of
the charge-density correlator on the transverse coordinates in order
to set the stage for the subsequent discussion.
In Sec.~\ref{NEUTRALITY} we describe the conditions which
the charge-density correlator must satisfy in order to
be consistent with nucleons which are color neutral.
We examine the properties of the resulting gluon number density 
in Sec.~\ref{GLUONNUM}.  
If the charge-density correlator is Gaussian and
local in the longitudinal coordinate, we show that the gluon number
density is always proportional to $1/\xf$.  
We derive a
sum rule for the integral of the gluon number density over
the transverse momenta, and apply it to the computation of  
the gluon distribution function at large values of the momentum
transfer.  Our result is closely related to the
Dokshitzer-Gribov-Lipatov-Altarelli-Parisi equation.
Sec.~\ref{GLUONNUM} concludes with a discussion of
the mean value of the transverse momentum-squared obtained
by our approach.
We illustrate
our results in Sec.~\ref{EXAMPLE} within the context of a 
specific nuclear model by Kovchegov~\cite{paper12}.
Finally, Sec.~\ref{CONC} contains our conclusions.
A discussion of our notation and conventions may be found
in Appendix~\ref{NOTATION}.  
The details of how to evaluate various integrals
arising in the text are contained in
Appendices~\ref{SIGMA},~\ref{MOMENTS}, and~\ref{INTEG1}.

%%%%%%%%%%%%%%%%%%%%%%%%%%%%%%%%%%%%%%%%%%%%%%%%%%%%%%%%%%%%%%%%
%%%%%%%%%%%%%%%%%%%%%%%%%%%%%%%%%%%%%%%%%%%%%%%%%%%%%%%%%%%%%%%%
%%%%%%%%%%%%%%%%%%%%%%%%%%%%%%%%%%%%%%%%%%%%%%%%%%%%%%%%%%%%%%%%
%%%%%%
%%%%%%       REVIEW OF THE MCLERRAN-VENUGOPALAN MODEL
%%%%%%
%%%%%%%%%%%%%%%%%%%%%%%%%%%%%%%%%%%%%%%%%%%%%%%%%%%%%%%%%%%%%%%%
%%%%%%%%%%%%%%%%%%%%%%%%%%%%%%%%%%%%%%%%%%%%%%%%%%%%%%%%%%%%%%%%
%%%%%%%%%%%%%%%%%%%%%%%%%%%%%%%%%%%%%%%%%%%%%%%%%%%%%%%%%%%%%%%%

\section{Review of the McLerran-Venugopalan Model}\label{Review}

%%%%%%%%%%%%%%%%%%%%%%%%%%%%%%%%%%%%%%%%%%%%%%%%%%%%%%%%%%%%%%%%
%%
%%              GENERAL CONSIDERATIONS
%%
%%%%%%%%%%%%%%%%%%%%%%%%%%%%%%%%%%%%%%%%%%%%%%%%%%%%%%%%%%%%%%%%

\subsection{General Considerations}\label{GENERAL}

The goal of the McLerran-Venugopalan (\MV) model is to 
compute the gluon distribution function at small values of 
the longitudinal momentum fraction 
$\xf \equiv q^{+}_{gluon} / Q^{+}_{nucleon}$ for a very large nucleus.  
We begin our discussion
with the assumptions which form the basis of the \MV\ model,
and outline the steps in the computation
of the gluon distribution function.  Having
described the ingredients which underlie the calculations, we
will explain the conditions which are required for this
picture to be applicable.

Consider a nucleus moving along the $z$ axis at nearly the
speed of light.  
The central assertion of the \MV\ model is that, in a specific 
kinematic regime, we may determine the gluon distribution function
of the nucleus as follows.  First, we are told to 
compute the classical vector potential
generated by the valence quarks.  
Since the valence quarks appear as a Lorentz-contracted 
pancake-shaped distribution of color charge, 
the current appearing in the classical Yang-Mills equations
takes the form (see Appendix~\ref{NOTATION} for a summary
of our notation and conventions)
\beqa
J^{+}(x) &=& \rho(x^{-},\xt), \cr
J^{-}(x) &=& 0, \cr 
\Jt(x) &=& {\bf 0}.
\label{QCDcurrent}
\eeqa
There is no $x^{+}$ dependence in Eq.~(\ref{QCDcurrent}) in the limit 
that the source ({\it i.e.}\ the nucleus) 
is moving at the speed of light.  
Because the intuitive parton model is 
realized in the 
light-cone gauge~\cite{Curci,CollinsSoper,LC1,paper31},
we should write the solution in that gauge, 
{\it i.e.}\ $A^{+}\equiv 0$.

By definition, the gluon number density is just the gluon
number operator evaluated in the (quantum) state of interest.
This may be expressed in terms of the two-point vector potential
correlation function as~\cite{paper35,CollinsSoper}
\beq
\gnum =
{ {q^{+}}\over{4\pi^3} }
\int_{-\infty}^{\infty}  \nts\nts dx^{-} \nts
\int d^2\xt
\int_{-\infty}^{\infty} \nts\nts dx^{\prime -} \nts
\int d^2\xprimet
\ts
e^{-iq^{+}(x^{-}-x^{\prime -})}
e^{i\qt\cdot(\xt-\xprimet)}
\TRcfun.
\label{GluonDensity}
\eeq
In the \MV\ model, the average over quantum fluctuations
implied by the angled brackets on the right hand 
side of~(\ref{GluonDensity}) is replaced by a classical
average over a suitably weighted ensemble of charge densities.
The ensemble should consist of  all of the physically realizable
states of the quarks (in color singlet combinations) within
the nucleons and of the nucleons within the nucleus.
Then, what we require
is an expression for the vector potential
in terms of the charge density plus
the probability distribution for determining
how likely a given charge configuation is.  
In the \MV\ model,
ensemble averages are computed according to the Gaussian weight
\beq
\int [d\rho] \exp \ts 
\Biggl\{ - \int dx^{-}\ts d^2\xt\int dx^{\prime -} \ts d^2\xprimet\ts
{\rm Tr}\ts\Bigl[\rho(x^{-},\xt)
                   \rho(x^{\prime -},\xprimet)\Bigr] 
{ {\delta(x^{-}-x^{\prime -}\ts)}
\over
{ \mu^2(x^{-}) }
} 
\curlyD^{-1}(\xt-\xprimet)
\Biggr\}.
\label{RhoWeight}
\eeq
This weight implies that the charge density has an average
value of zero.
There are, however, fluctuations about this average value, the size 
of which are governed by $\mu^2(x^{-})$, which is related to the 
average charge density squared per unit area per unit
thickness.\footnote{The authors of 
Ref.~\protect\cite{paper9} choose 
to replace the spatial variable $x^{-}$ 
with what they call the space-time rapidity 
$y$, defined by $y \equiv y_0 + \ln(x^-_0/x^-)$.  
Thus, they write $\mu^2(y)$ and refer to 
it as the average charge density squared
per unit area per unit rapidity.  We will 
stick to the description in terms of $x^{-}$ 
in this paper.}                           %%%%%%%%%%% END OF FOOTNOTE
The precise statement of the relationship appears in 
Sec.~\ref{NEUTRALITY}.  The longitudinal dependence of
Eq.~(\ref{RhoWeight}) is purely local in $x^{-}$, 
reflecting the fact that the large number of quarks which contribute
to the charge per unit area at a given transverse position typically
come from different nucleons:  hence, they should be uncorrelated.
The transverse space dependence of the \MV\ model as described in 
Refs.~\cite{paper1,paper2,paper3,paper4,paper9} is
\beq
\curlyD(\xt-\xprimet) = \delta^2(\xt-\xprimet).
\label{MVtrans}
\eeq
Ultimately, we intend to modify this dependence.
Therefore, we will write our expressions in terms of $\curlyD$
and insert~(\ref{MVtrans}) only when we wish to specifically
discuss the results of Ref.~\cite{paper9}.

Thus, given the solution for the vector potential in
terms of the charge density and an ansatz for the charge
density correlation function, we may compute an approximation
to the gluon distribution function.  We now describe the
conditions necessary for this procedure to yield a good
approximation to the result which would be obtained in a full QCD 
treatment.  

First, we note that the charge distribution for the nucleus
is taken to be a Lorentz-contracted pancake moving along the $z$
axis at (nearly) the speed of light.  
We wish to be in the regime where the partons do not
resolve its longitudinal structure.
This is only true for partons with very small longitudinal
momentum fractions. 
To determine how small the momentum should be, we note that the
corresponding length scale $x^{-}~\sim~1/q^{+}_{gluon}$ should
be larger than the thickness of the Lorentz-contracted nucleus.
This thickness is of order $A^{1/3} a / \gamma$, where $A$ is 
the number of nucleons, $a$
the radius of a single nucleon, and $\gamma$ is the (common) 
boost factor of the nucleons which comprise the nucleus.
Writing the mass of the nucleon to be $m$,
we see then that the condition 
on the momentum fraction reads
\beq
\xf \equiv
{  { q^{+}_{gluon} }\over{ Q^{+}_{nucleon}}  } \alt 
{ {A^{-1/3}}\over{ma} }.
\label{xCondition}
\eeq
Gluons with longitudinal momenta satisfying~(\ref{xCondition})
cannot resolve distances shorter than the thickness of the 
Lorentz-contracted nucleus they see. 

An immediate consequence of being unable to resolve nucleons
with different values of $x^{-}$ is that we may view the nucleons
as being piled up onto effectively the same value of $x^{-}$.
Thus, each unit of transverse area will, on average, contain
a large number of valence quarks.
Since there are a large number
of quarks contributing, the total contribution will be in a large
representation of the gauge group, 
allowing us to treat the source in the
the classical approximation.  
The large number of quarks also allows us to justify the use
of the Gaussian weight~(\ref{RhoWeight}) via the central
limit theorem.
The number of valence quarks per unit area in a flattened nucleus
of radius $R$ is $3A/\pi R^2 \sim 3 A^{1/3}/\pi a^2$,
since the nuclear radius scales with the number of nucleons as
$R \sim A^{1/3} a$.
Gluons with a transverse momentum $\qt^2$ see a transverse
area of about $\pi/\qt^2$.
Thus, to maintain the
condition that there is a large number of quarks in the region
being probed, we should only look at transverse momenta such that
\beq
\LQCD^2 \alt \qt^2 \alt 3 A^{1/3} a^{-2},
\label{qtCondition}
\eeq
that is, $\qt$ should be soft enough to correspond to a large
enough area, but
not so soft that the coupling becomes strong.  
A second factor limiting the maximum $\qt^2$ for which this
treatment is valid is the observation that
if the transverse momentum of the emitted
gluons is too large, then the eikonal approximation 
we are making breaks down.
That is, the no-recoil approximation that allowed us to 
treat the valence quarks as being localized on the light-cone fails.

The bottom line is that  in large enough nuclei
for small enough values of the longitudinal
momentum carried by the parton and 
moderately large tranverse momenta,
the number of quarks per unit area becomes large.
In this
regime, we expect that we may compute the gluon distribution
function  at lowest order from 
the classical solution to the Yang-Mills equations.

%%%%%%%%%%%%%%%%%%%%%%%%%%%%%%%%%%%%%%%%%%%%%%%%%%%%%%%%%%%%%%%%
%%
%%    NONABELIAN WEISACKER-WILLIAMS FIELDS
%%
%%%%%%%%%%%%%%%%%%%%%%%%%%%%%%%%%%%%%%%%%%%%%%%%%%%%%%%%%%%%%%%%

\subsection{Non-Abelian Weis\"acker-Williams Fields}

As outlined above, the \MV\
model makes use of the classical solution to the Yang-Mills
equations in the presence of a source of the form~(\ref{QCDcurrent}),
with $J^{+}(x) \equiv \rho(x^{-},\xt)$ the only nonzero
component, independent of the light-cone time $x^{+}$. 
Ultimately, we will take the dependence on the
light-cone distance $x^{-}$ to be a delta function,
corresponding to the pancake-shaped distribution 
described above.  
Initially, however, we will work with a finite spread
in $x^{-}$ (a multi-layer pancake), to avoid the ambiguities
that would otherwise arise in the commutator terms of the
Yang-Mills equations~\cite{paper9}.
Furthermore, certain quantum corrections 
may be included by a renormalization group 
treatment~\cite{paper9,paper11,paper39,paper40,paper43,paper34,paper44} 
which generates a specific
$x^{-}$ dependence, reflecting the 
presence of the Yukawa cloud
of quark pairs and gluons around the original source.
So, it behooves us to 
make provision for such a dependence.

The simplest route to the solution for the vector potential
in the light-cone gauge for a current of the form~(\ref{QCDcurrent})
is to begin by finding the solution in the covariant gauge,
and subsequently transform into the 
light-cone gauge~\cite{paper18,paper12}.
Therefore, we begin in the covariant gauge,
$\partial_{\mu}\widetilde{A}^{\mu}(x) = 0$, where the field
equations~(\ref{YangMills}) become
\beq
\partial^2\widetilde{A}^{\nu}
-2i \Bl \widetilde{A}^{\mu},\partial_{\mu}\widetilde{A}^{\nu} \Br
+i  \Bl \widetilde{A}_{\mu},\partial^{\nu}\widetilde{A}^{\mu} \Br
- \Bbl \widetilde{A}_{\mu}, \Bl \widetilde{A}^{\mu},
                               \widetilde{A}^{\nu} \Br\Bbr = J^{\nu}.
\eeq
We recall that in electrodynamics, the $\mu$th component of the
vector potential is generated by the $\mu$th component of the
current (in the covariant gauge).  Analogously,
we attempt to form
a solution where the only non-zero component is $\widetilde{A}^{+}$.
Imposing this condition, we find that this is possible,
provided that
\beq
\partial^2\widetilde{A}^{+} 
- 2i\Bl \widetilde{A}^{+}, \partial_{+}\widetilde{A}^{+} \Br
= \rho.
\label{QCDplus1}
\eeq
Since $\widetilde{A}^{-}$ and $\widetilde{A}^{i}$ vanish,
the condition that we are in covariant gauge reduces to
$\partial_{+} \widetilde{A}^{+} = 0$.  Hence $\widetilde{A}^{+}$
is independent of $x^{+}$ and Eq.~(\ref{QCDplus1}) reduces
to
\beq
\delt^2\widetilde{A}^{+}(x^{-},\xt) = \rho(x^{-},\xt),
\label{EasyQCD}
\eeq
Notice that Eq.~(\ref{EasyQCD}) is local in $x^{-}$.  
This is a consequence of the $x^{+}$-independence
of the vector potential, which reduces 
$\partial^2 = -2\partial^{+}\partial^{-} + \delt^2$
to just $\delt^2$.
The solution of~(\ref{EasyQCD}) is
\beq
\widetilde{A}^{+}(x^{-},\xt) = 
\int d^2\xprimet \ts 
G(\xt-\xprimet) \rho(x^{-},\xprimet),  
\label{QCDCovarSoln}
\eeq
where 
\beq
G(\xt-\xprimet) \equiv
{ {1}\over{4\pi} }
\ln \biggl( {
{\vert\xt-\xprimet\vert^2}
\over
{\lambda^2}
} \biggr).
\label{GreenFunction}
\eeq
The arbitrary length scale $\lambda$ which appears 
in~(\ref{GreenFunction}) 
reflects the lack of a scale in
the classical theory.
This same lack of scale leads to the infrared divergence
in perturbation theory.
Note that in the covariant gauge, the fact that $J$ and $A$ are 
color matrices
has introduced no real complications into the solution:  the
vector potential is still linear in the sources.

The non-Abelian features make themselves felt when we gauge
transform our solution to the light-cone gauge.  We write the
gauge transformation as
\beq
A^{\mu}(x) = \U(x) \widetilde{A}^{\mu}(x) \U^{-1}(x)
           -i\Bigl[ \partial^{\mu}\U(x) \Bigr]\U^{-1}(x).
\label{QCDGaugeTrans}
\eeq
Asking that the new gauge be the light-cone gauge, 
$A^{+}\equiv 0$, we find that
$\U(x)$ should satisfy
\beq
i\Bigl[\partial_{-}\U(x)\Bigr]  = \U(x)\widetilde{A}^{+}(x).
\label{Ueqn}
\eeq
The solution to Eq.~(\ref{Ueqn}) is the path-ordered exponential
\beq
\U(x) \equiv  
\overline\pexp\Biggl[ i \int_{-\infty}^{x^{-}} \nts d\xi^{-} 
              \ts\widetilde{A}^{+}({\xi^{-}},\xt) \Biggr],
\label{Udef}
\eeq
where the sense of the path-ordering $\overline{\cal P}$
is such that
\beq
\U(x) \equiv {\bf 1} +
\sum_{j=1}^{\infty} \ts i^j
\int_{-\infty}^{x^{-}} \nts d\xi_1^{-} 
\int_{-\infty}^{\xi_1^{-}} \nts d\xi_2^{-} \ts
\cdots
\int_{-\infty}^{\xi_{j-1}^{-}} \nts d\xi_j^{-} \ts\ts
\widetilde{A}^{+}(\xi_j^{-},\xt)\cdots
\widetilde{A}^{+}(\xi_2^{-},\xt)
\widetilde{A}^{+}(\xi_1^{-},\xt),
\eeq
{\it i.e.}\ the factors are arranged such that $x^{-}$ increases
as the factors of $\widetilde{A}^{+}$ are read from left to right.
Using~(\ref{QCDGaugeTrans}) and~(\ref{Udef}) to obtain the light-cone
gauge expressions yields
\beq
A^{-}(x) = 0
\eeq
and
\beqa
A^{j}(x) &=& 
-i \Bigl[\partial^j\U(x)\Bigr] \Uinv(x) \phantom{\Biggl]}
\cr &=&
\int_{-\infty}^{x^{-}} d\xi^{-} \ts\ts
\U(\xi^{-},\xt)
\Bigl[ \partial^j\widetilde{A}^{+}(\xi^{-},\xt) \Bigr]
\Uinv(\xi^{-},\xt).
\label{QCDsoln}
\eeqa
The Abelian limit may be obtained from Eq.~(\ref{QCDsoln}) by
allowing all quantities to commute.  Equivalently, we may
take only those contributions which are linear in the 
source.%, {\it i.e.}\ by setting $\U$ equal to unity.

It is instructive to recall the features of the solution in
electrodynamics for a point charge moving along the $x^3$ 
axis at the speed of light.  In this situation, Eq.~(\ref{QCDsoln})
reduces to
\beq
A^{j}(x) =
{{e}\over{2\pi}} \ts \Theta(x^{-}) 
{{x^{j}}\over{\xt^2}}.
\label{QEDsoln}
\eeq
The only non-vanishing 
components of the field tensor produced by the vector potential
in~(\ref{QEDsoln}) are
\beq
F^{{+}j}(x) = - {{e}\over{2\pi}} \ts \delta(x^{-}) \ts
{{x^{j}}\over{\xt^2}}.
\label{EB}
\eeq
These components of the field tensor correspond to the transverse
electric and magnetic fields:
the fact that $F^{{-}j} = 0$ indicates that these fields have
equal strength.  There are no longitudinal fields
generated in this limit.
An observer sitting at some fixed position  $(\bt,x^3)$ would 
see no fields except at the instant when the charge
made its closest approach.  At this time a $\delta$-function pulse
of transverse and mutually perpendicular electric
and magnetic fields would be seen.  The magnitude of the
pulse would be proportional to $1/b$.  The (unobservable) 
vector potential,
which was zero before the passage of the charge, takes on
a non-zero value for all times afterward.  Since the fields
vanish at these times, the late-time vector potential may be thought of
as some particular gauge transformation of the vacuum.

The overall features of the above description
continue to hold when we switch to QCD, although the
details are slightly altered by the presence of the path-ordered
exponentials, which introduce a nonlinear dependence on the
source.  The net effect of these factors is
to color-rotate the source in
a complicated fashion.  Nevertheless, the chromoelectric and
chromomagnetic fields are non-zero only at the instant of
closest approach by the charge, and the vector potential
switches from one gauge transform of the vacuum to a different
one at this instant.

%%%%%%%%%%%%%%%%%%%%%%%%%%%%%%%%%%%%%%%%%%%%%%%%%%%%%%%%%%%%%%%%
%%
%%             THE CORRELATION FUNCTION  
%%
%%%%%%%%%%%%%%%%%%%%%%%%%%%%%%%%%%%%%%%%%%%%%%%%%%%%%%%%%%%%%%%%

\subsection{Structure of the Correlation Function}

Having obtained the solution for $\At$ in terms of 
$\rho$ via Eqs.~(\ref{QCDCovarSoln}), (\ref{Udef}) and~(\ref{QCDsoln}), 
we now turn to the question of actually performing the color averaging.
The Gaussian weight~(\ref{RhoWeight}) argued for above leads to
the two-point charge density correlator
\beq
\langle \rho^a(x^{-},\xt) \rho^b(x^{\prime -},\xprimet) \rangle
= \delta^{ab} \ts\mu^2(x^{-}) \ts
\delta(x^{-}-x^{\prime -}) \ts\curlyD(\xt-\xprimet).
\label{RhoRho1}
\eeq
Thus, to compute
the correlation function
$\cfun$, we expand $\At$ in terms of $\rho$ with the
help of Eqs.~(\ref{QCDCovarSoln}), (\ref{Udef}), and~(\ref{QCDsoln}), 
and perform all possible contractions on pairs of $\rho$'s
using~(\ref{RhoRho1}).  At the end of this process, we
resum the series.  The details of this calculation may be found 
in Ref.~\cite{paper9}.  Here, we shall simply quote the result:
\beq
\cfun =
{{\delta^{ab}}\over{N_c}}  \ts
{ 
{\partial_i \partial^{\prime}_j L(\xt-\xprimet)}
\over
{L(\xt-\xprimet)} 
} \ts
\biggl\{
\exp\Bigl[ N_c \chi(x^{-},x^{\prime -}) L(\xt-\xprimet)\Bigr]
- 1 \biggr\}.
\label{MVCorrelator}
\eeq
Eq.~(\ref{MVCorrelator}) depends on two new functions.
The first, $\chi(x^{-},x^{\prime -})$, is the total charge-squared
per unit area at positions to the 
left of the leftmost of $(x^{-},x^{\prime -})$:
\beq
\chi(x^{-},x^{\prime -}) \equiv 
\int_{-\infty}^{\min(x^{-},x^{\prime -})}
d\xi^{-}
\ts  \mu^2(\xi^{-}).
\label{ChiDef}
\eeq
The appearance of such an expression is not surprising when
we recall that the value of the vector
potential depends on whether or not the charge has yet reached
its point of closest approach.  What $\chi(x^{-},x^{\prime -})$
measures is the amount of charge in those layers of the source
which have already passed {\it both}\ of the points which we
are comparing.
Although the range of integration extends to $x^{-}=-\infty$,
in practice this is cut off by the form of $\mu^2(x^{-})$,
which for a pancake-shaped charge distribution, should be
non-zero only in a relatively small range around the position
of the nucleus along the $x^{-}$ axis.

The second new function appearing in~(\ref{MVCorrelator})
is given by
\beqa
L(\xt-\xprimet) \equiv 
\int d^2 \xit \int d^2 \xiprimet \ts
\curlyD(\xit - \xiprimet) 
\Bigl[ G(\xt - \xit) G(\xprimet - \xiprimet)
&-&\half G(\xt-\xit) G(\xt-\xiprimet)
\cr 
&-&\half G(\xprimet-\xit) G(\xprimet-\xiprimet) \Bigr].
\label{Ldef}
\eeqa
Note that this function vanishes when $\xt=\xprimet$.
Thus, at very short distances 
the nonlinear terms in~(\ref{MVCorrelator}) drop out
and the behavior of the
correlation function is the same as in the Abelian
case:
\beq
\lim_{\vert\xt-\xprimet\vert\rightarrow 0}
\cfun =
\delta^{ab}
\chi(x^{-},x^{\prime -}) 
\Bigl[\partial_i \partial_j^\prime L(\xt-\xprimet)\Bigr].
\label{AbelianLimit}
\eeq
In this limit, all of the dependence on the transverse
coordinates is contained in the derivative of $L$:
\beqa
\partial_i \partial_j^{\prime}L(\xt-\xprimet) &=&
\int d^2 \xit \int d^2 \xiprimet \ts
\curlyD(\xit - \xiprimet) 
[\partial_i G(\xt - \xit)] [\partial_j G(\xprimet - \xiprimet)]
\cr \phantom{\Biggl[} & \equiv &
\delta_{ij}\curlyL(\xt-\xprimet)
+ \curlyL_{ij}(\xt-\xprimet).
\label{decomp}
\eeqa
To uniquely specify the decomposition appearing on the second
line of~(\ref{decomp}) we take $\curlyL_{ij}$ to be traceless.
Hence, the only piece of~(\ref{decomp}) which contributes
to the gluon number density is $\delta_{ij}\curlyL(\xt-\xprimet)$.

%%%%%%%%%%%%%%%%%%%%%%%%%%%%%%%%%%%%%%%%%%%%%%%%%%%%%%%%%%%%%%%%
%%
%%              THE ORIGINAL MV RESULT
%%
%%%%%%%%%%%%%%%%%%%%%%%%%%%%%%%%%%%%%%%%%%%%%%%%%%%%%%%%%%%%%%%%

\subsection{Result Assuming Purely Local Charge Density Correlations}

As mentioned earlier, the results presented in 
Ref.~\cite{paper9} utilize a purely local correlator for
the charge density correlation 
function, {\it i.e.}\ Eq.~(\ref{MVtrans}).
Inserting such a $\delta$-function dependence
into~(\ref{Ldef}) produces
\beq
L(\xt-\xprimet) = \gamma(\xt-\xprimet) - \gamma({\bf 0})
\eeq
where, formally,\footnote{The authors of Ref.~\protect\cite{paper9}
write $\gamma(\xt) = {{1}\over{\delt^4}}\delta^2(\xt)$, whereas
we obtain the convolution of ${1\over{\delt^2}}\delta^2(\xt)$ 
with itself.  To demonstrate the equivalence of these two quantities 
requires that we employ a consistent regularization scheme for the 
infrared divergences, for example by doing the entire computation 
in $2+\vareps$ dimensions.}           %%%%%%%%%%%% END OF FOOTNOTE
\beq
\gamma(\xt-\xprimet) =
\int d^2\xit \ts G(\xt - \xit) G(\xprimet - \xit).
\label{GammaFormal}
\eeq
This function contains a quadratic infrared divergence,
on top of the arbitrary length scale introduced by the 
logarithmic divergence of $G(\xt - \xprimet)$ itself.
Fortunately, it is only the
combination $\gamma(\xt-\xprimet) - \gamma({\bf 0})$ which
appears in the correlation
function.
Thus, the  quadratic part of the divergence drops out.
There is still a logarithmic dependence on the infrared
cutoff.
The authors of Ref.~\cite{paper9} 
take\footnote{This expression for $\gamma$ differs from
the one given in Eq. (26) of Ref.~\protect\cite{paper9} by a 
factor of 2.   This is in agreement with the requirement
$\delt^2\gamma(\xt) = {{1}\over{\delt^2}}\delta^2(\xt)$.}
\beq
\gamma(\xt-\xprimet) 
- \gamma({\bf 0}) =
{ {1}\over{16\pi} }
\vert\xt-\xprimet\vert^2
\ln\Bigl( \vert\xt-\xprimet\vert^2 \LQCD^2\Bigr)
\label{Gamma}
\eeq
where they have chosen on physical grounds to set this cutoff
to $\LQCD^{-1}$.
The derivative of~(\ref{Gamma}) is
\beqa
\partial_i \partial^{\prime}_j \gamma(\xt - \xprimet)
= &-&{{\delta_{ij}}\over{8\pi}} \ts
\biggl[
\ln\Bigl( \vert\xt-\xprimet\vert^2 \LQCD^2\Bigr) + 2 \biggr]
\phantom{\Biggl[} \cr 
&+& {{1}\over{4\pi}}
\Biggl[  {{1}\over{2}} \delta_{ij} 
- { {(x-x')_i (x-x')_j} \over { \vert\xt-\xprimet\vert^2 } }
\Biggr].
\label{dGamma}
\eeqa
Note that the terms on the second line of~(\ref{dGamma}) are traceless,
and do not contribute to the gluon number density~(\ref{GluonDensity}).

Because $\gamma(\xt-\xprimet)$ is formally
the inverse of $\delt^4$, when we form the trace 
$\partial_i \partial^{\prime}_i \gamma(\xt - \xprimet)$
that appears in~(\ref{MVCorrelator}), we should obtain
the inverse of $\delt^2$ [{\it i.e.}\ Eq.~(\ref{GreenFunction})].
Eq.~(\ref{dGamma}) satisfies this expectation, since the
extra ``$+2$'' appearing in the first line may be absorbed into
the arbitrary scale.  
What this means is that in a strict evaluation of 
Eq.~(\ref{MVCorrelator}), if we want the numerator factor
$\partial_i \partial_j^{\prime} \gamma(\xt-\xprimet)$
to consist of a single logarithmic term, then that logarithm
must contain a different scale from the one used in $L(\xt-\xprimet)$
itself.
On the other hand, 
in the qualitative treatment of the infrared employed in
Ref.~\cite{paper9}, both logarithms are assumed to have
the same scale ($\sim \LQCD$), producing
\beq
\TRcfun =
{ {4(N_c^2-1)} \over {N_c\vert\xt-\xprimet\vert^2}} \ts
\biggl[ 1 - \Bigl( \vert\xt
            -\xprimet\vert^2 \LQCD^2\Bigr)^{(N_c/16\pi)
                \chi(x^{-},x^{\prime -})
                \vert\xt-\xprimet\vert^2}
\biggr].
\label{RunAway}
\eeq
According to Eq.~(\ref{GluonDensity}), the gluon number
density depends on the Fourier transform of~(\ref{RunAway}).
Unfortunately, there is a problem with the correlation function
appearing in Eq.~(\ref{RunAway}):  it diverges like $(x^2)^{x^2}$
at large distances.  Thus, {\it the required Fourier transform
does not exist for any value of the momentum $\qt$!}
In Ref.~\cite{paper10},
the $\xt-\xprimet$ integration 
is cut off at $\LQCD^{-1}$,
on the grounds that the approximations made in 
obtaining~(\ref{RunAway}) are not valid at large distances.
While physically reasonable, such an approach 
can only be applied qualitatively.

It is apparent from the discussion 
leading to Eq.~(\ref{AbelianLimit}) that Eq.~(\ref{RunAway})
nevertheless correctly reproduces
the Abelian limit at short enough distances.
Explicitly,
\beq
\lim_{\vert\xt-\xprimet\vert\rightarrow 0}
\TRcfun =
{{N_c^2-1}\over{4\pi}} \ts
\chi(x^{-},x^{\prime -}) \ts
\ln\Bigl( \vert\xt-\xprimet\vert^2 \LQCD^2\Bigr).
\label{AbelianCorrel}
\eeq
The Fourier transform of this function goes like $1/\qt^2$.

At larger distances, $ 0 \ll \vert \xt -\xprimet\vert \ll \LQCD^{-1}$,
and for large enough nuclei, the second term of~(\ref{RunAway})
becomes negligible.  Under these circumstances, the correlation
function has a Fourier transform which behaves as $\ln\qt^2$.
To see how large the nucleus has to be to have such a regime,
consider the function $f(\xi) \equiv \xi^{\Omega\xi}$.
It has a minimum at $\xi = e^{-1}$, independent of $\Omega$,
and a minimum value of $e^{-\Omega/e}$.  Thus, to legitimately
drop the second term of~(\ref{RunAway}) we should have
\beq
\Omega = {{N_c \chi}\over{16\pi\LQCD^2}} \gg e.
\label{ReallyBig}
\eeq
This constraint is difficult to satisfy:  in fact, for experimentally
accessible nuclei, such as uranium, we actually have $\Omega < 1$.
Just to obtain $\Omega \sim e$ requires $\chi \sim 45 \LQCD^2$.
For the model described in Sec.~\ref{EXAMPLE}, this implies
a nucleus with of order $10^5$ nucleons.  
This is a somewhat larger nucleus than is required simply to
make the color charge density large enough for the approximations
described at the beginning of this section to be reasonable
[compare Eqs.~(\ref{qtCondition}) and~(\ref{ReallyBig})].
So while the authors of
Refs.~\cite{paper1,paper2,paper3,paper4,paper9,paper10}
point out that realistic nuclei are only marginally large enough,
one must go to even larger nuclei to justify dropping the
second term in~(\ref{RunAway}).

Our point is, that the range of momenta where the picture
described at the beginning of this section and in
Refs.~\cite{paper9,paper10,paper35} is applicable
is very narrow at best.
While it is unlikely that anything short of a full non-perturbative
solution to QCD can produce reliable predictions for the
region $\vert\xt-\xprimet\vert \agt \LQCD^{-1}$, we
may reasonably hope that 
if we construct a model which is
well-behaved in this region, then we will be able to trust
it closer to the boundary point 
$\vert\xt-\xprimet\vert = \LQCD^{-1}$ than if we use Eq.~(\ref{RunAway}),
with its unphysical large-distance behavior.
In the next section we will show that the source of the difficulties
encountered in Eq.~(\ref{RunAway}) is the inconsistency of the
charge density correlator~(\ref{RhoRho1}) with the intuitive
picture of color neutral nucleons.  When the correlator is
altered to reflect this expectation, we find that the resulting
gluon distribution is well-behaved and completely infrared finite.

%%%%%%%%%%%%%%%%%%%%%%%%%%%%%%%%%%%%%%%%%%%%%%%%%%%%%%%%%%%%%%%%
%%%%%%%%%%%%%%%%%%%%%%%%%%%%%%%%%%%%%%%%%%%%%%%%%%%%%%%%%%%%%%%%
%%%%%%%%%%%%%%%%%%%%%%%%%%%%%%%%%%%%%%%%%%%%%%%%%%%%%%%%%%%%%%%%
%%%%%%
%%%%%%       COLOR NEUTRALITY -- THE FIX TO ALL THAT AILS
%%%%%%
%%%%%%%%%%%%%%%%%%%%%%%%%%%%%%%%%%%%%%%%%%%%%%%%%%%%%%%%%%%%%%%%
%%%%%%%%%%%%%%%%%%%%%%%%%%%%%%%%%%%%%%%%%%%%%%%%%%%%%%%%%%%%%%%%
%%%%%%%%%%%%%%%%%%%%%%%%%%%%%%%%%%%%%%%%%%%%%%%%%%%%%%%%%%%%%%%%

\section{Color Neutrality}\label{NEUTRALITY}

As described in the previous section, the \MV\ model
employs the Gaussian weight~(\ref{RhoWeight}) for performing
ensemble averages.  Thus,
the ensemble average of the charge denisty vanishes~\cite{paper1}:
\beq
\langle \rho^a(x^{-},\xt) \rangle = 0.
\label{ZeroAvg}
\eeq
This says that at any given point, when averaging over all of the
different nuclei within the ensemble, we see no net charge.
Eq.~(\ref{ZeroAvg}) is gauge-invariant in the sense that
if we perform the {\it same}\ gauge transformation on the
fields present in all of the nuclei, the ensemble average
remains unchanged.

Let us consider the situation inside a single hadronic state,
which we will denote by $\vert h \rangle$.  This state could
consist of a single baryon or meson, or it could be some
collection of hadrons.  For such a state,
the charge density at a given point is not necessarily zero:
\beq
\langle h \vert \ts \rho^a(x^{-},\xt) \ts \vert h \rangle \ne 0.
\eeq
However, if we sum over all of the space occupied by the hadronic
state, we should obtain zero:
\beq
\langle h \vert 
\int dx^{-} d^2\xt \ts
\rho^a(x^{-},\xt) \ts \vert h \rangle
= 0.
\label{ColorNeutral}
\eeq
Eq.~(\ref{ColorNeutral}) says that any physically observable
hadronic state is a color singlet.  
Since this is true for {\it all}\ of the
hadronic states in our Hilbert space,
we may use the polarization trick to generalize~(\ref{ColorNeutral})
to 
\beq
\langle h \vert 
\int dx^{-} d^2\xt \ts
\rho^a(x^{-},\xt) \ts \vert h' \rangle
= 0.
\label{ColorNeutral2}
\eeq
Since we have phrased Eq.~(\ref{ColorNeutral2}) in terms
of color singlet physical states, it must hold independent of the
gauge chosen for $\rho^a(x^{-},\xt)$.

Now consider the bilinear combination
\beq
{\cal Z} \equiv 
\langle h \vert 
\int dx^{-} d^2\xt \int dx^{\prime -} d^2\xprimet 
\ts\rho^a(x^{-},\xt)
\rho^b(x^{\prime -},\xprimet) \ts\vert h \rangle.
\eeq
On one hand, by inserting a complete set of hadronic states
and applying Eq.~(\ref{ColorNeutral2}) we conclude that
${\cal Z}$ vanishes.
Since the ensemble utilized in the \MV\
model consists of a specific collection of hadronic states
({\it i.e.}\ nuclei with some fixed number of nucleons),
we further conclude that the ensemble average of ${\cal Z}$ vanishes.
On the other hand, we may also
compute the ensemble average of ${\cal Z}$
with the help of the correlation function~(\ref{RhoRho1}):
\beqa
\langle {\cal Z} \rangle 
&=& \int dx^{-} d^2\xt \int dx^{\prime -} d^2\xprimet 
\ts\langle \rho^a(x^{-},\xt)\rho^b(x^{\prime -},\xprimet) \rangle
\cr &=& \delta^{ab}
\int d^2\sigmat
\int_{-\infty}^{\infty} dx^{-} \ts\mu^2(x^{-}) 
\int d^2 \deltat \ts \curlyD(\deltat).
\label{TotChar}
\eeqa
The integral over $\sigmat$ is just the (non-zero) transverse area
of the nucleus.
Since $\mu^2(x^{-})$ is positive (by definition),
the only way to obtain a vanishing ensemble average is to require
that $\curlyD$ satisfy
\beq
\int d^2\deltat \ts\curlyD(\deltat) = 0.
\label{Dint}
\eeq
This relation is not true for the choice~(\ref{MVtrans})
employed in Ref.~\cite{paper9}.
Hence we conclude that this choice
is incompatible with the expectation of color-neutral
nucleons.  
In fact, this violation of color-neutrality
contributes to the poor infrared behavior of the \MV\
correlation function~(\ref{RunAway}).

Although the function $\curlyD(\deltat)$ is itself not gauge-invariant
(its transformation properties are determined by the condition
that the Gaussian weight be gauge-invariant),
the color-neutrality condition~(\ref{Dint}) was, nonetheless,
derived in a gauge-invariant fashion, with the help of
Eq.~(\ref{ColorNeutral2}).  The constraint represented
by Eq.~(\ref{Dint}) reflects the physical observation that the
nucleus does not carry a net color charge.

The fact that a color neutral charge distribution 
posseses an intrinsic scale explains why the infrared
behavior of the gluon number density is improved by 
enforcing~(\ref{Dint}).
We may see the
roots of this result by
considering Eq.~(\ref{QCDCovarSoln}),
the expression for the vector potential in the covariant
gauge.  As pointed out earlier, it depends on an arbitrary
length scale $\lambda$, because of the infrared divergence
in the inverse of $\delt^2$.  If we integrate both 
sides of~(\ref{QCDCovarSoln}) over $x^{-}$ and
take the expectation value for the hadronic state $\vert h \rangle$
[so that
we may apply~(\ref{ColorNeutral}) to the
term involving $\lambda$] we obtain
\beqa
\langle h \vert 
\int_{-\infty}^{\infty} dx^{-} \ts 
\widetilde{A}^{+}(x^{-},\xt) \ts \vert h \rangle 
&=& \langle h \vert
\int_{-\infty}^{\infty} dx^{-} \int d^2\xprimet \ts 
G(\xt-\xprimet) \ts \rho(x^{-},\xprimet) \ts\vert h \rangle 
\cr & = &
{ {1}\over{4\pi} }
\langle h \vert
\int_{-\infty}^{\infty} dx^{-} \int d^2\xprimet \ts 
\ln (\xt-\xprimet)^2 \ts
\rho(x^{-},\xprimet) \ts\vert h \rangle.
\label{ScaleReplaced}
\eeqa
Instead of an arbitrary scale, the logarithm is supplied with
the scale intrinsic to $\rho$, that is, the scale over 
which~(\ref{ColorNeutral}) is true.

Therefore, let us employ a charge-density correlation function
$\curlyD$ which satisfies~(\ref{Dint}).
Furthermore, let us make the reasonable assumption that $\curlyD$
is rotationally invariant.  Under these circumstances,
the function $L(\xt-\xprimet)$
is completely free of infrared difficulties!
In this situation, we find that~(\ref{Ldef}) may
be simplified to
\beqa
L(\xt-\xprimet) &=& 
{1\over{16\pi}} \ts
\int d^2\deltat \ts\ts
\curlyD(\deltat) \ts
\Bigl[
(\xt-\xprimet-\deltat)^2 \ln (\xt-\xprimet-\deltat)^2
- \deltat^2 \ln \deltat^2
\Bigr].
\label{Lgen}
\eeqa
(see Appendix~\ref{SIGMA}).
In addition, the expression for $\curlyL$ becomes
\beq
\curlyL(\xt-\xprimet) =
{{-1}\over{8\pi}}
\int d^2\deltat \ts\ts
\curlyD(\deltat) \ts
\ln(\xt-\xprimet-\deltat)^2 
\label{curlyLgen}
\eeq
In the next section we will study the gluon number density
implied by these two equations.

At various points in our subsequent discussion it will prove
convenient to take the form of $\curlyD(\deltat)$ to be
\beq
\curlyD(\deltat) = \delta^2(\deltat) - \Ihat(\deltat).
\label{TwoTerms}
\eeq
In Sec.~\ref{ASYMPT} we will see that to reproduce the
Abelian limit at short distances, $\curlyD(\deltat)$
must contain a delta function.  Likewise, such a delta function
is argued for in Refs.~\cite{paper1,paper2,paper3,paper4,paper9}.
Such a term is consistent with nucleons that contain
pointlike quarks.  In addition, we will see that such a structure
arises naturally in the model examined in Sec.~\ref{EXAMPLE}.
To satisfy Eq.~(\ref{Dint}), it is clear that the other
function appearing in~(\ref{TwoTerms}) must obey
\beq
\int d^2\deltat \ts\Ihat(\deltat) = 1.
\label{IHATint}
\eeq

We conclude this section by quantifying the connection between
$\mu^2(x^{-})$
and the total color charge per unit area.
We begin by defining $\CHI_\infty$ to be the
integral of $\mu^2(x^{-})$ over all $x^{-}$:
\beq
\CHI_\infty \equiv 
\int_{-\infty}^{\infty}
dx^{-}
\ts  \mu^2(x^{-}).
\label{CHIinf}
\eeq
Then, the trace of Eq.~(\ref{TotChar}) reads
\beq
\int dx^{-} d^2\xt \int dx^{\prime -} d^2\xprimet 
\ts\langle \rho^a(x^{-},\xt)\rho^a(x^{\prime -},\xprimet) \rangle
= (N_c^2 - 1) \CHI_\infty \pi R^2
\int d^2 \deltat \ts \curlyD(\deltat).
\label{TotChar2}
\eeq
Imagine that, instead of a color-neutral distribution of charge 
satisfying~(\ref{Dint}),  we consider a single quark.
Then, it would indeed be true that 
$\curlyD(\deltat) = \delta^2(\deltat)$,
and the charge-squared of that quark should be given by
\beq
g^2 C_F
\equiv  
(N_c^2 - 1) \CHI_\infty \pi R^2
\label{RealChi}
\eeq
where $C_F = (N_c^2-1)/2N_c$ is the Casimir factor for $SU(N)$.
Thus, we see that the color charge squared per unit area is
$(N_c^2-1)\CHI_\infty$, and that the precise
meaning of $\mu^2(x^{-})$ is actually the charge squared per 
unit area per unit thickness divided by $N_c^2-1$.

%%%%%%%%%%%%%%%%%%%%%%%%%%%%%%%%%%%%%%%%%%%%%%%%%%%%%%%%%%%%%%%%
%%%%%%%%%%%%%%%%%%%%%%%%%%%%%%%%%%%%%%%%%%%%%%%%%%%%%%%%%%%%%%%%
%%%%%%%%%%%%%%%%%%%%%%%%%%%%%%%%%%%%%%%%%%%%%%%%%%%%%%%%%%%%%%%%
%%%%%%
%%%%%%            GLUON NUMBER DENSITY
%%%%%%
%%%%%%%%%%%%%%%%%%%%%%%%%%%%%%%%%%%%%%%%%%%%%%%%%%%%%%%%%%%%%%%%
%%%%%%%%%%%%%%%%%%%%%%%%%%%%%%%%%%%%%%%%%%%%%%%%%%%%%%%%%%%%%%%%
%%%%%%%%%%%%%%%%%%%%%%%%%%%%%%%%%%%%%%%%%%%%%%%%%%%%%%%%%%%%%%%%
 
\section{Gluon Number Density}\label{GLUONNUM}

We begin our discussion of the properties of the gluon number
density resulting from a nucleus consisting of color-neutral
nucleons by inserting~(\ref{MVCorrelator}) into the
master formula~(\ref{GluonDensity}).  Employing sum and difference
variables for the transverse integrations we obtain
\beqa
\gnum =
{ {q^{+}}\over{2\pi^3} }
\ts {{N_c^2-1}\over{N_c}} 
\int d^2\sigmat   &&
\int d^2\deltat \ts
e^{i\qt\cdot\deltat}\ts
{{\curlyL(\deltat)}\over{L(\deltat)}} 
\cr \times
\int_{-\infty}^{\infty}  && \nts\nts dx^{-} \nts
\int_{-\infty}^{\infty} \nts\nts dx^{\prime -} 
e^{-iq^{+}(x^{-}-x^{\prime -})} 
\biggl\{
\exp\Bigl[ N_c \chi(x^{-},x^{\prime -}) \ts
    L(\deltat)\Bigr] - 1
\biggr\}.
\label{GlueStart}
\eeqa
The result of the $\sigmat$ integration appearing in 
Eq.~(\ref{GlueStart})
should be interpreted as $\pi R^2$, the transverse area of the
Lorentz-contraced nucleus.\footnote{See Appendix~\protect\ref{INTEG1}
for a more careful discussion of this point.}
We would like to perform the
transverse integrations.
All of the longitudinal dependence
is contained in the function $\chi(x^{-},x^{\prime -})$.

Recall that within the Gaussian ansatz~(\ref{RhoWeight}), the
function $\chi(x^{-},x^{\prime -})$ is given by
\beq
\chi(x^{-},x^{\prime -}) \equiv 
\int_{-\infty}^{\min(x^{-},x^{\prime -})}
d\xi^{-}
\ts  \mu^2(\xi^{-}).
\label{ChiDefAgain}
\eeq
In order to obtain a more useful expression for $\chi$,
let us define
\beq
\CHI(x^{-}) \equiv 
\int_{-\infty}^{x^{-}}
d\xi^{-}
\ts  \mu^2(\xi^{-}).
\label{CHIdef}
\eeq
Then, we see that Eq.~(\ref{ChiDefAgain}) has the form
\beq
\chi(x^{-},x^{\prime -}) = \CHI(x^{-})\ts\Theta(x^{\prime -}-x^{-})
                   + \CHI(x^{\prime-})\ts\Theta(x^{-}-x^{\prime -}).
\label{ChiForm}
\eeq
For convenience, let us assume that the entire nucleus is localized
around some positive value of $x^{-}$.  Then, we may insert 
Eq.~(\ref{ChiForm}) into~(\ref{GlueStart}) and do one of the
longitudinal integrals in each term, with the aid of the
identity
\beq
\int_{-\infty}^{\infty} dx \ts 
e^{i(k+i\vareps)x} \Theta(x-y) = 
{ {ie^{i(k+i\vareps)y}} \over {k+i\vareps} },\qquad y>0.
\label{Careful}
\eeq
The result is
\beqa
\gnum =
{ {q^{+}}\over{2\pi^3} }
\ts {{N_c^2-1}\over{N_c}} 
\ts \pi R^2 \ts  &&
\int d^2\deltat \ts
e^{i\qt\cdot\deltat}\ts
{{\curlyL(\deltat)}\over{L(\deltat)}} 
\cr \times &&
{{ 2\vareps }\over{(q^{+})^2 + \vareps^2} } 
\int_{-\infty}^{\infty} \nts\nts dx^{-} 
e^{-2\vareps x^{-} }
\biggl\{
\exp\Bigl[ N_c \ts\CHI(x^{-}) \ts
    L(\deltat)\Bigr] - 1
\biggr\}.
\label{Surprise}
\eeqa
To get a finite nonvanishing result, the remaining integral must
produce exactly one inverse power of $\vareps$.  To see how this
comes about, let us integrate the
key factors of~(\ref{Surprise}) by parts:
\beqa
{{ 2\vareps }\over{(q^{+})^2 + \vareps^2} } 
\int_{-\infty}^{\infty}  \nts\nts dx^{-} 
e^{-2\vareps x^{-}} \ts &&
\biggl\{
\exp\Bigl[ N_c \ts\CHI(x^{-}) \ts
    L(\deltat)\Bigr] - 1
\biggr\} =
\cr &&
-{  
{ e^{-2\vareps x^{-}} } 
\over
{ (q^{+})^2 + \vareps^2 }
}
\biggl\{
\exp\Bigl[ N_c \ts\CHI(x^{-}) \ts
    L(\deltat)\Bigr] - 1
\biggr\} 
\Biggl\vert_{-\infty}^{\infty}
\cr && +
\int_{-\infty}^{\infty} dx^{-} \ts
N_c \ts\mu^2(x^{-}) L(\deltat) 
\exp\Bigl[ N_c \ts\CHI(x^{-}) \ts L(\deltat)\Bigr]
{  
{ e^{-2\vareps x^{-}} } 
\over
{ (q^{+})^2 + \vareps^2 }
}.
\eeqa
Physically, we know that  for $x^{-} \rightarrow -\infty$,
$\CHI(x^{-})$  vanishes identically, whereas for 
$x^{-} \rightarrow \infty$,
$\CHI(x^{-})$ becomes proportional to the total charge squared per unit 
area.\footnote{If $\CHI(x^{-})$ grows at asymptotically large
values of $x^{-}$, then the total charge squared per unit area
is unbounded and the integral~(\protect\ref{Surprise}) diverges.}
Thus, the surface term does not contribute.
We may let $\vareps\rightarrow 0$ in the remaining integral
without mishap, since the remaining longitudinal integration
will turn out to be finite.  
Therefore, we find that
\beq
\gnum =
{{N_c^2-1}\over{2\pi^3}} 
\ts \pi R^2 \ts  
{{1}\over{q^{+}}} 
\int d^2\deltat \ts
e^{i\qt\cdot\deltat}\ts
\curlyL(\deltat)
\int_{-\infty}^{\infty} \nts\nts dx^{-} 
\mu^2(x^{-})
\exp\Bigl[ N_c \ts\CHI(x^{-}) \ts L(\deltat)\Bigr] .
\label{Glue3}
\eeq
The definition~(\ref{CHIdef}) for $\CHI$ implies 
that $\mu^2(x^{-}) = d\CHI/dx^{-}$.  Hence, we may perform the
remaining longitudinal integration to obtain the result
\beq
\gnum =
{{N_c^2-1}\over{N_c}} 
\ts { {\pi R^2}\over{2\pi^3} }
\ts {{1}\over{q^{+}}} 
\int d^2\deltat \ts
e^{i\qt\cdot\deltat}\ts
{{\curlyL(\deltat)}\over{L(\deltat)}} 
\biggl\{
\exp\Bigl[ N_c \ts\CHI_\infty L(\deltat)\Bigr] - 1
\biggr\}.
\label{GlueEnd}
\eeq
Remarkably, the gluon number density goes
like $1/q^{+}$ independent of the longitudinal charge profile!
The only nuclear dependence is on the total charge squared
per unit area via $\CHI_\infty$.
This conclusion depends upon the fact that the Gaussian from
chosen in~(\ref{RhoWeight}) is local in $x^{-}$.  To generate a
dependence other than $1/q^{+}$ apparently requires 
non-trivial correlations in $x^{-}$.  As pointed out in 
Ref.~\cite{paper9}, a complete renormalization group analysis 
of the longitudinal dependence would generate correlations
in $x^{-}$, although including such correlations was beyond
the scope of that paper.
Since the derivation leading to the form of
the correlation function given in Eq.~(\ref{MVCorrelator}) 
depends crucially on having locality in $x^{-}$, it is clear
that the inclusion of such correlations is a non-trivial problem.

%%%%%%%%%%%%%%%%%%%%%%%%%%%%%%%%%%%%%%%%%%%%%%%%%%%%%%%%%%%%%%%%
%%
%%            ASYMPTOTIC BEHAVIOR
%%
%%%%%%%%%%%%%%%%%%%%%%%%%%%%%%%%%%%%%%%%%%%%%%%%%%%%%%%%%%%%%%%%
 
\subsection{Transverse Asymptotic Behavior}\label{ASYMPT}

The asymptotic behavior at short or long transverse
distances of the 
gluon number density~(\ref{GlueEnd})
is governed by the behavior indicated by Eqs.~(\ref{Lgen}) 
and~(\ref{curlyLgen}) for the functions $L(\xt)$
and $\curlyL(\xt)$.
Consider first the situation at short distances.
It is straightforward to demonstrate that for small $x$
\beq
L(\xt) \sim 
\cases{ \xt^2, &
if $\curlyD(\deltat)$ is smooth at $\Delta=0$;\cr
\xt^2 \ln \xt^2,
& if $\curlyD(\deltat)$ contains $\delta^2(\deltat)$.\cr}
\label{plainLshort}
\eeq
Likewise, the short-distance form of $\curlyL$ is 
\beq
\curlyL(\xt) \sim 
\cases{ const., &
if $\curlyD(\deltat)$ is smooth at $\Delta=0$;\cr
\ln \xt^2,
& if $\curlyD(\deltat)$ contains $\delta^2(\deltat)$.\cr}
\label{curlyLshort}
\eeq
In either case, it follows that
\beq
\lim_{\xt\rightarrow {\bf 0}}
    L(\xt) \ts \curlyL(\xt) = 0.
\label{ImpliesSumRule}
\eeq
This observation will be used in the next subsection
where we will derive a sum rule for the integral
of the gluon number density over $\qt$.  
According to~(\ref{curlyLshort}), if we want to
reproduce the Abelian limit at large $\qt$ with a
gluon number density which goes like $1/\qt^2$,
$\curlyD(\deltat)$ must contain $\delta^2(\deltat)$.
Otherwise, the necessary logarithmic divergence in the
correlation function at short distances will not occur,
and the gluon number density will fall off faster than 
in an Abelian theory.

The situation at long distances is a bit more complicated.
For a rotationally symmetric correlation function $\curlyD$,
the color neutrality condition~(\ref{Dint}) may be rewritten
as
\beq
\int_{0}^{\infty} d\Delta \ts \Delta \ts \curlyD(\Delta) = 0.
\label{DintRotSym}
\eeq
Because this integral is convergent, at large distances
we have either
$\curlyD(\Delta)\sim 1/\Delta^{2+p}$ where $p > 0$
(and could be non-integral), or else $\curlyD(\Delta)$ 
oscillates.\footnote{If $\curlyD(\Delta)$ falls off
faster than a power of $\Delta$,  then the infrared behavior
is even better than described in this section.  The model described
in Sect.~\protect\ref{EXAMPLE} provides two examples of this type 
of correlator.}
We will not consider the second possibility any
further, as it 
corresponds to onion-like nucleons, with successive
layers of alternating color charge all the way to infinity.

For positive
$p$, we find from~(\ref{Lgen}) and~(\ref{curlyLgen})
that
\beq
L(\xt) \sim 
\cases{ \displaystyle{{\sgn\curlyD}\over{p^2(2-p)^2}}
                     \ts x^{2-p}, & if $p\ne 2$;\cr
\phantom{x} & \cr
\displaystyle{{\sgn\curlyD}\over{16}} \ln^2 \xt^2, & if $p=2$,\cr}
\label{plainLlong}
\eeq
where ``$\sgn\curlyD$'' stands for the sign of $\curlyD$ at
large distances.
The long-distance form of $\curlyL$ is 
\beq
\curlyL(\xt) \sim -{{\sgn\curlyD}\over{2p^2}} {{1}\over{x^p}}.
\label{curlyLlong}
\eeq
Combining these expressions into the integrand appearing 
in~(\ref{GlueEnd}) we see that
\beq
{{\curlyL(\deltat)}\over{L(\deltat)}} 
\biggl[
\exp\Bigl\{ N_c \ts\CHI_\infty L(\deltat)\Bigr\} - 1
\biggr]
\sim \cases{
\displaystyle{{(2-p)^2}\over{2 \Delta^2}}
\Biggl\{ 1 - \exp\biggl[
\displaystyle{{ N_c \CHI_\infty \sgn\curlyD }\over{p^2(2-p)^2}}
\Delta^{2-p}
\biggr]
\Biggr\}, & if $p\ne 2$;\cr
\phantom{x} & \cr
\displaystyle{{2}\over{\Delta^2 \ln^2\Delta^2 }} 
\Biggl\{ 1 - \exp\biggl[ 
\displaystyle{{ N_c \CHI_\infty \sgn\curlyD }\over{16}}
             \ln^2\Delta^2 \biggr]
\Biggr\}, & if $p=2$.
}
\label{cfunAsym}
\eeq
Since $\CHI_\infty$ is proportional to the 
charge-{\it squared} per unit area, it is positive.
Thus, we see that
whether or not the gluon number density remains bounded
will be determined by the sign of $\curlyD(\xt)$ at large distances 
in addition to the value of $p$.

First consider the situation when $\sgn\curlyD = -1$.
Such a correlation function corresponds to the notion
of a charge which is screened at large distances.
If $0<p<2$, then  the exponent in~(\ref{cfunAsym}) goes
to $-\infty$ and we are left with an integrand
which falls like $1/\Delta^2$.  This corresponds to a gluon number
density which grows like $\ln\qt^2$ at low momenta.
For $p>2$, the exponent goes to zero, and we end up with an
integrand which falls like $1/\Delta^p$.  The
Fourier transform of such a function goes like $q^{p-2}$
at small momenta.  Thus, we conclude that the gluon number
density is constant at small $\qt$ in this situation.
Finally, if $p=2$, the leading behavior 
of~(\ref{cfunAsym}) gives an integrand which
falls like $(\Delta\ln\Delta)^{-2}$, intermediate between the two
previous cases.
Thus, when $\sgn\curlyD = -1$, we are guaranteed
that the Fourier transform appearing in the gluon
number density exists.

On the other hand, when $\sgn\curlyD = +1$, we encounter
difficulties for $0 < p \le 2$.  In this situation, the
integrand blows up at large distances.
In order to have a physically sensible gluon number density
when $\sgn\curlyD = 1$, we must have $p>2$, in which case
we once again have an integrand which falls like $1/\Delta^p$.

In Ref.~\cite{paper2}, McLerran and Venugopalan argued
that for $\qt\rightarrow 0$, the gluon number density should
approach a constant.  However, as the authors of Ref.~\cite{paper9}
correctly observe, the subtleties associated
with using a $\delta$-function for the longitudinal dependence
source were not correctly taken into account in Ref.~\cite{paper2}.
Consequently,
these authors claim that the correct $\qt\rightarrow 0$ dependence
is $\ln\qt^2$.  In our refined treatment, where the nucleons
are forced to obey color-neutrality,
we see that either behavior is possible, depending upon the 
form of the two-point charge density correlation function at 
large separations.  If $\curlyD(\Delta)$ falls off faster 
than $1/\Delta^4$ at large distances, then the gluon number 
density approaches a constant as $\qt\rightarrow 0$, independent 
of the sign of $\curlyD$ in this region.  
On the other hand, if $\curlyD(\Delta)$ falls
off faster than $1/\Delta^2$ but more slowly than $1/\Delta^4$,
we obtain a gluon number density which grows like $\ln\qt^2$
for $\qt\rightarrow 0$.  In this case, we must have $\sgn\curlyD = -1$
to obtain a physically sensible result.

Our gluon number density possesses
two distinct noteworthy features at small transverse momenta.
The first, as described in the previous paragraph, is the
softening of the $\qt$ dependence from $1/\qt^2$.
This softening is present for all values of
$\CHI_\infty$, and it even occurs in an Abelian theory!  
As we have already noted, the Abelian expression for the
correlation function is simply proportional to $\curlyL(\deltat)$.
Hence, from Eq.~(\ref{curlyLlong}) we see that even the long-distance
Abelian correlator possesses the power-law behavior required to
slow the growth of the distribution function to less than $1/\qt^2$
at small transverse momenta.  The physical interpretation
of this result is simple:  a color-neutral
system of size $a$ is an
inefficient generator of radiation with wavelengths $\lambda \gg a$.
At the end of the next section we will discuss the second
noteworthy feature of our results:
our distribution functions depend on $\CHI_\infty$ at low $\qt$
in a manner consistent with
the picture of gluon recombination at high densities
envisioned by Gribov, Levin and Ryskin~\cite{GLR}.

%%%%%%%%%%%%%%%%%%%%%%%%%%%%%%%%%%%%%%%%%%%%%%%%%%%%%%%%%%%%%%%%
%%
%%           TRANSVERSE MOMENTUM SUM RULE
%%
%%%%%%%%%%%%%%%%%%%%%%%%%%%%%%%%%%%%%%%%%%%%%%%%%%%%%%%%%%%%%%%%
 
\subsection{Transverse Momentum Sum Rule}\label{SUMRULE}

Now that we have established the conditions under which the
Fourier transform appearing  in~(\ref{GlueEnd}) is
well-defined, let us see more explicitly how the
non-Abelian corrections alter the transverse
momentum dependence of the lowest order gluon number
density.   In the Abelian limit, Eq.~(\ref{GlueEnd}) becomes
\beq
\gnum\Biggl\vert_{{\rm lowest}\atop{\rm order}} =
(N_c^2-1)
\ts { {\CHI_\infty\pi R^2}\over{2\pi^3} }
\ts {{1}\over{q^{+}}} 
\int d^2\deltat \ts
e^{i\qt\cdot\deltat}\ts
\curlyL(\deltat).
\label{F0def}
\eeq
Let us subtract~(\ref{F0def}) from~(\ref{GlueEnd}) and
integrate over all transverse momenta.  The only $\qt$-dependence
contained in either expression is $e^{i\qt\cdot\deltat}$:
hence, the $\qt$ integration produces $4\pi^2\delta^2(\deltat)$.
This allows us to trivially perform the $\deltat$ integration,
with the result
\beqa
\int d^2\qt \ts
\Biggl\{ 
\gnum\Biggl\vert_{{\rm all}\atop{\rm orders}} \nts\nts
- && \ts \gnum\Biggl\vert_{{\rm lowest}\atop{\rm order}} 
\Biggr\}
= \phantom{\Biggl[} \cr \phantom{\Biggl[^{[}} 2 \ts &&
{{N_c^2-1}\over{N_c}} 
\ts {{R^2}\over{q^{+}}} 
\lim_{\deltat \rightarrow {\bf 0}} 
{{\curlyL(\deltat)}\over{L(\deltat)}} \ts
\biggl\{ 
\exp \Bigl[ N_c \CHI_\infty L(\deltat)\Bigr] 
 - 1 - N_c \CHI_\infty L(\deltat)
\biggr\}.
\eeqa
Since the position-space function $L(\deltat)$ vanishes at
zero separation, we may evaluate the limit by
expanding the exponential, producing
\beq
\int d^2\qt \ts
\Biggl\{ 
\gnum\Biggl\vert_{{\rm all}\atop{\rm orders}} \nts\nts
-  \ts \gnum\Biggl\vert_{{\rm lowest}\atop{\rm order}} 
\Biggr\}
= 
 N_c(N_c^2-1) 
\CHI_\infty^2 R^2
\ts {{1}\over{q^{+}}} 
\lim_{\deltat \rightarrow {\bf 0}} 
    L(\deltat) \ts \curlyL(\deltat).
\label{Area}
\eeq
But, according to Eq.~(\ref{ImpliesSumRule}),
the limit appearing on the right hand side of~(\ref{Area})
vanishes.  Thus, we are left with the transverse
momentum sum rule
\beq
\int d^2\qt \ts
\Biggl\{ 
\gnum\Biggl\vert_{{\rm all}\atop{\rm orders}} \nts\nts
-  \ts \gnum\Biggl\vert_{{\rm lowest}\atop{\rm order}} 
\Biggr\}
= 0.
\label{SumRule}
\eeq
Eq.~(\ref{SumRule}) states that the non-Abelian contributions
have no effect on the total number of gluons:  we could have
obtained the same number of gluons by ignoring the non-linear
terms in the vector potential.  What these contributions
actually do is to move gluons from one value of the transverse
momentum to another.  Thus, the total energy in the gluon field
is affected by the non-Abelian terms.
Note that this conclusion 
is independent of the form of $\curlyD(\deltat)$.

If we now assume that $\curlyD(\deltat)$ is of the 
form~(\ref{TwoTerms}), with the smooth part of the correlation
function $C(\deltat)$ 
being a positive monotonic function, we can show
from Eqs.~(\ref{Lexp})--(\ref{XiDef}) that $L(\deltat)$
and $\curlyL(\deltat)$
are both monotonically decreasing functions of $\Delta$.
Since $L(0)=0$, this means that $L(\deltat)$ is negative.
Comparing Eqs.~(\ref{GlueEnd}) and~(\ref{F0def}) we see that
the integrand for the all orders gluon number density differs
from the lowest-order result by the factor
\beq
{
{ \exp[N_c\CHI_\infty L(\deltat)] - 1 }
\over
{ N_c \CHI_\infty L(\deltat) }
},
\label{KeyFactor}
\eeq
which is less than or equal to unity for $L(\deltat) \le 0$.
Since $L(\deltat)$ becomes more negative at large $\deltat$,
we see that the non-Abelian corrections actually serve
to {\it reduce}\ the degree of correlations in the infrared,
resulting in a depletion in the number of low-$\qt$
gluons relative to the Abelian result.
This depletion can be quite drastic.  For example, when
$p<2$, we have an Abelian distribution which 
diverges like $1/\qt^{2-p}$ at small $\qt$ whereas the 
non-Abelian terms moderate this to $\ln\qt^2$!
On the other hand, in the ultraviolet the number
of gluons is unchanged:  the factor in~(\ref{KeyFactor})
is very nearly~1.  Hence, our transverse momentum sum rule
tells us that there
must be an enhancement in the number of gluons at intermediate
momenta.  Indeed, we will see this explicitly in the model
presented in Sec.~\ref{EXAMPLE}.
Furthermore,~(\ref{KeyFactor}) implies that the effect
becomes more pronounced as $\CHI_\infty$ increases.
This is consistent with the idea of saturation presented
in~\cite{GLR}, which is framed in terms of parton recombination.
When $\CHI_\infty$ is large, gluons begin to overlap, and
can readily merge via the non-Abelian terms in the equations
of motion, which become more and more important 
in determining the correlation function~(\ref{MVCorrelator})
as the density is increased.  
If we further imagine that each source factor $\rho(\xt)$ contributes
some characteristic amount of
transverse momentum, then at the end
of the day we find fewer gluons with low $\qt$, and a corresponding
enhancement at intermediate~$\qt$.

%%%%%%%%%%%%%%%%%%%%%%%%%%%%%%%%%%%%%%%%%%%%%%%%%%%%%%%%%%%%%%%%
%%
%%            CONNECTION TO DGLAP
%%
%%%%%%%%%%%%%%%%%%%%%%%%%%%%%%%%%%%%%%%%%%%%%%%%%%%%%%%%%%%%%%%%

\subsection{Gluon Structure Functions and the DGLAP 
Equation}\label{DGLAPsec}

The usual gluon structure functions resolved at the scale $\Qprobe^2$,
namely $g_A(\xf,\Qprobe^2)$, may be obtained by 
supplying the trivial factors needed to convert $q^{+}$ into $\xf$
and integrating 
Eq.~(\ref{GlueEnd}) over transverse momenta less than or
equal to $\Qprobe$:
\beqa
g_A(\xf,\Qprobe^2) & \equiv &
\int_{\vert\qt\vert\le\Qprobe}\nts\nts\nts  d^2\qt \ts
{{dN}\over{d\xf d^2\qt}}
\phantom{\Biggl[}
\cr \phantom{\Biggl[^{[}} & = &
{{N_c^2-1}\over{N_c}} 
\ts { {\pi R^2}\over{2\pi^3} }
\ts {{1}\over{\xf}} 
\int_{\vert\qt\vert\le\Qprobe}\nts\nts\nts  d^2\qt \ts
\int d^2\deltat \ts
e^{i\qt\cdot\deltat}\ts
{{\curlyL(\deltat)}\over{L(\deltat)}} 
\biggl\{
\exp\Bigl[ N_c \ts\CHI_\infty L(\deltat)\Bigr] - 1
\biggr\}.
\label{StrFnDef}
\eeqa
Suppose that we consider measuring the gluon distribution function
at large $\Qprobe^2$. 
Then, we may use the transverse momentum sum rule~(\ref{SumRule})
to replace the all-orders expression on the right hand side
of~(\ref{StrFnDef}) by its Abelian counterpart:
\beq
g_A(\xf,\Qprobe^2) =
{{1}\over{\xf}} 
\int_{\vert\qt\vert\le\Qprobe}\nts\nts\nts  d^2\qt \ts
\int d^2\deltat \ts
e^{i\qt\cdot\deltat}\ts
\curlyL(\deltat)
\ts (N_c^2-1)
\ts { {\CHI_\infty \pi R^2}\over{2\pi^3} }.
\label{Easier}
\eeq
As explained in the discussion surrounding Eq.~(\ref{RealChi}),
the combination $(N_c^2-1)\CHI_\infty \pi R^2$ is the total
charge squared of the nucleus, which 
for a nucleus containing $A$ nucleons is simply $3A g^2 C_F$.

Eq.~(\ref{Easier}) is useful because
although we are unable to perform the Fourier transform appearing
in~(\ref{StrFnDef}) for the all-orders result 
with an arbitrary correlation function $\curlyD(\deltat)$, we
are able to do so at lowest order.  
Recall that
$\curlyL$ is the convolution of $\curlyD$ with a logarithm
[see Eq.~(\ref{curlyLgen})].  Its Fourier transform is just
the product of the transforms of these two functions.
Therefore, Eq.~(\ref{Easier}) simplifies to
\beq
g_A(\xf,\Qprobe^2) =
{{1}\over{\xf}} 
\int_{\vert\qt\vert\le\Qprobe}\nts\nts\nts  d^2\qt \ts\ts
{{\widetilde\curlyD(\qt)}\over{2\qt^2}}
\ts { {3A g^2 C_F}\over{2\pi^3} }.
\label{preFactorized}
\eeq
Finally, if the correlation function is rotationally invariant,
we may do the angular integration to obtain
\beq
g_A(\xf,\Qprobe^2) =
3A C_F 
\ts {{1}\over{\xf}} 
\int_{0}^{\Qprobe^2}
{{dq^2}\over{q^2}}
\widetilde\curlyD(q)
{ {\alpha_s}\over{\pi} }.
\label{Factorized}
\eeq
Thus, for large enough values of $\Qprobe$, we see that
the number of gluons at a given value of $\xf$ scales
with the number of nucleons.  
Thanks to the sum rule, this statement is true to
all orders in the coupling constant:  any non-Abelian
corrections to~(\ref{Factorized}) vanish in 
the $\Qprobe^2 \rightarrow \infty$ limit.

The expression in~(\ref{Factorized}) is also
closely connected to the Dokshitzer-Gribov-Lipatov-Altarelli-Parisi
(DGLAP) equation~\cite{DGLAP1,DGLAP2}.
Denoting the gluon distribution function by $g(\xf,\Qprobe^2)$,
the quark distribution function by $q_i(\xf,\Qprobe^2)$ for
flavor $i$, and the antiquark distribution function by 
$\bar{q}_i(\xf,\Qprobe^2)$, the DGLAP evolution equation
for the gluon distribution function of a single nucleon reads
\beqa
{{\partial g(\xf,\Qprobe^2)}
\over
{\partial\ln\Qprobe^2}}
=
{{\alpha_s(\Qprobe^2)}\over{2\pi}}
\int_{\xf}^{1}
{{d\xi}\over{\xi}}
&&\Biggl[
P_{gq}\biggl( {{x}\over{\xi}} , \alpha_s(\Qprobe^2) \biggl)
\sum_{i} \Bigl[ q_i(\xi,\Qprobe^2) + \bar{q}_i(\xi,\Qprobe^2) \Bigr]
\cr && +
P_{gg}\biggl( {{x}\over{\xi}} , \alpha_s(\Qprobe^2) \biggl)
g(\xi, \Qprobe^2) \Biggr].
\label{DGLAPequation}
\eeqa
The functions $P_{gq}$ and $P_{gg}$ appearing 
in Eq.~(\ref{DGLAPequation}) are the usual
Altarelli-Parisi splitting functions\cite{DGLAP2}.  
At lowest order, we 
set $g(\xi,\Qprobe^2)=0$ on the right hand
side of~(\ref{DGLAPequation}), in accordance with the
premise of the \MV\ model that the gluons are generated
only by the valence quarks.
Thus, the only splitting function we require is
(to leading order)
\beq
P_{gq}(x) = C_F \biggl[ {{1+(1-x)^2}\over{x}} \biggr].
\label{Pgq}
\eeq
Inserting~(\ref{Pgq}) into~(\ref{DGLAPequation}) we obtain
\beq
{{\partial g(\xf,\Qprobe^2)}
\over
{\partial\ln\Qprobe^2}}
=
C_F {{\alpha_s(\Qprobe^2)}\over{2\pi}}
{{1}\over{\xf}}
\int_{\xf}^{1} d\xi
\Bigl[ {1+(1-x)^2} \Bigr]
\sum_{i} \Bigl[ q_i(\xi,\Qprobe^2) + \bar{q}_i(\xi,\Qprobe^2) \Bigr].
\label{DGLAPvalence}
\eeq
Since the \MV\ model is only supposed to apply at small $\xf$, 
we take the $\xf\rightarrow 0$ limit of~(\ref{DGLAPvalence}).
In this limit, we end up with the sum of the
quark and antiquark distribution functions  for all flavors
integrated over all values of $\xf$:  this is just 3 for a 
single nucleon.  Hence
\beq
{{\partial g(\xf,\Qprobe^2)}
\over
{\partial\ln\Qprobe^2}}
=
3C_F {{\alpha_s(\Qprobe^2)}\over{\pi}}
{{1}\over{\xf}}.
\label{DGLAPsmallx}
\eeq
We now see that the DGLAP equation at small $\xf$,
Eq.~(\ref{DGLAPsmallx}), is very similar to 
the \MV\ result of Eq.~(\ref{Factorized}).
Remarkably,
the \MV\ gluon distribution function is almost $A$ times
the DGLAP result for a single nucleon.  
Until now, we
have been ambiguous about the precise scale at which we should
evaluate the strong coupling.  This comparison tells us that
we ought to let it run, evaluating it at $q^2$ and keeping it
inside the integral in Eq.~(\ref{Factorized}).  
By doing this, we effectively incorporate certain quantum
corrections in our otherwise classical treatment.
The only other difference between Eqs.~(\ref{Factorized})
and~(\ref{DGLAPsmallx}) is the appearance of $\widetilde\curlyD(q)$
in the \MV\ expression.  Since the color neutrality 
condition~(\ref{Dint}) also tells us that $\widetilde\curlyD(0) = 0$,
we see that the presence of this factor serves to regulate
the integral at small values of $q^2$.  
On the other hand, 
as described in the discussion following Eq.~(\ref{curlyLshort}),
if we believe that the nucleon should contain point-like
quarks, then $\curlyD(\deltat)$ contains $\delta^2(\deltat)$
and $\widetilde\curlyD(q) \sim 1$ at large momenta.

%%%%%%%%%%%%%%%%%%%%%%%%%%%%%%%%%%%%%%%%%%%%%%%%%%%%%%%%%%%%%%%%
%%
%%            MOMENTS OF THE DISTRIBUTION FUNCTION
%%
%%%%%%%%%%%%%%%%%%%%%%%%%%%%%%%%%%%%%%%%%%%%%%%%%%%%%%%%%%%%%%%%

\subsection{Average Transverse Momentum-Squared}\label{MOMENTTEXT}

Because the gluon number density we have obtained in
Eq.~(\ref{GlueEnd}) is well-defined, we are able to
explicitly compute the average transverse momentum-squared
$\langle \qt^2 \rangle$ associated with this distribution.  
In performing this calculation, we have assumed that 
\beq
\curlyD(\deltat) = \delta^2(\deltat) - \Ihat(\Delta).
\label{TwoTermsOneMoreTime}
\eeq
That is, we choose a rotationally invariant correlator
which is consistent with nucleons that contain point-like quarks.
Furthermore, we work to next-to-leading order in the
momentum expansion, since the first
non-trivial nuclear dependence enters in at that level.

As we show in Appendix~\ref{MOMENTS}, the gluon number
density may be cast into the form
\beq
{{dN}\over{d\xf}} =
{{N_c^2-1}\over{\pi^2}} \ts
\pi R^2 \ts
{{1}\over{\xf}} \ts
\Qprobe^2
\int_{-\infty}^{\infty} dx^{-} \mu^2(x^{-})
\int d^2\xt \ts
{{J_1(\Qprobe x)}\over{\Qprobe x}}  \ts
\curlyL(\xt) e^{N_c \CHI(x^{-})L(\xt)}.
\label{prescale}
\eeq
As we have remarked earlier, all of the non-Abelian effects
are contained in the exponential factor.  We should like
some way to organize the expansion of~(\ref{prescale})
in powers of $1/\Qprobe$.  
In the limit $\Qprobe\rightarrow\infty$, the Bessel function
appearing in this expresion tends to a $\delta$-function.
This suggests that the $\xt$ integration will be dominated
by the small $\xt$ region.  To connect an expansion in $\xt$
with the desired expansion in $1/\Qprobe$ we rescale the
integral in~(\ref{prescale}) via $\xt \equiv \yt/\Qprobe$:
\beq
{{dN}\over{d\xf}} =
{{N_c^2-1}\over{\pi^2}} \ts
\pi R^2 \ts
{{1}\over{\xf}} \ts
\int_{-\infty}^{\infty} dx^{-} \mu^2(x^{-})
\int d^2\yt \ts
{{J_1(y)}\over{y}}  \ts
\curlyL\biggl({{\yt}\over{\Qprobe}}\biggr) 
\exp\Biggl[ N_c \CHI(x^{-})L\biggl({{\yt}\over{\Qprobe}}\biggr)\Biggr].
\label{rescaled}
\eeq
In Appendix~\ref{SIGMA}, we argue that for a correlation function
of the form~(\ref{TwoTermsOneMoreTime}), the functions $L(\xt)$
and $\curlyL(\xt)$ are simply some power series in $x$
plus $\ln x^2$ times some other power 
series in $x$.\footnote{If $\Ihat(\Delta)$ is derived from
a rotationally invariant function of the {\it vector}\ $\deltat$,
then the odd powers of $x$ will be absent in this expansion.}
In the case of $L(\xt)$ both series start at $x^2$ rather
than $x^0$.  Furthermore, we observe that the intrinsic
length scale which must appear in the logarithms 
should be of order the nucleon size $a$.  This same
length scale serves to make the expansion parameter dimensionless.
Looking at Eq.~(\ref{rescaled}), we see that by
expanding everything except for the Bessel function,
it becomes apparent that the final result will consist of
terms of the form $[1/(a\Qprobe)^m] \ln^n(a\Qprobe)^2$.
Because the series for $L(\xt)$ begins at $x^2$, the
smallest inverse power of $a\Qprobe$ which may multiply
$\ln^n(a\Qprobe)^2$ is $1/(a\Qprobe)^{2n-2}$.
Furthermore, if we keep track of the powers of $\CHI$,
we see that each additional power of $\CHI$ comes with 
an additional factor of $1/\Qprobe^2$.  
Thus, the leading order term is the same as in an Abelian
theory.
Since $\CHI \sim A^{1/3}$,
we conclude that the non-Abelian corrections are enhanced
in very large nuclei:  the subleading terms become 
important if $N_c \CHI_\infty \sim Q^2$.

We now proceed directly to our results, deferring the 
details of this somewhat lengthy calculation to Appendix~\ref{MOMENTS}.
At next-to-leading order, we find that
\beq
{{dN}\over{d\xf}} = 
{{1}\over{\xf}}
{{\Qtotal^2}\over{4\pi^2}}
\biggl[ 1 - {{N_c \CHI_\infty}\over{4\pi\Qprobe^2}} \biggr]
\ln\biggl({{\Qprobe}\over{\Qprobe_0}}\biggr)^2,
\label{zeroth}
\eeq
where
\beqa
\ln\biggl({{1}\over{\Qprobe_0}}\biggr)^2 & \equiv &
2(\gamma_E - \ln 2)
+\int d^2\deltat \ts C(\Delta)\ln\Delta^2,
\label{LogMom}
\eeqa
$\gamma_E$ is Euler's constant,
and $\Qtotal^2 = 3Ag^2C_F$ is the total charge-squared of the nucleus.
As advertised, the scale appearing in the logarithm 
does not enter
in as an additional parameter which must be
determined from the data.
Instead, given a particular model for implementing
the color neutrality condition~(\ref{Dint}), we obtain a specific
value for the  momentum scale, Eq.~(\ref{LogMom}).
The integral appearing in Eq.~(\ref{LogMom}) is closely
related to the value of $\curlyL(\xt)$, {\it i.e.}
\beq
\int d^2\deltat \ts C(\Delta)\ln\Delta^2 = 
\lim_{\xt\rightarrow{\bf 0}}
\Bigl[ 8\pi\curlyL(\xt) + \ln\xt^2 \Bigr].
\eeq
As we have already remarked in the previous subsection,
at leading order in $\Qprobe$,
$dN/d\xf$ simply proportional to the number of nucleons:
all of the non-Abelian contributions are suppressed for
$\Qprobe \rightarrow \infty$.   
Although at first glance
the leading order contribution to Eq.~(\ref{zeroth})
looks very different from the result presented in
the previous subsection, the two expressions actually do agree.
To see this, insert the
Fourier integral representation for $\widetilde\curlyD(\qt)$
into Eq.~(\ref{preFactorized}), perform the $\qt$ integration,
and subsequently integrate by parts.  
Calculation of the sub-leading term requires the full treatment
given in Appendix~\ref{MOMENTS}.

Finally, the next-to-leading order
expression we obtain for $\langle \qt^2 \rangle$ reads
\beq
\langle \qt^2 \rangle =
{
{ \Qprobe^2 - 4\pi C(0) }
\over
{ \ln(\Qprobe/\Qprobe_0)^2 }
}
+{{N_c\CHI_\infty} \over{8\pi}}
\biggl[
\ln\biggl({{\Qprobe}\over{\Qprobe_0}}\biggr)^2
- 2 \biggr].
\label{quadratic}
\eeq
We see that at leading order, Eq.~(\ref{quadratic})
is independent of the size of the nucleus.
A mild dependence on the modelling of the nucleons
enters in through the appearance of $C(0)$.
The non-Abelian terms make their first
appearance at next-to-leading order.
As previously, the scale of the logarithms is not
arbitrary, but rather determined by the form of $C(\Delta)$,
according to Eq.~(\ref{LogMom}).

%%%%%%%%%%%%%%%%%%%%%%%%%%%%%%%%%%%%%%%%%%%%%%%%%%%%%%%%%%%%%%%%
%%%%%%%%%%%%%%%%%%%%%%%%%%%%%%%%%%%%%%%%%%%%%%%%%%%%%%%%%%%%%%%%
%%%%%%%%%%%%%%%%%%%%%%%%%%%%%%%%%%%%%%%%%%%%%%%%%%%%%%%%%%%%%%%%
%%%%%%
%%%%%%                 KOVCHEGOV'S MODEL
%%%%%%
%%%%%%%%%%%%%%%%%%%%%%%%%%%%%%%%%%%%%%%%%%%%%%%%%%%%%%%%%%%%%%%%
%%%%%%%%%%%%%%%%%%%%%%%%%%%%%%%%%%%%%%%%%%%%%%%%%%%%%%%%%%%%%%%%
%%%%%%%%%%%%%%%%%%%%%%%%%%%%%%%%%%%%%%%%%%%%%%%%%%%%%%%%%%%%%%%%

\section{A Model with Color-Neutral Nucleons}\label{EXAMPLE}

To illustrate some of the features of our improved treatment, we
turn to a specific model introduced by Kovchegov~\cite{paper12}.
In this model, we imagine a nucleus of radius $R$, containing
$A$ nucleons of radius $a\sim\LQCD^{-1}$.  
Each ``nucleon'' is made up of a 
quark-antiquark pair.  The view in the rest frame of the nucleus
is as follows:
The quark and antiquark for the $i$th
nucleon is assumed to be located with equal probability anywhere
within a distance $a$ of the center of the nucleon.  The nucleons
are assumed to be located with equal probability anywhere within
a distance $R$ of the center of the nucleus. In this manner, confinement
of the valence partons within the individual nucleons is ensured.
For the sake of comparison, we will also present results where
the quarks are distributed within the nucleons using a Gaussian
weight instead of the uniform weight
employed in Ref.~\cite{paper12}.

Kovchegov begins with the laboratory frame 
charge density 
\beq
\rho(x^{-},\xt) = 
g \sum_{i=1}^{N} (T^{a}) (T^{a}_i) 
\Bigl[ \delta(x^{-}-x^{-}_i) \delta^2(\xt-\xt_i)
      -\delta(x^{-}-x^{\prime -}_i) \delta^2(\xprimet-\xprimet_i)
\Bigr]
\label{Kden}
\eeq
for the ultrarelativistic nucleus. 
In Eq.~(\ref{Kden}) the quark belonging to the $i$th nucleon
is located at $(x^{-}_i,\xt_i)$ whereas the $i$th antiquark
is located at $(x^{\prime -}_i,\xprimet_i)$.  The $T^a$'s
are generators in the SU($N_c$) color space, while the $T^a_i$'s
are similar generators in the color space of each nucleon.
Kovchegov explicitly constructs the light-cone gauge vector
potential generated by this distribution, 
and uses that result to derive an effective two-dimensional
charge distribution which leads to this vector potential,
{\it i.e.}\ he writes
\beq
\rho(x^{-},\xt) \equiv \delta(x^{-})\rho_2(\xt)
\eeq
and computes $\rho_2(\xt)$.  From this explicit form for $\rho_2(\xt)$,
one may then average over the positions of the quarks and nucleons
as described above to obtain the two point correlator
$ \langle \rho_2^a(\xt) \rho_2^b(\xprimet) \rangle $.  
This quantity is 
connected to the three-dimensional correlator
of Eq.~(\ref{RhoRho1}) via
\beq
\langle \rho^a(x^{-},\xt) \rho^b(x^{\prime -},\xprimet) \rangle
= \delta(x^{-}-x^{\prime -})\ts
\langle \rho_2^a(\xt) \rho_2^b(\xprimet) \rangle
\label{2to3}
\eeq
In this way, Kovchegov's model specifies the $\xt$-dependence
of the charge density correlator.

The result obtained in Ref.~\cite{paper12} is of the form
\beq
\langle \rho_2^a(\xt) \rho_2^b(\xprimet) \rangle
= {{g^2}\over{N_c}} \ts A\delta^{ab}
\ts \Bigl[ \Isng(\xt,\xprimet)
           -\Ismth(\xt,\xprimet) \Bigr].
\label{KovForm}
\eeq
In Eq.~(\ref{KovForm}), the function
$\Isng(\xt,\xprimet)$ is the contribution
that results in the averaging when a quark overlaps with 
a quark (or an antiquark overlaps with an antiquark), 
whereas $\Ismth(\xt,\xprimet)$
is the contribution that results when a quark overlaps with
an antiquark.  Explicitly,
\beq
\Isng(\xt,\xprimet) = 
\int {{d^2\rt \ts dz}\over{\hbox{$4\over3$}\pi R^3}}
\int {{d^2\xit \ts d\zeta}\over{\hbox{$4\over3$}\pi a^3}}
\ts\ts \delta^2(\xt-\rt-\xit) \ts \delta^2(\xprimet-\rt-\xit)
\label{Isng}
\eeq
and
\beq
\Ismth(\xt,\xprimet) = 
\int {{d^2\rt \ts dz}\over{\hbox{$4\over3$}\pi R^3}}
\int {{d^2\xit \ts d\zeta}\over{\hbox{$4\over3$}\pi a^3}}
\int {{d^2\xiprimet \ts d\zeta'}\over{\hbox{$4\over3$}\pi a^3}}
\ts\ts \delta^2(\xt-\rt-\xit) \ts \delta^2(\xprimet-\rt-\xiprimet).
\label{Ismth}
\eeq
In Eqs.~(\ref{Isng}) and~(\ref{Ismth}), $(\rt,z)$ locate
the position of the nucleon with respect to the (arbitrary)
origin while $(\xit^{(\prime)},\zeta^{(\prime)})$  locate the
(anti)quark with respect to the center of the nucleon.  Note
that the $\delta$ functions are {\it two}-dimensional, whereas
the integrations are {\it three}-dimensional.  
If we use a Gaussian instead of a uniform distribution,
we should make the replacement
\beq
{ {1} \over { {{4}\over{3}}\pi a^3 } }
\rightarrow 
{
{ 1 }
\over
{ (2\pi a^2)^{3/2} }
} \ts 
\exp\Biggl[
- { {\xit^2+\zeta^2}\over{2a^2} } 
\Biggr] 
\eeq
for the quark %positions and similarly for the nucleon 
positions.\footnote{There is no {\it a priori}
reason to assume that the parameters $a$ and $R$ appearing
in the uniform and Gaussian distributions are the same.
When we need to make the distinction clear, we will use a subscript
$U$ on the uniform parameters and a $G$ on the Gaussian
parameters.}
A detailed evaluation
of all of the integrals relating to the uniform distribution
is presented in Appendix~\ref{INTEG1}.
In the Gaussian case, the integrations are straightforward,
and we shall simply present the results.

\begin{figure}[h]

\vspace*{15cm}
\includegraphics{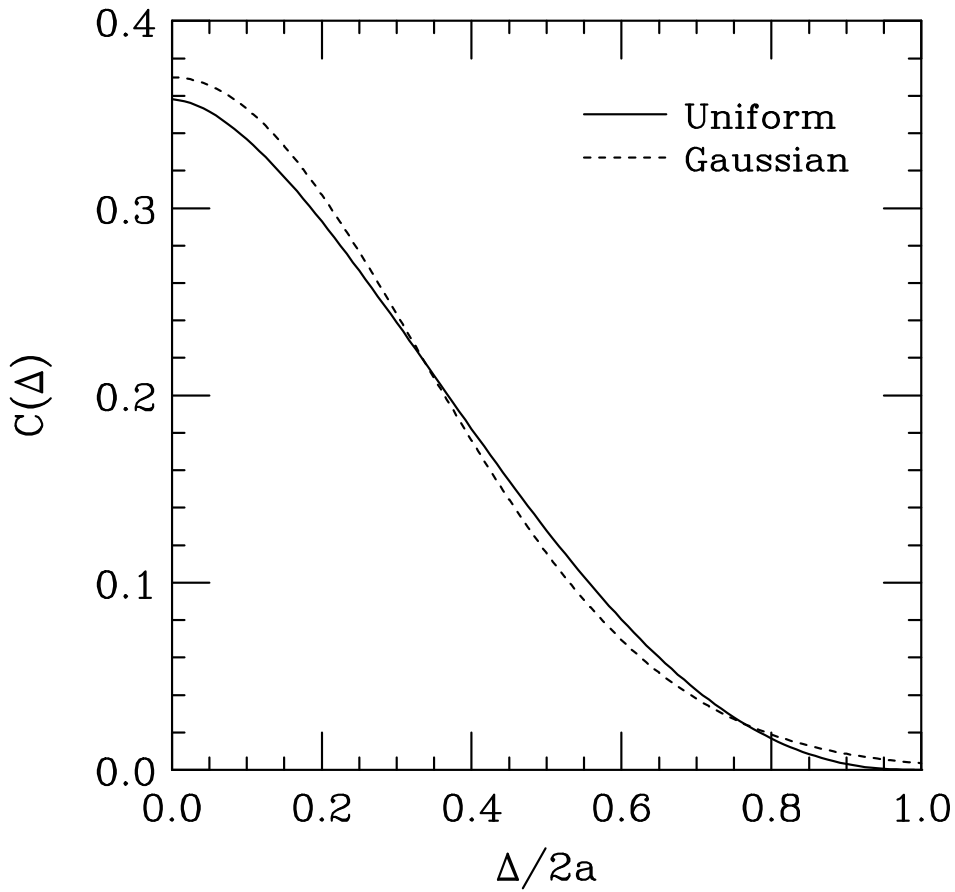}

\caption[]{The smooth part of the two-point charge density
correlation function in Kovchegov's model~\protect\cite{paper12}, 
as given in
Eqs.~(\protect\ref{IHATu}) (uniform quark/nucleon distribution)
and~(\protect\ref{IHATg}) (Gaussian quark/nucleon distribution).
The nucleon size parameters $a$ have been chosen according to
Eq.~(\protect\ref{aGdef}) so that resulting gluon number densities
match in the ultraviolet limit.
In the uniform case, $C(\Delta)$ vanishes for $\Delta > 2a$.
In the Gaussian case, there is a Gaussian tail in this region.
}
\label{IhatPlot}
\end{figure}

The function defined by Eq.~(\ref{Isng}) is obviously proportional 
$\delta^2(\xt-\xprimet)$, the precise $\xt$ dependence postulated
by the authors of Ref.~\cite{paper9}.  It is also obvious that
Eq.~(\ref{Ismth}) contains no such singular behavior:  it is,
in fact, a smooth function of the coordinates.  For
physical applications where we will be looking at scales much
smaller than the nucleon radius $a$, 
this smooth function may be neglected in
comparison to the $\delta$-function terms,
as was done in Ref.~\cite{paper12}. 
However, as we noted in Sec.~\ref{NEUTRALITY}, taking only
the $\delta$-function term produces a correlator which
is incompatible with the idea of color-neutral nucleons.
Thus, when we look at scales
approaching $a$, it becomes crucial to include
the contribution from Eq.~(\ref{Ismth}).
Eq.~(\ref{KovForm}) obviously implies that $\curlyD(\xt-\xprimet)$
has the form introduced in Eq.~(\ref{TwoTerms}).
By sorting out the various prefactors associated with~(\ref{2to3}),
(\ref{KovForm}), and~(\ref{sng3}), we may
determine the value of $\CHI_\infty$ from the requirement that
the two contributions to~(\ref{TwoTerms}) each have unit integral.
We find that
\beqa
\CHI_\infty &=& {{g^2 } \over{N_c}} \ts
{{A}\over{\pi R^2}}.
\label{mu-sqr}
\eeqa
Eq.~(\protect\ref{mu-sqr})
implies a total charge squared of $2A g^2 C_F$,
instead of $3A g^2 C_F$ as was used in Sec.~\protect\ref{DGLAPsec},
because this model uses ``nucleons'' which are $q\bar{q}$ pairs
instead of $qqq$ triplets.  
The function $\Ihat(\xt-\xprimet)$ which enforces color neutrality
on scales near and beyond the nucleon radius is given by
\beqa
\Ihat(\xt-\xprimet)  &=&
{ {9}\over{4\pi^2 a^6} } \thinspace\thinspace
\int_{\vert\xit\vert \le a} 
\negthinspace\negthinspace\negthinspace\negthinspace 
d^2\xit
\int_{\vert\xiprimet\vert \le a} 
\negthinspace\negthinspace\negthinspace\negthinspace 
d^2\xiprimet \thinspace\thinspace
\sqrt{a^2-\smash\xit^2} \sqrt{a^2-{\smash\xiprimet}^2}
\thinspace
\delta^2(\xt-\xprimet-\xit+\xiprimet)
\cr\nonumber\\[0.2cm] &=&
{ {9\Theta(1-X)} \over {16\pi a^2} }
\thinspace
\Bigl[
(2+X^2)\sqrt{1-X^2}
-X^2(4-X^2)\tanh^{-1}\sqrt{1-X^2}
\Bigr]
\label{IHATu}
\eeqa
for a uniform distribution of quarks and nucleons and
\beq
\Ihat(\xt-\xprimet) =
{ {e^{-X^2}}\over{4\pi a^2} }
\label{IHATg}
\eeq
for the Gaussian distribution.
In Eqs.~(\ref{IHATu}) and~(\ref{IHATg}) we have
defined the dimensionless
distance measure $X\equiv \vert \xt-\xprimet \vert /(2a)$
(for details, see Appendix~\ref{INTEG1}).
These two functions have been plotted in Fig.~\ref{IhatPlot}.
In the uniform case, $C(\xt-\xprimet)$ vanishes identically
for $\vert \xt-\xprimet \vert \ge 2a$, whereas in the
Gaussian case it has, not surprisingly, a Gaussian tail.
Both functions are finite at the origin.

Now that we have specified the functional form of $\curlyD(\deltat)$,
we may return to Eqs.~(\ref{Lgen}) and~(\ref{curlyLgen})
and evaluate the inputs to the correlation function, $L(\xt-\xprimet)$
and $\curlyL(\xt-\xprimet)$.
In the uniform case we obtain
\beqa
L(\xt-\xprimet) &=& 
{ {a^2} \over {4\pi} } \ts
\Biggl[ 
\biggl(   {{93}\over{100}}
         +{{373}\over{200}} X^2
         -{{491}\over{800}} X^4
         -{{1}\over{64}} X^6 \biggr)\sqrt{1-X^2} 
\cr && \qquad
-\biggl(  {{2}\over{5}}
          + 2 X^2
          -{{1}\over{4}} X^6
          +{{1}\over{64}} X^8 \biggr)\tanh^{-1}\sqrt{1-X^2}
\Biggr]
\ts\Theta(1-X)
\cr\nonumber\\[0.01cm] &-& 
{ {a^2} \over {20\pi} }
\Biggl[
{{93}\over{20}} 
+ \ln\biggl( {{X^2}\over{4}} \biggr)
\Biggr]
\label{plainLu}
\eeqa
and
\beq
\curlyL(\xt-\xprimet) = 
{ {1-X^2} \over {32\pi} } 
\thinspace
\Bigl[ (8+8X^2-X^4) \tanh^{-1}\sqrt{1-X^2}
      -(14+X^2)\sqrt{1-X^2} \Bigr] \ts\Theta(1-X).
\label{curlyLu}
\eeq
The asymptotic forms of these functions are
%% see page x1100
\beq
L(\xt-\xprimet) \sim
\cases{
\displaystyle{{a^2 X^2}\over{4\pi}}\ts
\Biggl[ \ln\biggl( \displaystyle{{X^2}\over{4}} \biggr) 
         + \displaystyle{{3}\over{2}} \Biggr],&
$X \ll 1$; \cr
\phantom{x} & \cr
-\displaystyle{{a^2}\over{20\pi}}\ts
\Biggl[ \ln\biggl( \displaystyle{{X^2}\over{4}} \biggr) 
         + \displaystyle{{93}\over{20}} \Biggr],&
$X \gg 1$;\cr}
\eeq
and
\beq
\curlyL(\xt-\xprimet) \sim
\cases{
-\displaystyle{{1}\over{8\pi}}\ts
\Biggl[ \ln\biggl( \displaystyle{{X^2}\over{4}} \biggr)
        + \displaystyle{{7}\over{2}} \Biggr],&
$X \ll 1$; \cr
\phantom{x} & \cr
\quad 0\enspace\hbox{(exactly),}& 
$X \gg 1$.\cr}
\label{curlyLasymU}
\eeq
Because $\curlyL(\xt-\xprimet)$ vanishes at large distances,
the Fourier transform of the correlation function~(\ref{MVCorrelator})
constructed from~(\ref{plainLu}) and~(\ref{curlyLu})
will remain finite at zero (transverse) momentum.

If instead we employ Gaussian distributions for the quarks
we obtain
\beq
L(\xt-\xprimet) =
{{a^2}\over{4\pi}}\ts
\Bigl[ (1+X^2) \Ei(-X^2) + \exp(-X^2) - \ln(X^2) - (1+\gamma_E) \Bigr]
\label{plainLg}
\eeq
and
\beq
\curlyL(\xt-\xprimet) = 
-{{1}\over{8\pi}}\ts \Ei(-X^2).
\label{curlyLg}
\eeq
Eqs.~(\ref{plainLg}) and~(\ref{curlyLg}) contain the
exponential integral function evaluated at negative values
of the argument.  We have employed the conventions of 
Ref.~\cite{PhysicistsFriend}, {\it i.e.}
\beq
\Ei(-z) \equiv - \int_{x}^{\infty} {{dt}\over{t}}\ts e^{-t},\qquad
\hbox{($z>0$)}.
\eeq
Also appearing in the first of these two equations is Euler's 
constant $\gamma_E$.
Asymptotically, we have
\beq
L(\xt-\xprimet) \sim
\cases{
\displaystyle{{a^2 X^2}\over{4\pi}}\ts
\Bigl[ \ln( X^2 ) + \gamma_E - 2 \Bigr], &
$X \ll 1$; \cr
\phantom{x} & \cr
-\displaystyle{{a^2}\over{4\pi}}\ts
\Bigl[ \ln(X^2) - (1+\gamma_E) \Bigr],&
$X \gg 1$;\cr}
%% see pages x1097 and x1098
\eeq
and
\beq
\curlyL(\xt-\xprimet) \sim
\cases{
-\displaystyle{{1}\over{8\pi}}\ts
\Bigl[ \ln( X^2 ) + \gamma_E \Bigr], &
$X \ll 1$; \cr
\phantom{x} & \cr
\quad\displaystyle{{1}\over{8\pi}}\ts
{{\exp(-X^2)}\over{X^2}}, &
$X \gg 1$.\cr}
%% see page x1096
\label{curlyLasymG}
\eeq
In this case, $\curlyL(\xt-\xprimet)$ has an exponential
tail at large distances, and we end up with a correlation
function which falls like $\exp(-X^2)/[X^2\ln(X^2)]$ in
the infrared.  Again, the gluon number density will be
finite at zero transverse momentum.

\begin{figure}[h]

\vspace*{15cm}
\includegraphics{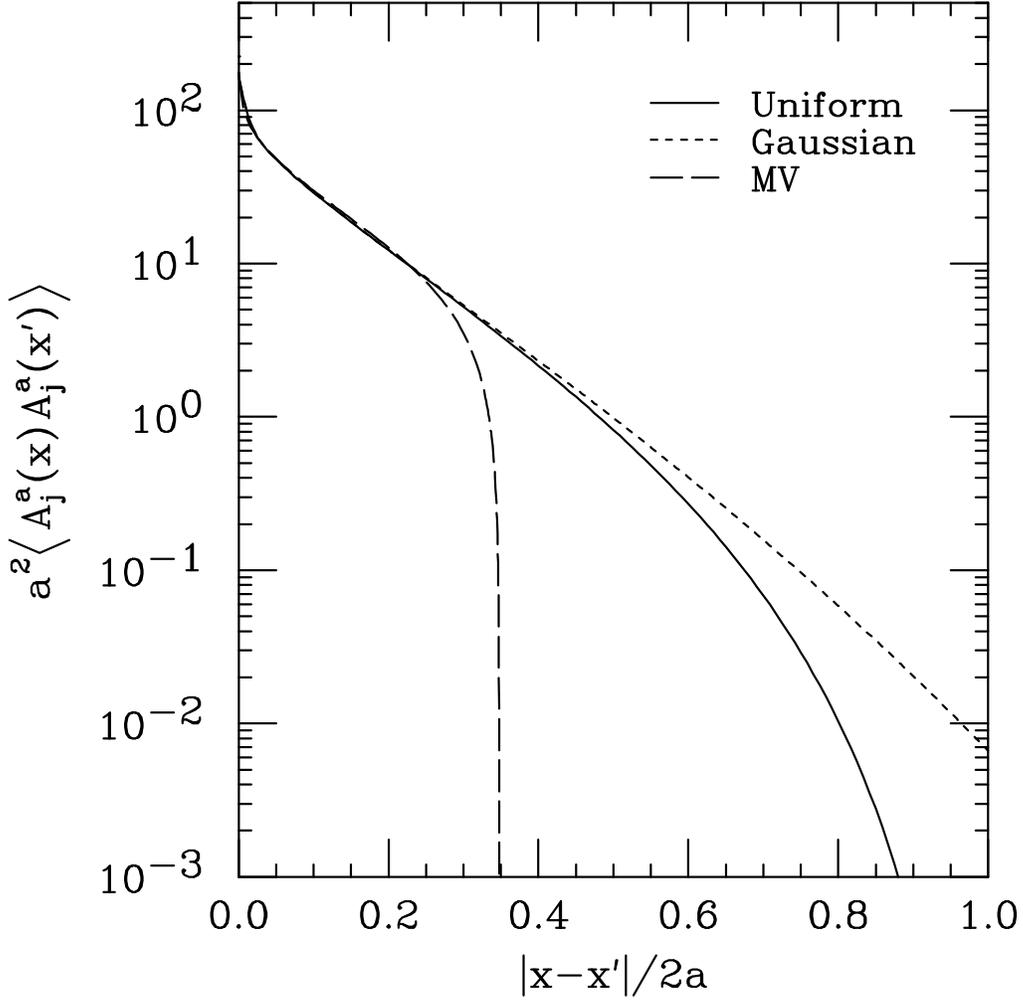}

\caption[]{The trace of the two-point vector potential correlation
function in position space.  
The Fourier transform of this function
is proportional to the gluon number density 
[{\it c.f.}\ Eq.~(\protect\ref{GluonDensity})].
The nucleon size parameters $a$ and $\LQCD$
have been chosen according to
Eqs.~(\protect\ref{aGdef}) and~(\protect\ref{LQCDdef})
so that these functions match in the ultraviolet 
($\vert\xt-\xprimet\vert\rightarrow 0$)
limit.   The longitudinal coordinates $(x^{-},x^{\prime -})$
have been fixed at a place where $a^2 \chi = 20$.
Plotted are the results of Ref.~\protect\cite{paper9}
(labelled ``MV'') as well as results using Kovchegov's 
model~\protect\cite{paper12} with a uniform or Gaussian
distribution of quarks and nucleons.
}
\label{AAxPlot}
\end{figure}

\begin{figure}[h]

\vspace*{15cm}
\includegraphics{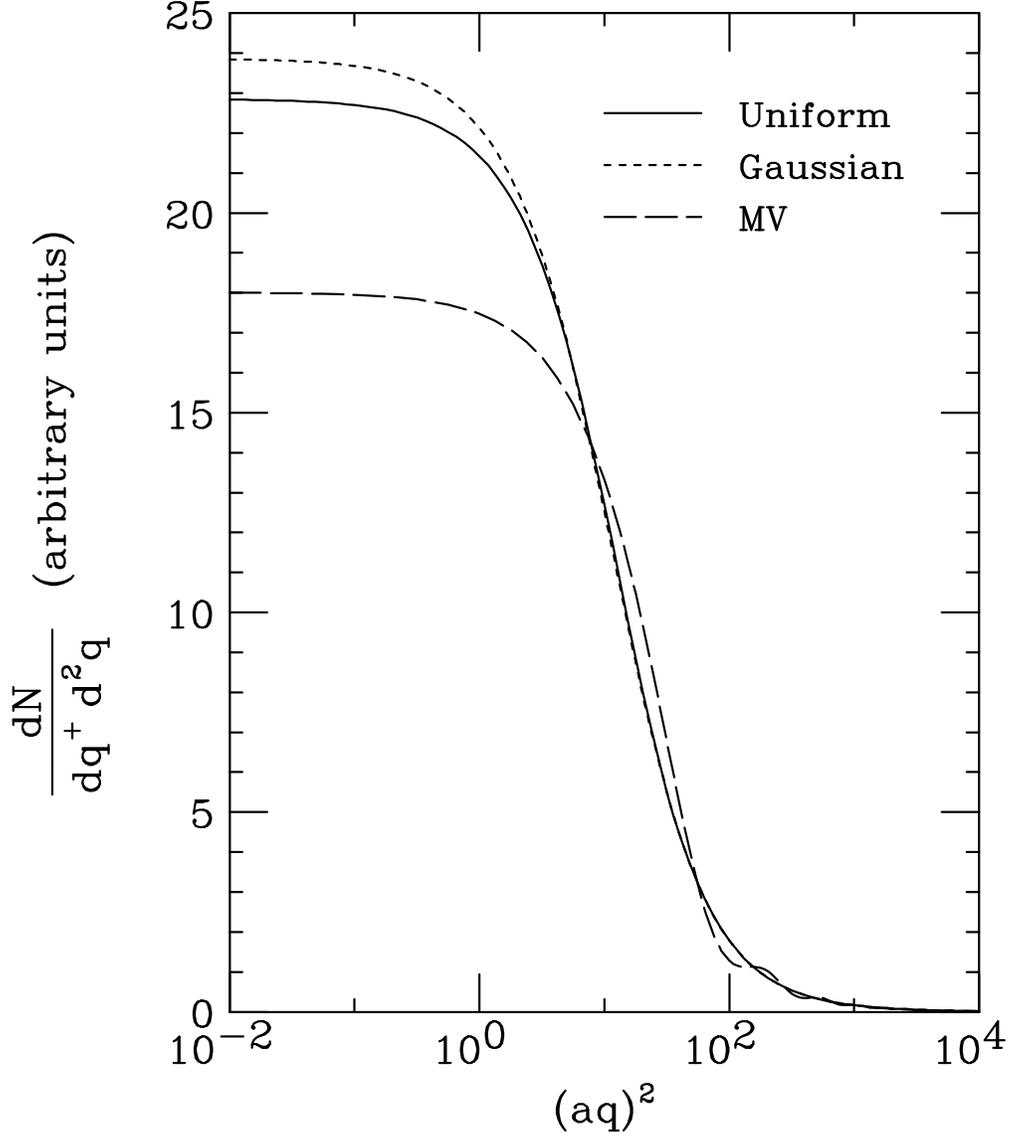}

\caption[]{Plot of the gluon number density~(\protect\ref{GlueEnd})
at fixed $q^{+}$.  
The nucleon size parameters $a$ and $\LQCD$
have been chosen according to
Eqs.~(\protect\ref{aGdef}) and~(\protect\ref{LQCDdef})
so that these functions match in the ultraviolet limit.   
The total charge squared per unit area has
been set to $a^2 \CHI_\infty = 20$.
Plotted are the results of Ref.~\protect\cite{paper9}
(labelled ``MV'') as well as results using Kovchegov's 
model\protect\cite{paper12} with a uniform or Gaussian
distribution of quarks and nucleons.
}
\label{FqPlotLinear}
\end{figure}

\begin{figure}[h]

\vspace*{15cm}
\includegraphics{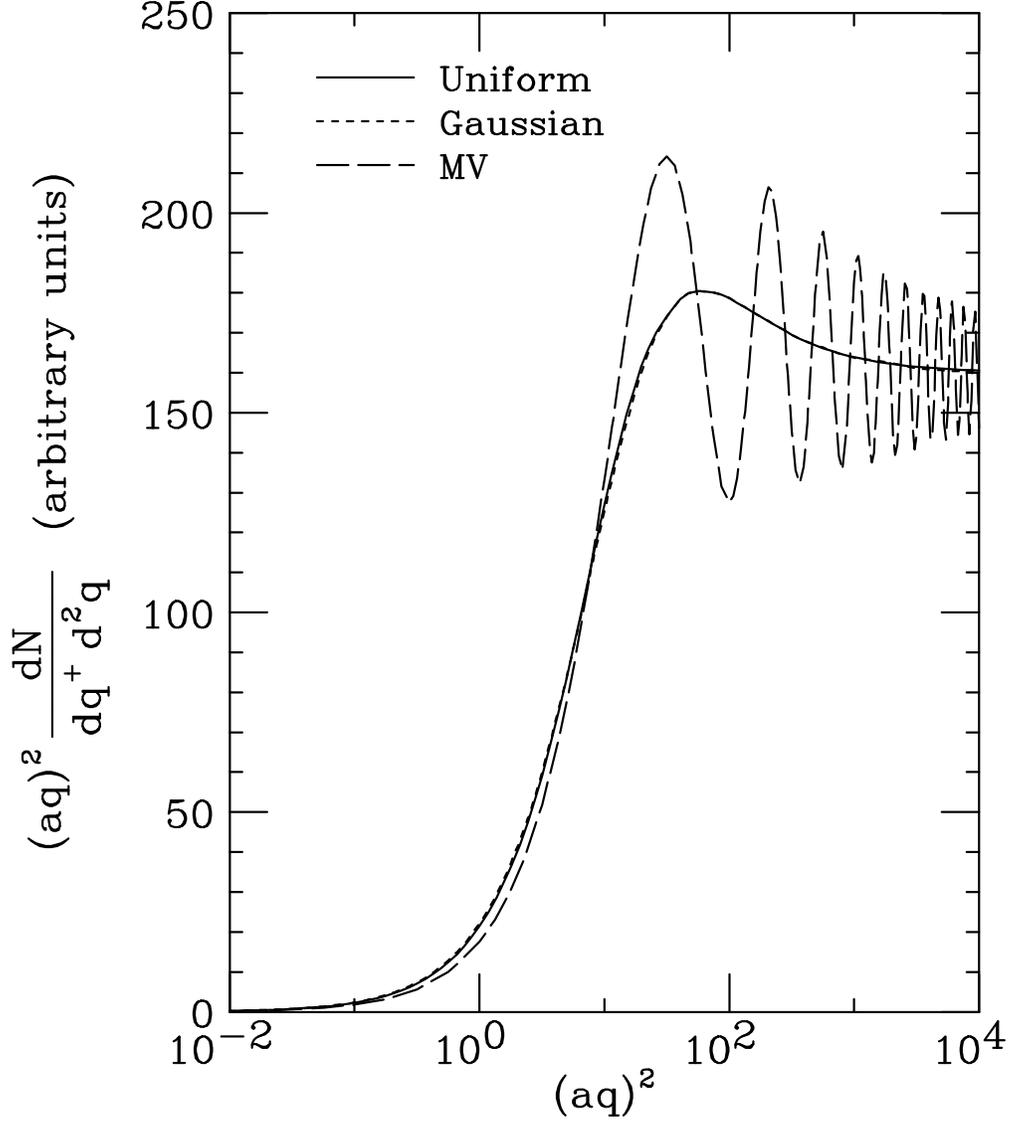}

\caption[]{Plot of $(aq)^2$ times the gluon
number density~(\protect\ref{GlueEnd})
at fixed $q^{+}$.
This combination
has been chosen to enhance the high momentum part of the
distribution and show the approach to the predicted $1/\qt^2$
ultraviolet behavior.
The nucleon size parameters $a$ and $\LQCD$
have been chosen according to
Eqs.~(\protect\ref{aGdef}) and~(\protect\ref{LQCDdef})
so that these functions match in this limit.
The total charge squared per unit area has
been set to $a^2 \CHI_\infty = 20$.
Plotted are the results of Ref.~\protect\cite{paper9}
(labelled ``MV'') as well as results using Kovchegov's 
model~\protect\cite{paper12} with a uniform or Gaussian
distribution of quarks and nucleons.
}
\label{qSqrFqPlot}
\end{figure}

\begin{figure}[h]

\vspace*{15cm}
\includegraphics{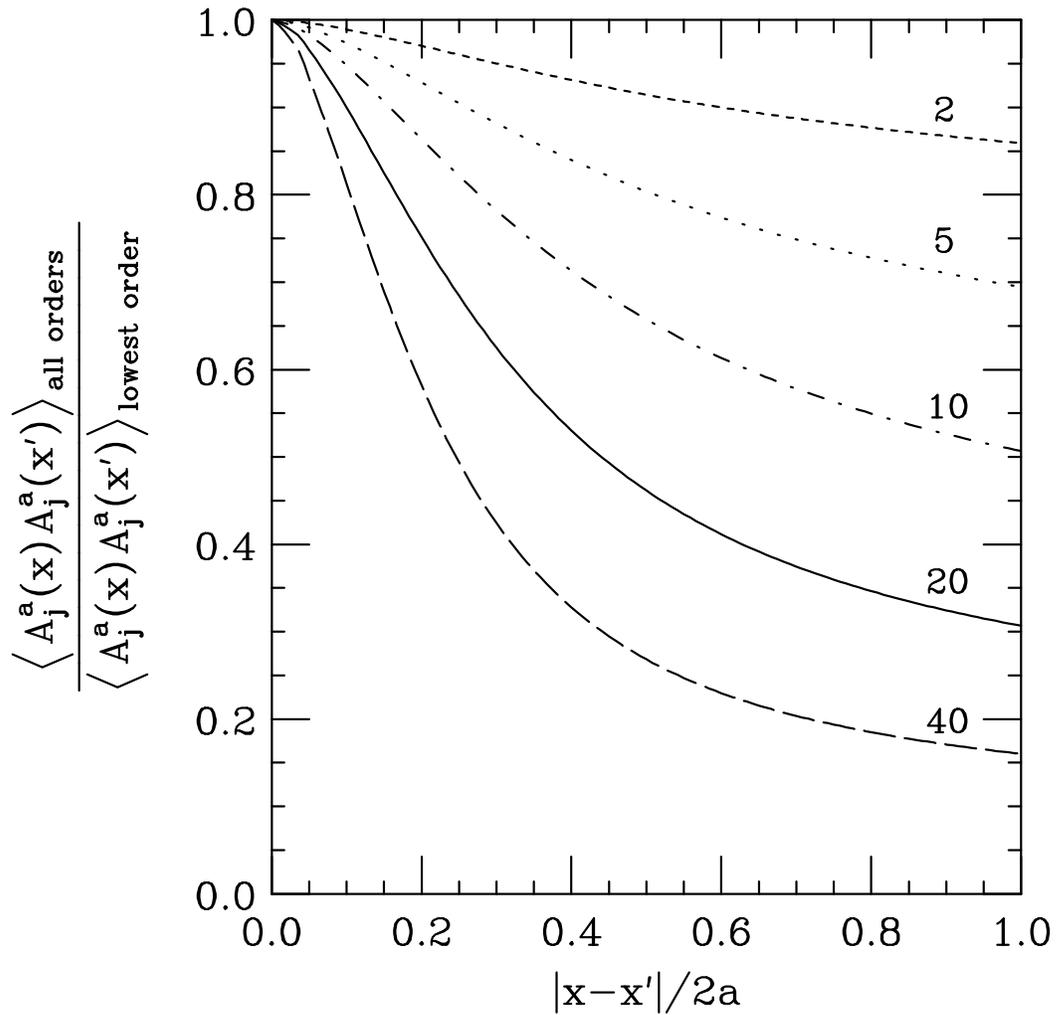}

\caption[]{The effect of the non-Abelian terms on the trace
of the correlation function.  Plotted is the ratio of the
all-orders result to the lowest-order (purely Abelian) result
using the uniform version of Kovchegov's model~\protect\cite{paper12}.
The different curves are for fixed values of the longitudinal
coordinates $(x^{-},x^{\prime -})$ such that
$a^2 \chi = 2$, 5, 10, 20, and 40.
}
\label{HO-AAxPlotRatio}
\end{figure}

\begin{figure}[h]

\vspace*{15cm}
\includegraphics{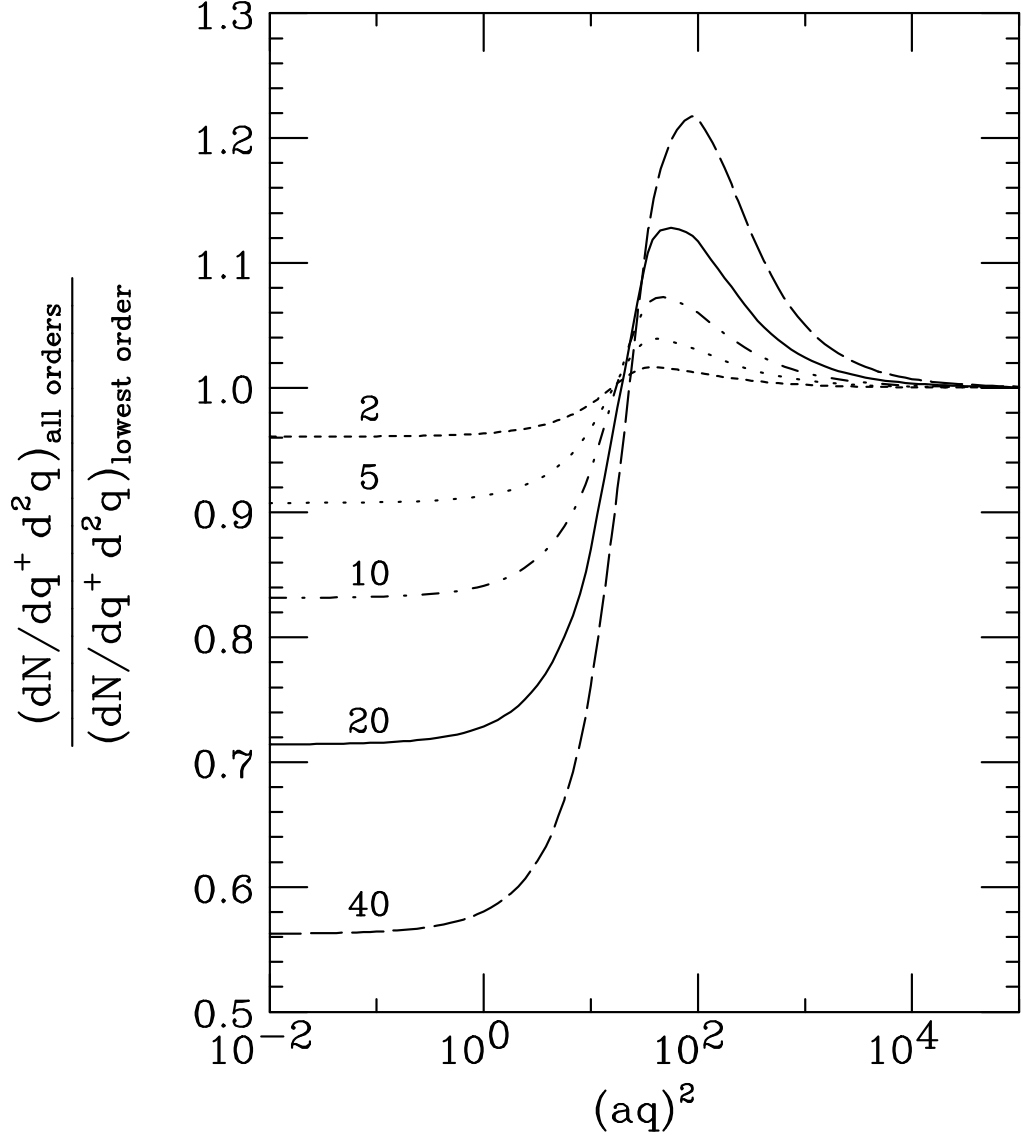}

\caption[]{The effect of the non-Abelian terms in momentum
space.  Plotted is the ratio of the all-orders gluon number
density to the 
lowest-order (purely Abelian) gluon number density
at fixed $q^{+}$
using the uniform version of Kovchegov's model~\protect\cite{paper12}.
The different curves are for $a^2 \CHI_\infty = 2$, 5, 10, 20, and 40.
}
\label{HO-FqPlotRatio}
\end{figure}

We have compiled a series of 
plots (Figs.~\ref{AAxPlot}--\ref{HO-FqPlotRatio})
to aid in the comparison of the two models to each other
and to the results of Ref.~\cite{paper9}.
Each version of the correlation function effectively
depends on two parameters, which we take to be
$\CHI_\infty$ and $a_U$, $a_G$, or $\LQCD$.
In preparing these plots, we have adjusted these parameters
so that the ultraviolet limit of~(\ref{MVCorrelator}) is
identical for all three cases.   Recall that in the ultraviolet
limit, $\CHI_\infty$ appears multiplicatively
[see Eq.~(\ref{AbelianLimit})].  Thus, matching in the ultraviolet
really means matching the values of $\curlyL$ at short distances.
From~(\ref{F0def}), (\ref{curlyLasymU}) 
and~(\ref{curlyLasymG}), we determine that
\beqa
a_G &=& 2 \exp \Bigl( \hbox{${\gamma_E}\over{2}$} 
                   - \hbox{${7}\over{4}$} \Bigr) a_U
\phantom{\biggl[} \cr &\approx& 0.464 a_U
\label{aGdef}
\eeqa
and
\beqa
\LQCD &=& \hbox{${1}\over{4}$} e^{7/4} a_U^{-1}
\phantom{\biggl[} \cr &\approx& 1.44 a_U^{-1}.
\label{LQCDdef}
\eeqa

In Fig.~(\ref{AAxPlot}), we have plotted the trace of
the correlation function~(\ref{MVCorrelator})
in position space
versus the dimensionless distance $X$.
In drawing these curves, we
have assumed that the longitudinal coordinates
have been fixed at some value such 
that $a^2 \chi(x^{-},x^{\prime -}) = 20$. 
Notice that the curve for the \MV\
correlator given in Eq.~(\ref{RunAway}) begins to 
diverge significantly from the other two curves near
$X=0.3$.  At $X \approx 0.347$,
corresponding to $\vert \xt -  \xprimet \vert = \LQCD^{-1}$,
the \MV\ correlator vanishes.  Beyond this point, it rockets
off to $-\infty$ faster than any exponential.
In contrast, both the uniform and Gaussian curves are well-behaved.
The correlation function generated from the uniform quark/nucleon
distribution vanishes for $X \ge 1$.

Next, we perform the 
Fourier transform to form the 
gluon number density, Eq.~(\ref{GlueEnd}).
The results are plotted in
Fig.~(\ref{FqPlotLinear}) for a fixed value of 
the longitudinal momentum $q^{+}$.
In order to define the Fourier transform in the
MV case, we have done what was suggested in 
Ref.~\cite{paper10} and simply cut off the $\deltat$ integration at
$\LQCD^{-1}$.  All three curves have generally the same over-all
shape:  a plateau at small values of $\qt$, a sharp decrease
at intermediate values of $\qt$, and a tail for large values of $\qt$.
The differing values for $\qt \rightarrow {\bf 0}$ may
easily understood from the position space functions in 
Fig.~\ref{AAxPlot}.  The \MV\ curve has the smallest number of
zero momentum gluons because of
the abrupt cut-off imposed at $\Delta = \LQCD^{-1}$.  
There are more
zero momentum gluons in the uniform case, since the corresponding
position space function extends out to $\vert\xt - \xprimet\vert = 2a$.
Finally, the exponential tail at large distances in
the Gaussian case produces even more zero momentum gluons.
At large momenta, the gluon number density is supposed
to go like $1/\qt^2$.  To illustrate this transition, we have plotted
the gluon number density multiplied by $(aq)^2$
in Fig.~\ref{qSqrFqPlot}.
In this figure, the uniform and Gaussian curves are virtually
indistinguishable.  Beyond $(aq)^2 = 10^4$, they remain flat.
The oscillations visible in the \MV\ curve are a result of the sharp
cutoff of the Fourier transform integral.  As $(aq)^2$ grows, 
they become 
more rapid, but decrease in amplitude.

Next, we illustrate
the effects of the non-Abelian contributions to the correlation
function.  For Fig.~\ref{HO-AAxPlotRatio}, we generate the ratio
\beq
{ {\TRcfun_{\rm all\ns orders}} \over {\TRcfun_{\rm lowest\ns order}} }
=
{ 
{  e^{N_c\chi L(\xt-\xprimet)} - 1 }
\over
{N_c \chi L(\xt-\xprimet)} 
}.
\label{xRatio}
\eeq
From this expression, we see that the relative importance
of the non-Abelian terms depends on the magnitude of $\chi$:
if $\chi \rightarrow 0$, then this ratio goes to 1.
Fig.~\ref{HO-AAxPlotRatio} plots this ratio for various
values of $a^2\chi$ ranging from 2 to 40 using the uniform
version of Kovchegov's model.\footnote{If we employ 
Eq.~(\protect\ref{mu-sqr}) and set $R \approx A^{1/3} a$,
we find that a typical magnitude for $a^2\chi$ is approximately 2.5
for uranium.}
We see that the non-Abelian terms have the effect of
suppressing the magnitude of the correlation functions at
large distances relative to the Abelian result.  
In momentum space, this translates into a depletion of
low momentum gluons.

Finally, we present Fig.~\ref{HO-FqPlotRatio}.  This is a
plot of the all-orders gluon density~(\ref{GlueEnd})
divided by the lowest-order gluon density~(\ref{F0def})
at fixed $q^{+}$ 
for various values of $\CHI_\infty$.  
We see that for small values of $\qt$, there are fewer gluons
as $\CHI_\infty$ increases.  At very large values of $\qt$, there is
no change.   On the other hand, at small values of $\qt$, 
this ratio is less than one, signalling that in this region
the number of gluons is no longer simply proportional
to the number of nucleons.  The amount of suppression increases
with increasing $\CHI_\infty$, consistent with gluon recombination
scenario envisioned in Ref.~\cite{GLR}.
Because of the transverse momentum sum rule~(\ref{SumRule}),
the area under each of the curves should be equal.  Thus,
we see a pile up of gluons at intermediate values of $\qt$.
These features of the gluon number density depend only on
choosing a correlator consistent
with color screening at large distances ($\sgn\curlyD(\Delta) = -1$
for large $\Delta$).

We close this section by evaluating the parameters
appearing in the expression for the average transverse
momentum-squared within this model.  According to Eq.~(\ref{LogMom}),
the value of the scale associated with the momentum logarithms 
may be constructed from Eqs.~(\ref{curlyLasymU})
and~(\ref{curlyLasymG}):
\beq
\Qprobe_0 =
\cases{
{1\over2}\exp(-\gamma_E+{7\over4}) \ts a_U^{-1}
\approx 1.615 a_U^{-1},\phantom{\Bigl[}  &
uniform distribution; \cr
\phantom{{1\over2}+{7\over4}} \exp(-{\gamma_E\over2}) \ts a_G^{-1}
\approx 0.749 a_G^{-1}, \phantom{\Bigl[} &
Gaussian distribution.\cr}
\label{logscales}
\eeq
If we relate $a_U$ to $a_G$ in the manner specified by the
ultraviolet matching condition~(\ref{aGdef}), these two
expressions become equal.
On the other hand, the value of the smooth part of the correlation
function at the origin appearing in Eq.~(\ref{quadratic})
in the two models is
\beq
4\pi C(0) =
\cases{
\displaystyle{{9}\over{2 a_U^2}}, &
uniform distribution; \cr
\phantom{x} & \cr
\displaystyle{{1}\over{a_G^2}}, &
Gaussian distribution.\cr}
%% see page x1205
\label{Corigin}
\eeq
These two quantities differ by a factor of
${{1}\over{18}}\exp({7\over2}-\gamma_E) \approx 1.03$ 
when Eq.~(\ref{aGdef}) is applied.

\begin{figure}[h]

\vspace*{15cm}
\includegraphics{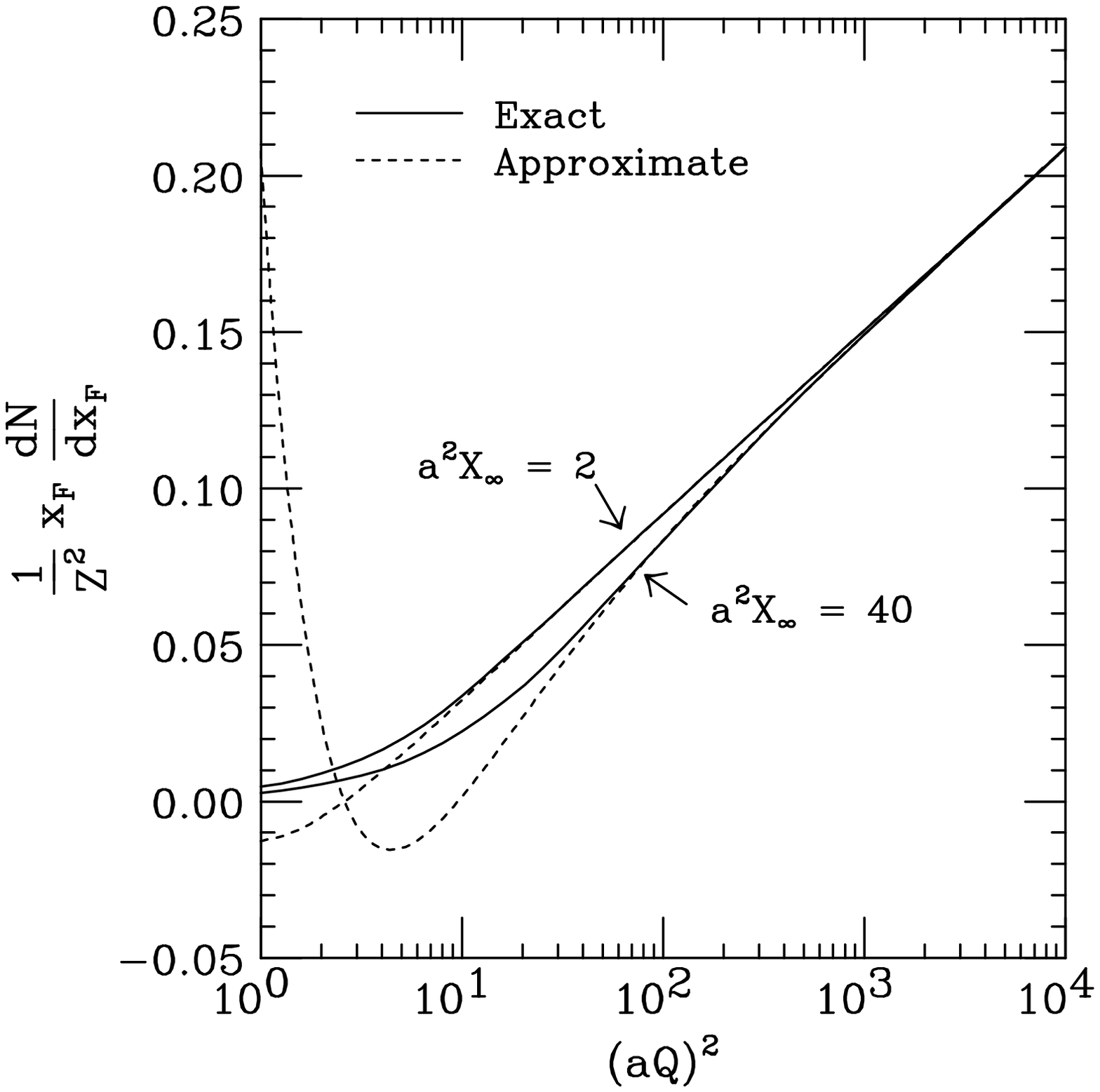}
\includegraphics{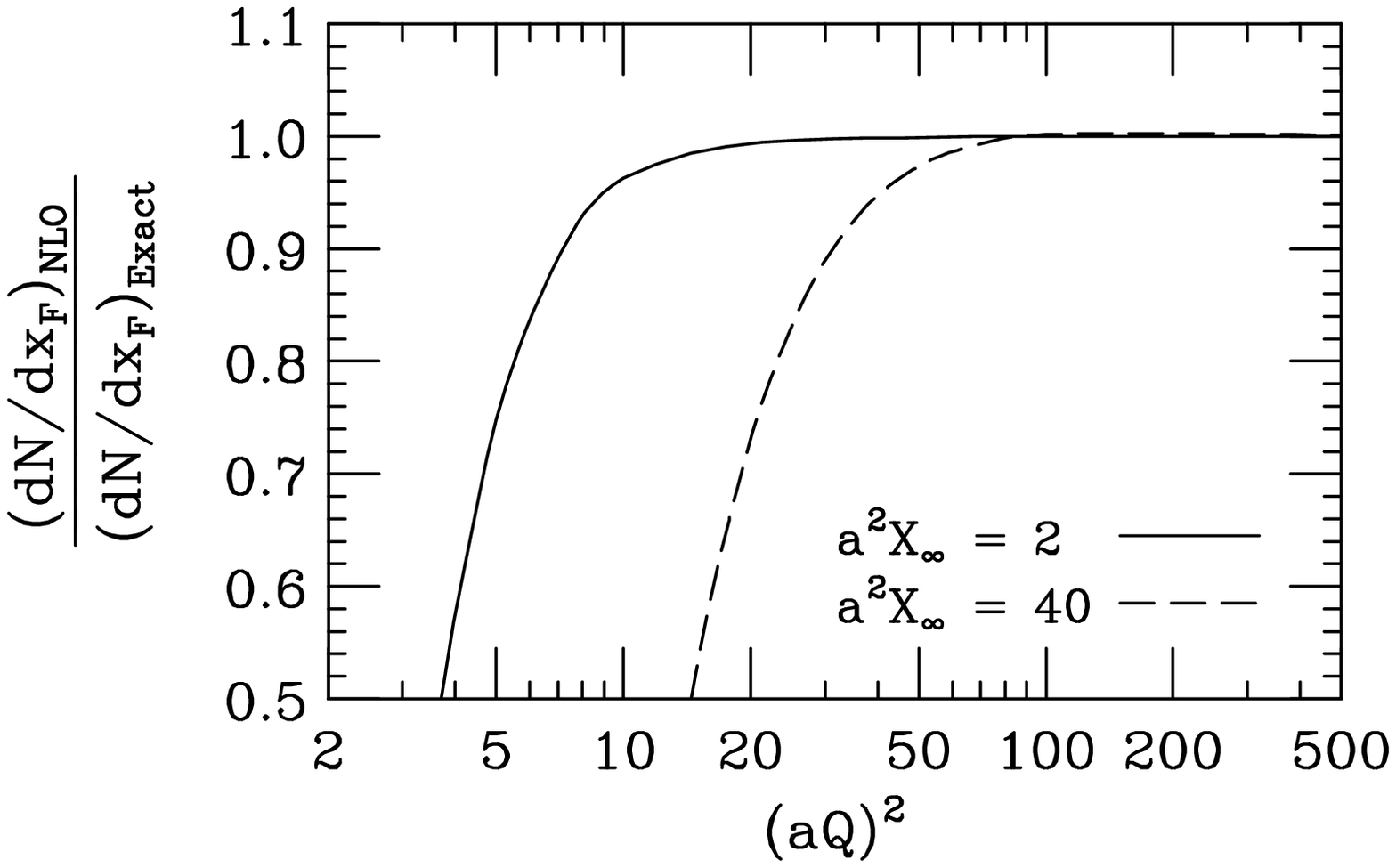}
\vspace{4.2cm}

\caption[]{Top:  Gluon structure function 
$\xf g(\xf,\Qprobe^2) = \xf dN/d\xf$
divided by the total color charge-squared for a large nucleus.
Plotted is a comparison of the 
next-to-leading order approximation~(\protect\ref{zeroth})
with the exact result obtained by numerical integration 
of~(\protect\ref{StrFnDef}) for two different values of $\CHI_{\infty}$.
Bottom:  The ratio of the next-to-leading order approximation
to the exact result in the region where the approximation begins
to break down.
}
\label{dNdqPlus}
\end{figure}

\begin{figure}[h]

\vspace*{15cm}
\includegraphics{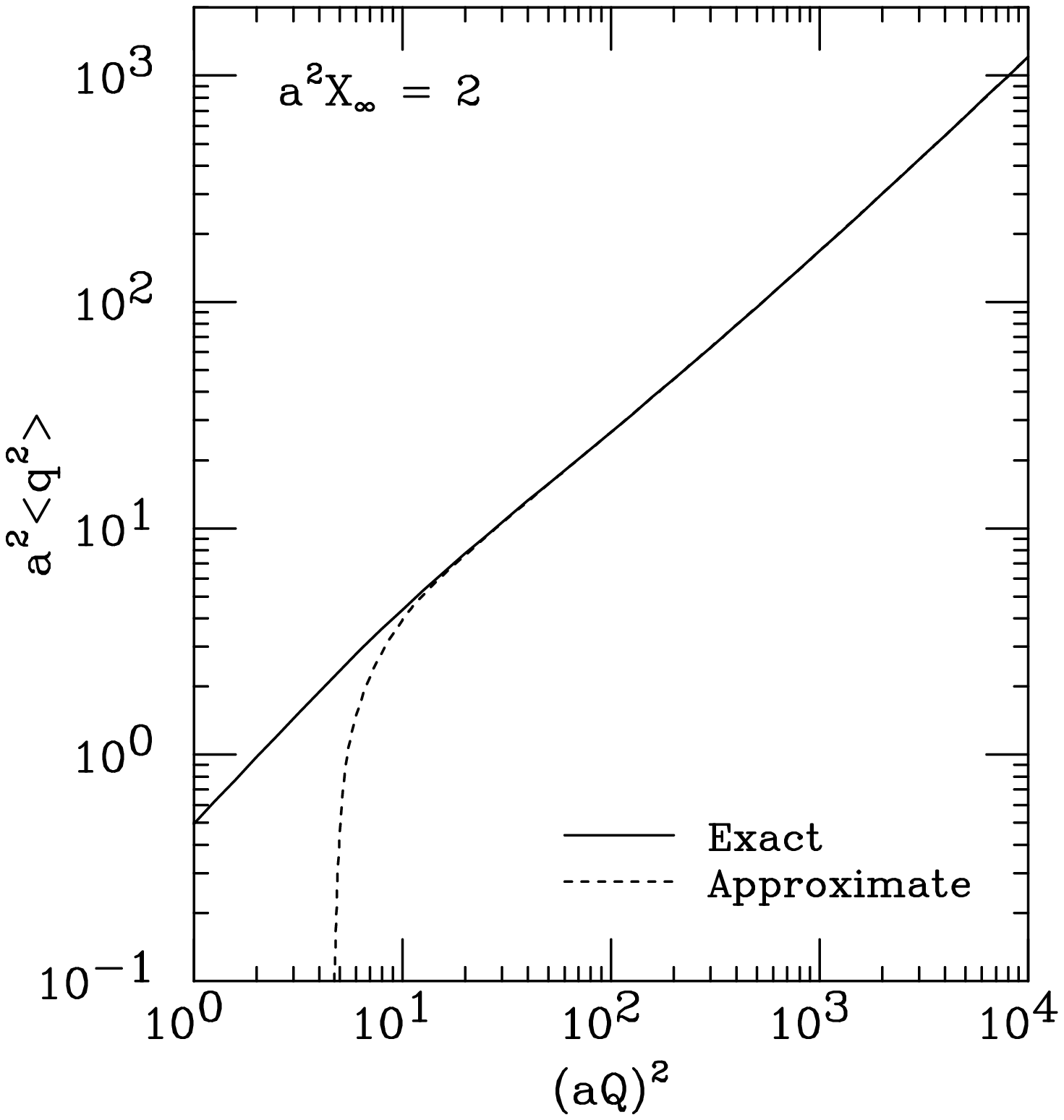}
\includegraphics{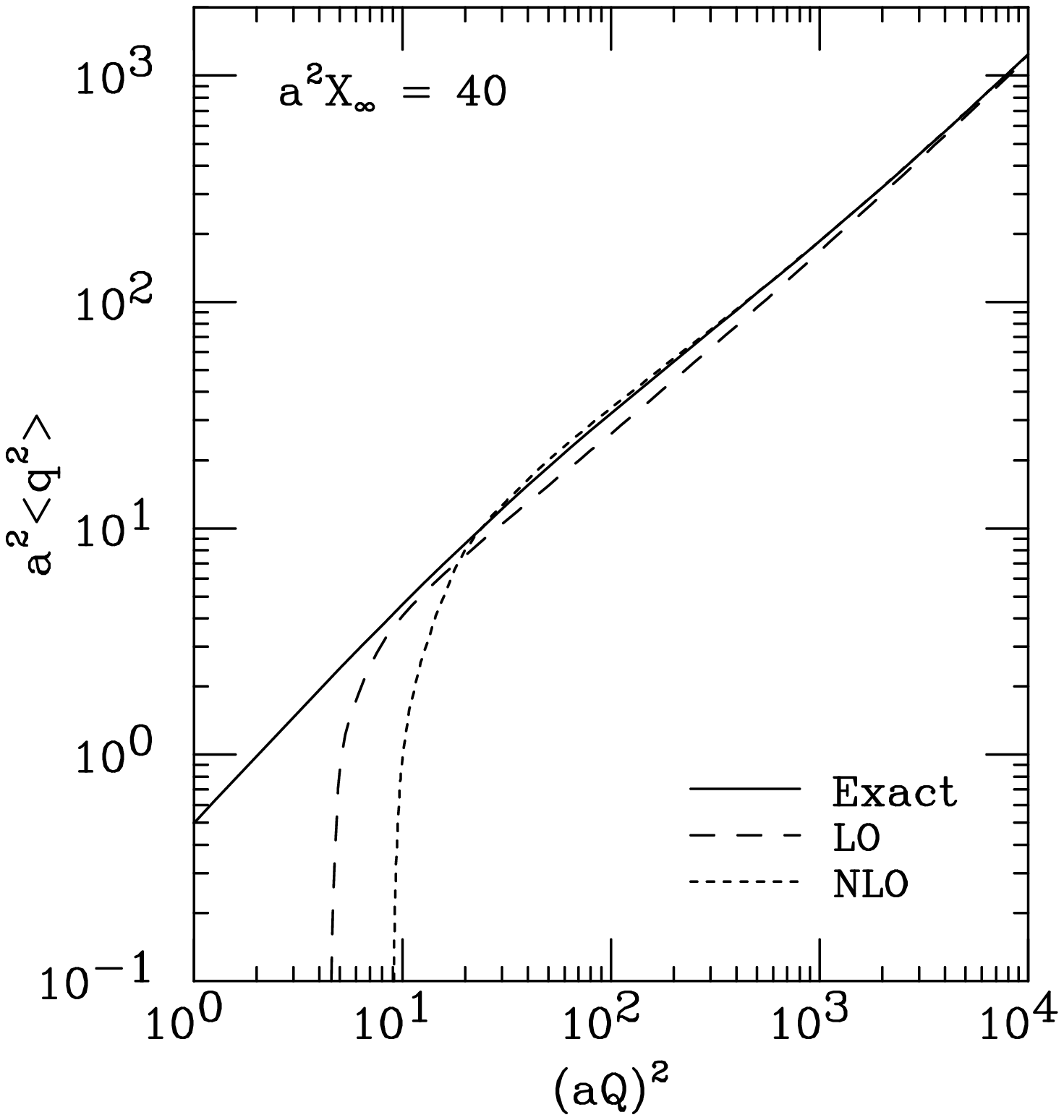}
\includegraphics{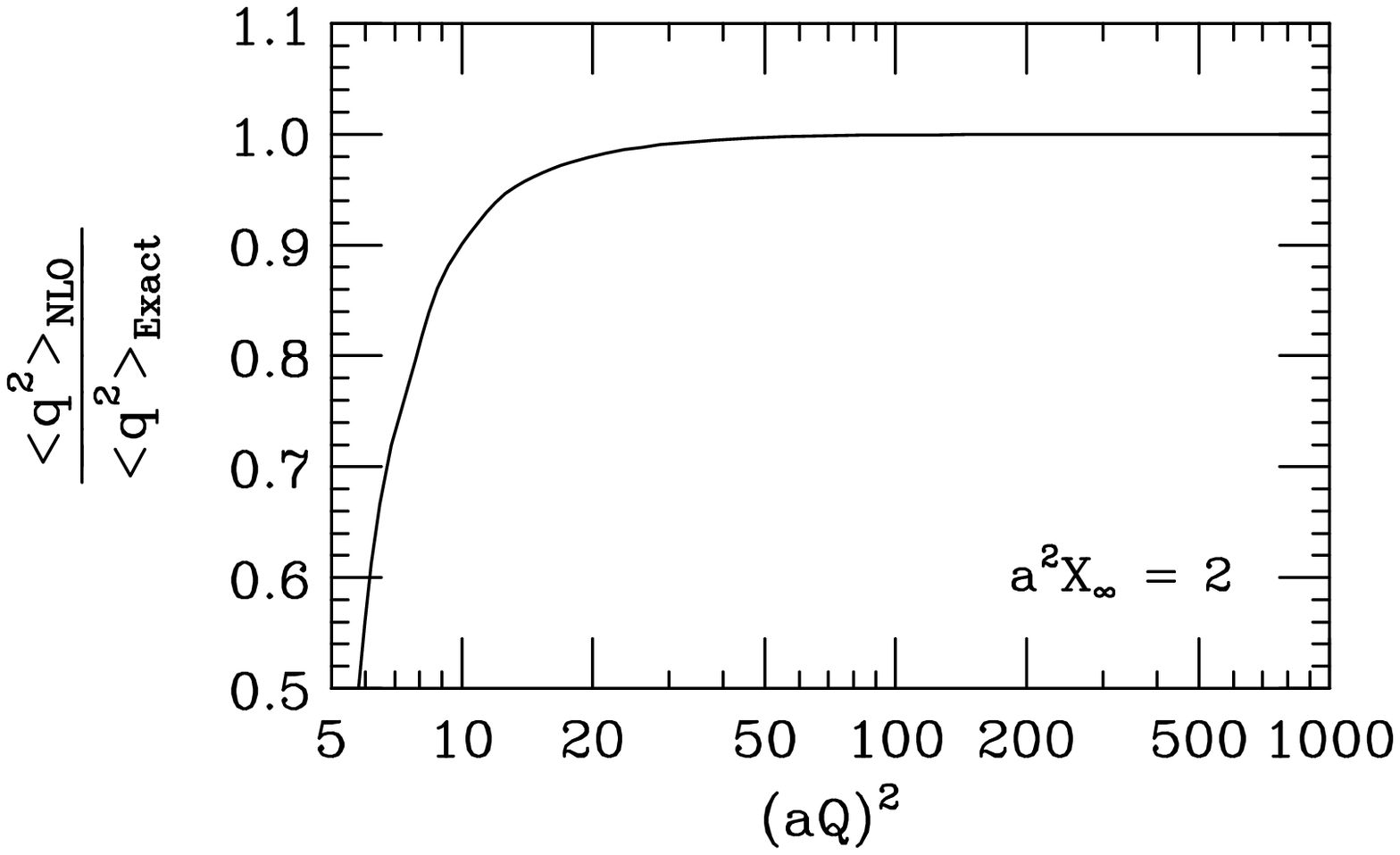}
\includegraphics{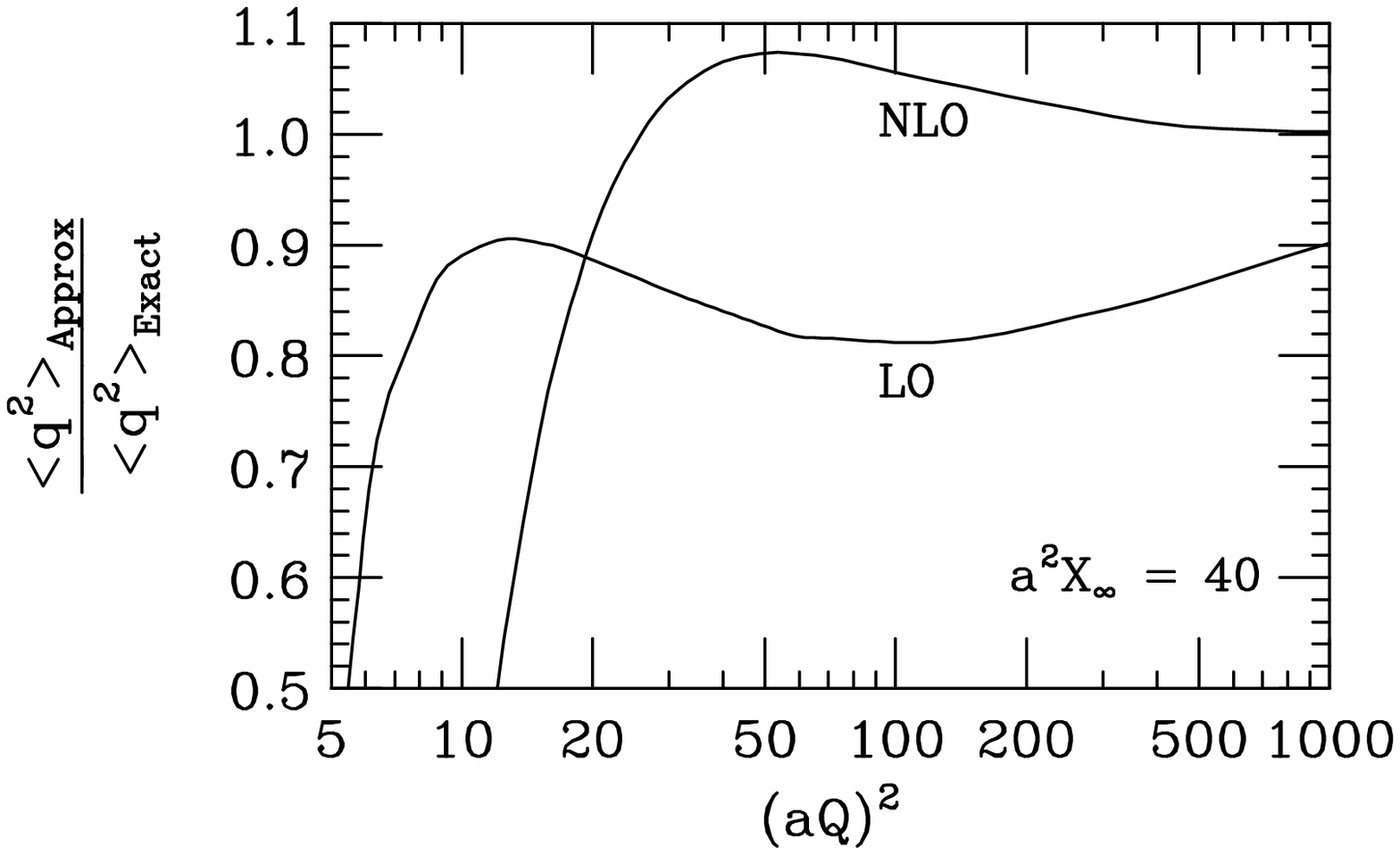}
\vspace{2.0cm}

\caption[]{Top:  Mean transverse momentum-squared as a function
of the parton resolution scale $\Qprobe^2$.  
Plotted is a comparison of the next-to-leading order
approximation~(\ref{quadratic}) with the exact result 
obtained by numerical integration for $a^2\CHI_\infty=2$,
and a comparison of the leading order, next-to-leading order,
and exact results for $a^2\CHI_\infty=40$.
Bottom:  The ratio of the approximate results to the 
exact results in the region where the approximation begins
to break down.
}
\label{Q2average}
\end{figure}

In Figs.~\ref{dNdqPlus} and~\ref{Q2average} we compare
the approximate results for $dN/d\xf$ and $\langle \qt^2 \rangle$
obtained in Sec.~\ref{MOMENTTEXT} with ``exact'' curves generated
by numerical integration of the full non-linear expression
for the gluon number density.  Since $\Qprobe_0$ does not
depend on whether we use a uniform or Gaussian distribution
for the quarks, and $\Ihat(0)$ only weakly so, we have employed
the uniform distribution in these plots, since its numerical
integral is more convenient to set up.
Fig.~\ref{dNdqPlus} is a plot
of $\xf dN/d\xf$ divided by the total charge-squared.  
In order to see more easily how the next-to-leading order
approximation~(\ref{zeroth}) is performing, we have
also plotted the ratio of the approximate to exact result
in the region where the approximation begins to diverge
from the full all-orders value.
We see that our approximation gives an excellent description
of the full result for $(a\Qprobe)^2$ as low as about 10
when $a^2\CHI_\infty =2$.
As could have been anticipated by studying 
Fig.~\ref{HO-FqPlotRatio}, for $a^2\CHI_{\infty}=40$
our approximation begins to break down a bit sooner, at around
$(a\Qprobe^2) = 50$ or so.

Fig.~\ref{Q2average} compares the exact and approximate 
curves for $\langle \qt^2 \rangle$ versus $\Qprobe^2$.
The cases $\CHI_{\infty}=2$ and $\CHI_{\infty}=40$ differ
by so little that they are virtually indistinguishable if
presented on the same plot.  This is a consequence of the
leading behavior $\langle \qt^2 \rangle \propto \Qprobe^2$.
For $a^2\CHI_\infty=40$ we have shown the leading order
approximation as well as the complete next-to-leading order
curve generated from Eq.~(\ref{quadratic}), since
this value of $\CHI_\infty$ is large enough for the difference
between the two approximations to be visible.
We have also plotted the ratio of the approximate to exact
results in the ``interesting'' region.  The behavior of
the $a^2\CHI_\infty=40$ result may be understood in terms
of Fig.~\ref{HO-FqPlotRatio}:  for large values of $\CHI_\infty$,
there is a pile-up of gluons at intermediate values of $\qt^2$.
This effect is purely non-Abelian in nature.
Now, the leading order approximation is completely independent
of any of the non-Abelian contributions to the gluon number
density.  Thus, it is not surprising that the leading order
expression falls short of the full result at moderate $\Qprobe^2$.
The first subleading contribution happens to overcorrect
by a small amount ($\sim 5\%$).  Presumably the next term in
the expansion will be negative in this region.
In any event, the next-to-leading order expression is good
to within a few percent for $(a\Qprobe)^2$ as low as about
20 for both values of $a^2\CHI_\infty$.

%%%%%%%%%%%%%%%%%%%%%%%%%%%%%%%%%%%%%%%%%%%%%%%%%%%%%%%%%%%%%%%%
%%%%%%%%%%%%%%%%%%%%%%%%%%%%%%%%%%%%%%%%%%%%%%%%%%%%%%%%%%%%%%%%
%%%%%%%%%%%%%%%%%%%%%%%%%%%%%%%%%%%%%%%%%%%%%%%%%%%%%%%%%%%%%%%%
%%%%%%
%%%%%%    CONCLUSIONS
%%%%%%
%%%%%%%%%%%%%%%%%%%%%%%%%%%%%%%%%%%%%%%%%%%%%%%%%%%%%%%%%%%%%%%%
%%%%%%%%%%%%%%%%%%%%%%%%%%%%%%%%%%%%%%%%%%%%%%%%%%%%%%%%%%%%%%%%
%%%%%%%%%%%%%%%%%%%%%%%%%%%%%%%%%%%%%%%%%%%%%%%%%%%%%%%%%%%%%%%%

\section{Conclusions}\label{CONC}

We have improved the McLerran-Venugopalan model by introducing
a constraint on the charge-density correlation function to 
ensure that it is consistent with a nucleus which is, as
a whole, color neutral.  We find that imposing color-neutrality
eliminates the divergences in the gluon number density present
in the original \MV\ model at small values of the transverse
momentum, provided that the transverse part of the charge-density
correlation function, $\curlyD(\Delta)$ is rotationally invariant
and falls off faster than $1/\Delta^4$ at large distances.
In this situation, the gluon number density approaches a 
constant value as $\qt\rightarrow{\bf 0}$.
To obtain a gluon number density which goes as $\ln\qt^2$
at small transverse momenta, we must have $\curlyD(\Delta)$
fall off more slowly than $1/\Delta^4$.  This is permissible
only if we impose the additional
constraint that $\curlyD(\Delta) < 0$ at large distances.  Then
we can use a $\curlyD(\Delta)$ which falls off nearly as
slowly as $1/\Delta^2$.

Because we have an expression which is mathematically
well-behaved, we are able to demonstrate that the gluon
distribution function within the \MV\ model is proportional
to $1/\xf$ to all orders in the coupling constant, independent
of the functional form of $\mu^2(x^{-})$.
This conclusion hinges upon the choice of a purely local
dependence on the longitudinal distance in the charge-density
correlator.  The inclusion of quantum 
corrections~\cite{paper9,paper11,paper39,paper40,paper43,paper34,paper44}
is expected to change this situation.

We have derived a transverse sum rule, Eq.~(\ref{SumRule}),
for the gluon number
density.  This sum rule indicates that the total number of
gluons may be computed as if the theory were purely Abelian:
the only effect of the non-Abelian terms is to shift gluons
from one value of transverse momentum to another.  
As a consequence
of this sum rule, we have shown that the gluon structure
function in a nucleus at large $\Qprobe^2$
is simply given by $A$ times the result
of the DGLAP equation for a single nucleon.  
We have shown that if we employ a charge density correlator
which is consistent with charge screening at large distances,
then we have saturation~\cite{GLR}:  as the density of
color charge $\CHI_\infty$ is increased, the number of low
momentum gluons grows more slowly than the number of nucleons.

We have also presented
relatively simple expressions for $dN/d\xf$ [Eq.~(\ref{zeroth})]
and the mean transverse
momentum-squared [Eq.~(\ref{quadratic})] as a function
of $\Qprobe$, accurate to order $1/\Qprobe^2$ and
$\Qprobe^0$ respectively within this model.  
We are able to compute the scales of the logarithms
in terms of a single model-dependent function $\Ihat(\Delta)$ 
which describes the manner in which charge neutrality is
approached at scales near and beyond the nucleon size.
Rather remarkably, the complicated non-linear structure
of the full gluon distribution function may be understood
in terms of the $\Qprobe^2$ expansion.  That is, only
the Abelian terms enter in at leading order in this expansion,
with the non-Abelian contributions making their first appearance
at next-to-leading order.

%%%%%%%%%%%%%%%%%%%%%%%%%%%%%%%%%%%%%%%%%%%%%%%%%%%%%%%%%%%%%%%%
%%%%%%%%%%%%%%%%%%%%%%%%%%%%%%%%%%%%%%%%%%%%%%%%%%%%%%%%%%%%%%%%
%%%%%%%%%%%%%%%%%%%%%%%%%%%%%%%%%%%%%%%%%%%%%%%%%%%%%%%%%%%%%%%%
%%%%%%
%%%%%%      ACKNOWLEDGMENTS
%%%%%%
%%%%%%%%%%%%%%%%%%%%%%%%%%%%%%%%%%%%%%%%%%%%%%%%%%%%%%%%%%%%%%%%
%%%%%%%%%%%%%%%%%%%%%%%%%%%%%%%%%%%%%%%%%%%%%%%%%%%%%%%%%%%%%%%%
%%%%%%%%%%%%%%%%%%%%%%%%%%%%%%%%%%%%%%%%%%%%%%%%%%%%%%%%%%%%%%%%

\acknowledgements

This research is supported in part by
the Natural Sciences and Engineering Research Council of Canada
and the Fonds pour la Formation de Chercheurs 
et l'Aide \`a la Recherche of Qu\'ebec.
GM would like to thank Guy Moore numerous insightful 
discussions %and seemingly endless patience
during the course of this work.
CSL would like to thank Larry McLerran for instructive
discussions about his model.

\newpage
\appendix

%%%%%%%%%%%%%%%%%%%%%%%%%%%%%%%%%%%%%%%%%%%%%%%%%%%%%%%%%%%%%%%%
%%%%%%%%%%%%%%%%%%%%%%%%%%%%%%%%%%%%%%%%%%%%%%%%%%%%%%%%%%%%%%%%
%%%%%%%%%%%%%%%%%%%%%%%%%%%%%%%%%%%%%%%%%%%%%%%%%%%%%%%%%%%%%%%%
%%%%%%
%%%%%%    APPENDIX:  NOTATION AND CONVENTIONS
%%%%%%
%%%%%%%%%%%%%%%%%%%%%%%%%%%%%%%%%%%%%%%%%%%%%%%%%%%%%%%%%%%%%%%%
%%%%%%%%%%%%%%%%%%%%%%%%%%%%%%%%%%%%%%%%%%%%%%%%%%%%%%%%%%%%%%%%
%%%%%%%%%%%%%%%%%%%%%%%%%%%%%%%%%%%%%%%%%%%%%%%%%%%%%%%%%%%%%%%%

\section{Notation and Conventions}\label{NOTATION}

We begin by defining the light-cone coordinates $x^{\pm}$:
\beq
x^{\pm} \equiv 
{ {x^0 \pm x^3} \over {\sqrt{2}} } .
\eeq
The transverse coordinates $x^1$ and $x^2$ form a two-dimensional
vector which we denote simply as $\xt$, without the usual subscript
``T'' or ``$\perp$'' to avoid excessive clutter.  
Our metric has the signature $({-},{+},{+},{+})$.  Thus,
the light-cone dot product reads 
$
q^{\mu} x_{\mu} = -q^{+}x^{-} - q^{-}x^{+} + \qt\cdot\xt,
$
and $x_{\pm} = -x^{\mp}$.  We will think of $x^{+}$ as the time 
coordinate, and $x^{-}$ as the longitudinal distance coordinate ($z$).

Occasionally it will be helpful to switch to sum and difference
coordinates.  We will employ the convention
\beq
\sigmat \equiv \hbox{$1\over2$} (\xt + \xprimet),
\qquad\deltat \equiv \xt - \xprimet.
\label{SumDiff}
\eeq
This asymmetric pair of definitions 
has the desirable property of a unit Jacobian.

The classical Yang-Mills equations are
\beq
D_\mu F^{\mu\nu} = J^{\nu},
\label{YangMills}
\eeq
where we have employed matrix form, {\it i.e.}
$ J^{\nu} \equiv T^a J^{a\nu}$, etc.
The $T^a$ are the normalized Hermitian generators of SU($N_c$)
in the fundamental representation, satisfying 
$2 \ts\trace(T^a T^b) = \delta^{ab}$.
The covariant derivative is
\beq
D_\mu F^{\mu\nu} \equiv
\partial_\mu F^{\mu\nu} - i \Bl A_\mu,F^{\mu\nu} \Br
\eeq
and the field strength is defined in terms of the potential as
\beq
F^{\mu\nu} \equiv 
\partial^{\mu} A^{\nu} 
-\partial^{\nu} A^{\mu} 
-i \Bl A^{\mu},A^{\nu} \Br.
\eeq
These definitions assume that the coupling constant $g$ has
been absorbed into $J^{\nu}$ [see, for example, Eq.~(\ref{Kden})].
Note that this is not the same convention as employed 
by \MV\ in Refs.~\cite{paper1,paper2,paper3,paper4,paper9},
where they choose to explicitly extract the factor of $g$ from the
current.
We define the charge density $\rho(x^{-},\xt)$ appearing
in Eq.~(\ref{QCDcurrent}) such that the total charge $\Qtotal$
of the nucleus is given by
\beq
\Qtotal \equiv \int dx^{-} d^2\xt \ts \rho(x^{-},\xt).
\eeq

Because our intuitive picture of the parton model is most
simply realized in the light-cone gauge~\cite{paper31},
we elect to work primarily in that gauge.
The light-cone gauge is defined by the condition
\beq 
A^{+}(x) \equiv 0.  
\label{LightConeGaugeCondition}
\eeq
However, solution of the classical Yang-Mills equations for
the type of source we consider is simplest in the covariant
(Lorentz) gauge.  
For clarity,
gauge-dependent quantities will be written
with a tilde in the covariant gauge:  {\it e.g.}\ $\widetilde{A}$.
The covariant gauge is specified by 
\beq
\partial_\mu \widetilde{A}^{\mu}(x) \equiv 0.
\label{CovariantGaugeCondition}
\eeq

%%%%%%%%%%%%%%%%%%%%%%%%%%%%%%%%%%%%%%%%%%%%%%%%%%%%%%%%%%%%%%%%
%%%%%%%%%%%%%%%%%%%%%%%%%%%%%%%%%%%%%%%%%%%%%%%%%%%%%%%%%%%%%%%%
%%%%%%%%%%%%%%%%%%%%%%%%%%%%%%%%%%%%%%%%%%%%%%%%%%%%%%%%%%%%%%%%
%%%%%%
%%%%%%      APPENDIX:  GENERAL FORM FOR L
%%%%%%
%%%%%%%%%%%%%%%%%%%%%%%%%%%%%%%%%%%%%%%%%%%%%%%%%%%%%%%%%%%%%%%%
%%%%%%%%%%%%%%%%%%%%%%%%%%%%%%%%%%%%%%%%%%%%%%%%%%%%%%%%%%%%%%%%
%%%%%%%%%%%%%%%%%%%%%%%%%%%%%%%%%%%%%%%%%%%%%%%%%%%%%%%%%%%%%%%%

\section{The Function $L(\xt-\xprimet)$}\label{SIGMA}

%%%%%%%%%%%%%%%%%%%%%%%%%%%%%%%%%%%%%%%%%%%%%%%%%%%%%%%%%%%%%%%%
%%
%%         GENERAL CHARGE CORRELATOR
%%
%%%%%%%%%%%%%%%%%%%%%%%%%%%%%%%%%%%%%%%%%%%%%%%%%%%%%%%%%%%%%%%%

\subsection{General Color-Neutral Correlator}

In this appendix we present the calculation of $L(\xt-\xprimet)$
for a general correlator $\curlyD(\deltat)$.  Our starting point
is Eq.~(\ref{Ldef}) written in terms of the sum and difference
variables defined as in~(\ref{SumDiff}):
\beqa
L(\xt-\xprimet) \equiv 
\int d^2 \deltat \int d^2 \sigmat \ts\ts
\curlyD(\deltat) 
&\Bigl[& G(\xt - \sigmat - \half\deltat) 
         G(\xprimet - \sigmat + \half\deltat) 
\cr
&-&\half G(\xt-\sigmat - \half\deltat) 
         G(\xt-\sigmat + \half\deltat)
\cr 
&-&\half G(\xprimet-\sigmat-\half\deltat) 
         G(\xprimet-\sigmat+\half\deltat) \Bigr],
\label{L1}
\eeqa
All three terms of~(\ref{L1})
may be evaluated by considering the integral
\beq
{\cal S}(\xt,\xprimet,\deltat) \equiv 
{{1}\over{16\pi^2}} \int d^2\sigmat \ts
\ln \biggl( {
{\vert\xt-\sigmat-\half\deltat\vert^2}
\over
{\lambda^2}
} \biggr)
\ln \biggl( {
{\vert\xprimet-\sigmat+\half\deltat\vert^2}
\over
{\lambda^2}
} \biggr).
\label{S1}
\eeq
To deal with the logarithms, we insert the identity
\beq
\ln\biggl({{B}\over{A}}\biggr) 
= \int_0^{\infty} {{dz}\over{z}} \Bigl(e^{-Az}-e^{-Bz}\Bigr).
\eeq
Because we will ultimately insert 
our result for ${\cal S}$ into~(\ref{L1}),
we may drop any terms which do not depend on $\deltat$, thanks
to the charge neutrality condition~(\ref{Dint}).  Thus, the
$\lambda$-dependent bits of~(\ref{S1}) do not contribute, and
we are left with
\beq
{\cal S}(\xt,\xprimet,\deltat) = 
{{1}\over{16\pi^2}} 
\int_0^{\infty} {{da}\over{a}} \int_0^{\infty} {{db}\over{b}}
\int d^2\sigmat \ts
\exp\Bigl[-a(\sigmat-\xt+\half\deltat)^2
          -b(\sigmat-\xprimet-\half\deltat)^2 \Bigr].
\eeq
The $\sigmat$ integration is now Gaussian, and easily performed,
yielding
\beqa
{\cal S}(\xt,\xprimet,\deltat) = 
{{1}\over{16\pi}} 
\int_0^{\infty} {{da}\over{a}} \int_0^{\infty} {{db}\over{b}} \ts
{{1}\over{a+b}}\ts
\exp\biggl\{&-&a(\xt-\half\deltat)^2
          -b(\xprimet+\half\deltat)^2 
\cr       &+&{  { [a(\xt-\half\deltat)+b(\xprimet+\half\deltat)]^2 }
            \over{a+b} }
\biggr\}.
\eeqa
The next step is to insert
\beq
1 = \int_0^{\infty} {{d\tau}\over{\tau}} 
\ts\delta\biggl(1 - {{a+b}\over{\tau}}\biggr)
\eeq
and rescale $a\rightarrow \tau a$, $b \rightarrow \tau b$.
This allows us to simplify the exponent, producing
\beqa
{\cal S}(\xt,\xprimet,\deltat) &=& 
{{1}\over{16\pi}} 
\int_0^{\infty} {{d\tau}\over{\tau^2}} 
\int_0^{\infty} {{da}\over{a}} 
\int_0^{\infty} {{db}\over{b}} \ts
\delta(1-a-b)
\exp\Bigl\{-\tau a(1-a)(\xt-\xprimet-\deltat)^2 
\Bigr\}.  
\eeqa
The $b$ integration is now trivial, and the $a$ integration
becomes trivial after the variable change
\beq
\tau = {{u}\over{a(1-a)}}.
\eeq
Thus, we obtain
\beq
{\cal S}(\xt,\xprimet,\deltat) = 
{{1}\over{16\pi}} 
\int_0^{\infty} {{du}\over{u^2}} 
\exp\Bigl\{-u(\xt-\xprimet-\deltat)^2 
\Bigr\},
\eeq
which, when inserted into~(\ref{L1}) yields
\beq
L(\xt-\xprimet) =
{{1}\over{16\pi}} \int_0^{\infty}  {{du}\over{u^2}}
\int d^2\deltat \ts\ts
\curlyD(\deltat) 
\Bigl[ e^{-u(\xt-\xprimet-\deltat)^2 }
       -e^{-u\deltat^2} \Bigr].
\label{LgenAgain}
\eeq
Although the $u$ integration in~(\ref{LgenAgain}) appears to be
divergent, for a rotationally invariant correlation
function $\curlyD$, the integral actually is finite.
To see this, note that for small $u$ the integrand of~(\ref{LgenAgain})
looks like
\beq
{{du}\over{u}}\ts
\Bigl[2 (\xt-\xprimet)\cdot\deltat - (\xt-\xprimet)^2\Bigr] \ts
\curlyD(\deltat) 
+ \hbox{finite terms}.
\label{smallu}
\eeq
The term containing $(\xt-\xprimet)^2$ vanishes when we perform 
the $\deltat$ integration, because of~(\ref{Dint}).
If $\curlyD$ is rotationally invariant, then the angular part
of the $\deltat$ integration will cause the 
$(\xt-\xprimet)\cdot\deltat$ 
term to vanish as well.

Because the integral over $u$ is finite, we may employ
a trick reminiscent of dimensional regulation and write
\beqa
L(\xt-\xprimet) = 
{{1}\over{16\pi}} 
\int d^2\deltat 
\int_0^{\infty}  {{du}\over{u^{2-\eps}}}
\ts\ts
\curlyD(\deltat) 
\Bigl[ e^{-u(\xt-\xprimet-\deltat)^2 }
       -e^{-u\deltat^2} \Bigr],
\label{plainL3}
\eeqa
where we understand that the limit $\eps\rightarrow 0$ is to be
taken on the right hand side as soon as it is safe to do so.
The $u$ integration may be performed by using
the analytic continuation of 
\beq
\int_0^\infty {{du}\over{u^{\kappa+1}}} \ts\ts
[e^{-\beta u} - e^{-\alpha u}]
= {{1}\over{\kappa}}\ts \Gamma(1-\kappa) [\alpha^\kappa - \beta^\kappa]
\label{grad3.434.1}
\eeq
(Eq.~(3.434.1) of Ref.~\cite{PhysicistsFriend}).
After applying~(\ref{grad3.434.1}) to~(\ref{plainL3}) and
performing some algebra we obtain
\beqa
L(\xt-\xprimet) &=& 
{{1}\over{16\pi}} \ts
{{1}\over{\eps}}\ts {{\Gamma(1-\eps)}\over{1-\eps}}
\int d^2\deltat \ts
\curlyD(\deltat) \ts\ts  
\Bigl[ 2 \deltat\cdot(\xt-\xprimet) - (\xt-\xprimet)^2 \Bigr]
\cr\nonumber\\[0.04cm] &+&
{1\over{16\pi}} \ts
\int d^2\deltat \ts\ts
\curlyD(\deltat) \ts
\Bigl[
(\xt-\xprimet-\deltat)^2 \ln (\xt-\xprimet-\deltat)^2
- \deltat^2 \ln \deltat^2
\Bigr]
\label{plainL4}
\eeqa
The pole term is the manifestation of the behavior
described in~(\ref{smallu}).  For a rotationally invariant
function $\curlyD(\deltat)$, it vanishes.
Thus, we are left with just 
\beqa
L(\xt-\xprimet) &=& 
{1\over{16\pi}} \ts
\int d^2\deltat \ts\ts
\curlyD(\deltat) \ts
\Bigl[
(\xt-\xprimet-\deltat)^2 \ln (\xt-\xprimet-\deltat)^2
- \deltat^2 \ln \deltat^2
\Bigr].
\label{LgenApx}
\eeqa

Differentiation of~(\ref{LgenApx}) to obtain explicit expressions for
the two functions defined in Eq.~(\ref{decomp}) produces
\beq
\curlyL(\xt-\xprimet) =
{{-1}\over{8\pi}}
\int d^2\deltat \ts\ts
\curlyD(\deltat) \ts
\ln(\xt-\xprimet-\deltat)^2 
\label{curlyLgenApx}
\eeq
and 
\beq
\curlyL_{ij}(\xt-\xprimet) = 
{{-1}\over{4\pi}} 
\int d^2\deltat \ts\ts
\curlyD(\deltat) \ts
{
{(x-x'-\Delta)_i(x-x'-\Delta)_j}
\over
 {(\xt-\xprimet-\deltat)^2 }
}.
\label{traceless}
\eeq
Recall that $\curlyL_{ij}$ was defined to be the traceless
piece of $\partial_i\partial^\prime_j L(\xt-\xprimet)$.
It does not contribute to the gluon number density.

%%%%%%%%%%%%%%%%%%%%%%%%%%%%%%%%%%%%%%%%%%%%%%%%%%%%%%%%%%%%%%%%
%%
%%         ROTATIONALLY INVARIANT CHARGE CORRELATOR
%%
%%%%%%%%%%%%%%%%%%%%%%%%%%%%%%%%%%%%%%%%%%%%%%%%%%%%%%%%%%%%%%%%

\subsection{Rotationally Invariant Correlator}

If $\curlyD(\deltat)$ is rotationally invariant, we may perform
the angular integration appearing in Eq.~(\ref{LgenApx}).
Letting $\psi$ be the angle between $\xt$ and
$\deltat$ we have
\beqa
L(\xt) = 
{1\over{16\pi}} \ts
\int_0^{\infty} && d\Delta \ts\ts\Delta\ts
\curlyD(\Delta) 
\phantom{\Biggl[} \cr \phantom{\Biggl[} \times &&
\nts \int_0^{2\pi} \nts d\psi \ts
\Bigl[ 
(x^2 - 2 x \Delta \cos\psi + \Delta^2) \ln (x^2-2x\Delta\cos\psi + \Delta^2)
- \Delta^2 \ln \Delta^2
\Bigr]
\eeqa
The non-trivial angular integrals appearing in this expression
are
\beq
\int_0^{2\pi} \nts d\psi \ts
\ln (x^2-2x\Delta\cos\psi + \Delta^2) = 
\cases{2\pi\ln x^2 & if $x>\Delta$;\cr
       2\pi\ln \Delta^2 & if $x<\Delta$,\cr}
\label{amusing1}
\eeq
and
\beq
\int_0^{2\pi} \nts d\psi \ts
\cos\psi \ts \ln (x^2-2x\Delta\cos\psi + \Delta^2) = 
\cases{-2\pi\Delta/x & if $x>\Delta$;\cr
       -2\pi x/\Delta & if $x<\Delta$.\cr}
\label{amusing2}
\eeq
Eqs.~(\ref{amusing1}) and~(\ref{amusing2}) force us to split up
the radial integration into two pieces:
\beqa
L(\xt) &=& 
{{1}\over{8}} \int_0^{x} d\Delta\ts\ts\Delta\ts\curlyD(\Delta)
\Bigl[ (x^2+\Delta^2) \ln x^2 - \Delta^2 \ln \Delta^2 + 2\Delta^2 \Bigr]
\cr &+&
{{x^2}\over{8}} \int_x^{\infty} d\Delta\ts\ts\Delta\ts\curlyD(\Delta)
\Bigl[ \ln \Delta^2 + 2 \Bigr].
\label{TwoForms2}
\eeqa
Similarly, we may perform the angular integration appearing
in~(\ref{curlyLgenApx}) to obtain
\beq
\curlyL(\xt) = 
-{{1}\over{4}} \int_0^{x} d\Delta\ts\ts\Delta\ts\curlyD(\Delta)
\ln x^2
-{{1}\over{4}} \int_x^{\infty} d\Delta\ts\ts\Delta\ts\curlyD(\Delta)
 \ln \Delta^2.
\label{TwoFormsCurly}
\eeq
Let us further assume that 
$\curlyD(\deltat) = \delta^2(\deltat) - \Ihat(\Delta)$, 
where $\Ihat(\Delta)$ is smooth at the origin.  The color-neutrality
requirement implies that
\beq
\int d^2\deltat \ts\Ihat(\Delta) = 1,
\qquad\hbox{or}\qquad
\int_0^{\infty} d\Delta \ts\Delta \Ihat(\Delta) = {{1}\over{2\pi}} .
\label{IhatOK}
\eeq
Making this ansatz allows us to rewrite Eqs.~(\ref{TwoForms2})
and~(\ref{TwoFormsCurly}) as
\beq
L(\xt) = -{1\over{8\pi}} x^2 
+ {1\over{16\pi}} x^2 \ln\biggl({{x^2}\over{\Xi^2}} \biggr)
+ {1\over8} \int_0^{x} d\Delta \ts\Delta\Ihat(\Delta)
\Biggl[ (x^2+\Delta^2)\ln\biggl( {{\Delta^2}\over{x^2}} \biggr)
       + 2(x^2-\Delta^2) \Biggr]
\label{Lexp}
\eeq
and
\beq
\curlyL(\xt) =
-{1\over{8\pi}} \ln\biggl({{x^2}\over{\Xi^2}} \biggr)
+ {1\over4} \int_0^{x} d\Delta \ts \Delta \Ihat(\Delta)
            \ln\biggl( {{x^2}\over{\Delta^2}} \biggr),
\label{curlyLexp}
\eeq
where
\beq
\ln\Xi^2 \equiv 2\pi
\int_0^{\infty} d\Delta \ts\Delta\Ihat(\Delta)\ln \Delta^2.
\label{XiDef}
\eeq
Because of Eq.~(\ref{IhatOK}), the integral defining the
length scale $\Xi$ in Eq.~(\ref{XiDef}) will be finite
for any reasonable function $\Ihat(\Delta)$.  Furthermore,
the value of $\Xi$ will be closely tied to the scale
implicit in $\Ihat(\Delta)$, namely the nucleon size $a\sim\LQCD^{-1}$.
Since we have stipulated that $\Ihat(\Delta)$ be smooth
at $\Delta=0$, we conclude that Eqs.~(\ref{Lexp}) and~(\ref{curlyLexp})
are telling us that $L(\xt)$ and $\curlyL(\xt)$ 
may be recast in the form of a power series in $x$ plus
$\ln x^2$ times a power series in $x$.  Furthermore, the
power series for $\curlyL(\xt)$ begins at $x^0$, whereas
the first non-vanishing contribution to the series for $L(\xt)$
is $x^2$.
This observation will be useful in organizing the computation
of the mean transverse momentum-squared associated with our
gluon distribution.

%%%%%%%%%%%%%%%%%%%%%%%%%%%%%%%%%%%%%%%%%%%%%%%%%%%%%%%%%%%%%%%%
%%%%%%%%%%%%%%%%%%%%%%%%%%%%%%%%%%%%%%%%%%%%%%%%%%%%%%%%%%%%%%%%
%%%%%%%%%%%%%%%%%%%%%%%%%%%%%%%%%%%%%%%%%%%%%%%%%%%%%%%%%%%%%%%%
%%%%%%
%%%%%%      APPENDIX:  MOMENTS OF THE DISTRIBUTION FUNCTION
%%%%%%
%%%%%%%%%%%%%%%%%%%%%%%%%%%%%%%%%%%%%%%%%%%%%%%%%%%%%%%%%%%%%%%%
%%%%%%%%%%%%%%%%%%%%%%%%%%%%%%%%%%%%%%%%%%%%%%%%%%%%%%%%%%%%%%%%
%%%%%%%%%%%%%%%%%%%%%%%%%%%%%%%%%%%%%%%%%%%%%%%%%%%%%%%%%%%%%%%%

\section{Computation of the Average Transverse 
Momentum-Squared}\label{MOMENTS}

In this appendix we will outline the computation
of the value of $\langle \qt^2 \rangle$,
the average transverse momentum-squared for the gluon 
distribution~(\ref{GlueEnd}).  By definition 
\beq
\langle \qt^2 \rangle \equiv
{
\displaystyle{\int_{\vert\qt\vert\le\Qprobe}\nts\nts\nts 
              d^2\qt\ts \qt^2 {{dN}\over{d\xf d^2\qt}} }
\over
\displaystyle{\int_{\vert\qt\vert\le\Qprobe}\nts\nts\nts 
              d^2\qt\ts {{dN}\over{d\xf d^2\qt}} }
}.
\label{Q2def}
\eeq
We will do our computation through next-to-leading order.
Since the numerator of~(\ref{Q2def}) grows like $\Qprobe^2$
in the limit $\Qprobe \rightarrow \infty$, this means we work
to order $\Qprobe^0$.  Likewise, since the denominator of~(\ref{Q2def})
grows like $\ln\Qprobe^2$ for large $\Qprobe$, we determine
it through order $1/Q^2$. 

For the purposes of this calculation we
will assume that $\curlyD(\deltat)$
is of the form
\beq
\curlyD(\deltat) = \delta^2(\deltat) - \Ihat(\Delta),
\label{TwoTermsTwo}
\eeq
where the function $\Ihat(\Delta)$ is smooth at the origin.
As discussed in Sec.~\ref{ASYMPT}, $\curlyD(\deltat)$
should contain $\delta^2(\deltat)$ if the results are
to agree with an Abelian theory in the ultraviolet.
This is consistent with having point-like quarks in the
nucleons.  Beyond specifying that $\Ihat(\Delta)$ is
rotationally invariant and has unit integral 
[to satisfy~(\ref{Dint})], we will not make any additional
assumptions about this function.

%%%%%%%%%%%%%%%%%%%%%%%%%%%%%%%%%%%%%%%%%%%%%%%%%%%%%%%%%%%%%%%%
%%
%%            NUMERATOR
%%
%%%%%%%%%%%%%%%%%%%%%%%%%%%%%%%%%%%%%%%%%%%%%%%%%%%%%%%%%%%%%%%%

\subsection{Calculation of $\langle\qt^2\rangle dN/d\xf$ to Order $Q^0$}

We begin with the numerator of~(\ref{Q2def}),
employing Eq.~(\ref{Glue3}) for the
integrand, since it has a simpler transverse structure than 
Eq.~(\ref{GlueEnd}).  Thus, our starting point is
\beq
\langle \qt^2 \rangle
{{dN}\over{d\xf}} = 
{{N_c^2-1}\over{2\pi^3}} \ts
\pi R^2 \ts
{{1}\over{\xf}}
\int_{-\infty}^{\infty} dx^{-} \mu^2(x^{-})
\int d^2\xt
\int_{\vert\qt\vert\le\Qprobe}\nts\nts\nts d^2\qt\ts
\qt^2 e^{i\qt\cdot\xt} \curlyL(\xt) e^{N_c \CHI(x^{-})L(\xt)}.
\label{numer1}
\eeq
If we were to evaluate the $\qt$ integral as it stands,
we would obtain a mildly complicated
combination of Bessel and Lommel functions.  Alternatively,
we may replace $\qt^2 e^{i\qt\cdot\xt}$ by $-\delt^2 e^{i\qt\cdot\xt}$.
Then,
the momentum integration appearing in~(\ref{numer1}) is
easily performed with the help of Ref.~\cite{PhysicistsFriend}:
\beq
\int_{\vert\qt\vert\le\Qprobe}\nts\nts\nts d^2\qt\ts
e^{i\qt\cdot\xt} = 
2\pi Q^2 \ts
{{J_1(\Qprobe x)}\over{\Qprobe x}}.
\label{Bessel}
\eeq
Observe that if the $\qt$ integration were to extend to infinity
instead of being cut off at $\Qprobe$, then the integral appearing
in Eq.~(\ref{Bessel}) would have produced a delta function
instead of a Bessel function.  Hence, we conclude that
\beq
\lim_{\Qprobe \rightarrow \infty} \Qprobe^2  \ts
{{J_1(\Qprobe x)}\over{\Qprobe x}} =
2\pi \delta^2(\xt).
\label{KeyLimit}
\eeq
Eq.~(\ref{KeyLimit}) will serve as a useful diagnostic 
tool in determining the order in $Q$ of each contribution as
it is encountered.  That is, if the application of Eq.~(\ref{KeyLimit})
to a given expression results in a finite result, we will
conclude that there are no contributions to that expression
which grow as $\Qprobe\rightarrow\infty$.  On the other hand,
if the result of applying Eq.~(\ref{KeyLimit}) is divergent,
then we will have to do the integral involving the Bessel
function exactly for finite $\Qprobe$, ensuring that we
correctly determine not only the contributions
which grow as $\Qprobe\rightarrow\infty$, but also the 
subleading terms as well.

After inserting~(\ref{Bessel}) into~(\ref{numer1}),
we integrate by parts to shift $\delt^2$ off of the Bessel
function.  Thus, we arrive at
\beq
\langle \qt^2 \rangle
{{dN}\over{d\xf}} =
-{{N_c^2-1}\over{\pi^2}} \ts
\pi R^2 \ts
{{1}\over{\xf}} \ts
\Qprobe^2
\int_{-\infty}^{\infty} dx^{-} \mu^2(x^{-})
\int d^2\xt \ts
{{J_1(\Qprobe x)}\over{\Qprobe x}}  \ts
\delt^2\Bigl[
\curlyL(\xt) e^{N_c \CHI(x^{-})L(\xt)}
\Bigr].
\label{numer2}
\eeq
The derivatives appearing in Eq.~(\ref{numer2}) yield
\beqa
\delt^2\Bigl[
\curlyL(\xt) e^{N_c \CHI(x^{-})L(\xt)}
\Bigr] =
\biggl\{ && \Bigl[\delt^2\curlyL(\xt)\Bigr] 
\cr && + 
N_c \CHI(x^{-}) \curlyL(\xt) \Bigl[\delt^2 L(\xt)\Bigr] 
\cr \phantom{\biggl[} && +
2N_c \CHI(x^{-}) \Bigl[\partial_i\curlyL(\xt)\Bigr]
                 \Bigl[\partial_i L(\xt)\Bigr]
\cr && +
N_c^2 \CHI^2(x^{-}) \curlyL(\xt)
                    \Bigl[\partial_i L(\xt)\Bigr]^2
\biggr\}e^{N_c \CHI(x^{-})L(\xt)}.
\label{derivative}
\eeqa
The next steps are straightforward but tedious.  
Basically we have to
examine each term bit-by-bit and do the $\xt$ integration
either after applying Eq.~(\ref{KeyLimit}) (for the constant
contributions) or by doing the actual integral with the Bessel
function intact (for the contributions that grow 
as $\Qprobe \rightarrow \infty$).
Rather than go through all of this mathematics in detail, let
us only mention the highlights.

The first term in Eq.~(\ref{derivative}),
up to the ubiquitous exponential factor, is purely Abelian in origin.
Differentiation of Eq.~(\ref{curlyLgen}) tells us that
\beq
\delt^2 \curlyL(\xt) = -\hbox{$1\over2$} \curlyD(\xt).
\eeq
Thus, the contribution from this term is especially easy to
compute, as there is no $\deltat$ integration.  The singular
part of $\curlyD(\xt)$ leads to a contribution proportional
to $\Qprobe^2$, whereas the smooth part produces a constant
piece proportional to $\Ihat(0)$.

The second and third terms of~(\ref{derivative}) are purely
non-Abelian in origin.  At lowest order ({\it i.e.}\ with
the exponential set to unity), they are quadratic in the
charge-squared.  At the end of the day, the longitudinal
dependence of these terms reads
\beqa
\int_{-\infty}^{\infty} dx^{-} \mu^2(x^{-})\CHI(x^{-})
&=&
\int_{-\infty}^{\infty} dx^{-} 
\int_{-\infty}^{x^{-}} d\xi^{-} \mu^2(x^{-})\mu^2(\xi^{-})
\cr &=&
{{1}\over{2}} \CHI_\infty^2.
\label{convertCHI}
\eeqa
Thus, we see the result will be independent of the functional
form of $\mu^2(x^{-})$, as would have been manifest from the
beginning had we chosen to begin our 
calculation with Eq.~(\ref{GlueEnd})
instead of Eq.~(\ref{Glue3}).

Because $\delt^2 L(\xt) = -2 \curlyL(\xt)$, the second term
of~(\ref{derivative}) contains $\curlyL^2(\xt)$.  As a consequence,
we encounter the transverse integrals
\beq
\transint_1 \equiv 
\int d^2\xt \ts
{{J_1(\Qprobe x)}\over{\Qprobe x}}  \ts
\ln^2 \xt^2
\label{trans1}
\eeq
and
\beq
\transint_2 \equiv 
\int d^2\xt \ts
{{J_1(\Qprobe x)}\over{\Qprobe x}}  \ts
\ln\xt^2 \ln(\xt-\deltat)^2.
\label{trans2}
\eeq
The first of these two integrals reduces to
\beq
\transint_1 =
{{8\pi}\over{\Qprobe}}
\int_{0}^{\infty} dx \ts
J_1(\Qprobe x) \ts \ln^2 x.
\label{trans1a}
\eeq
This integral may be evaluated by 
observing that
Eq.~(6.561.14) of Ref.~\cite{PhysicistsFriend}
yields the identity
\beq
\int_{0}^{\infty} dx \ts
J_1(\Qprobe x) \ts x^\vareps
= {{1}\over{\Qprobe}}
\biggl({{\Qprobe}\over{2}}\biggr)^{-\vareps}
{ {\Gamma(1+\vareps/2)}\over{\Gamma(1-\vareps/2)}}.
\label{trans1b}
\eeq
Because $\vareps$ is arbitrary, we may expand 
both sides of Eq.~(\ref{trans1b}) 
and equate corresponding powers of $\vareps$
to read off the integral of $J_1(\Qprobe x)$ times 
a logarithm to an arbitrary positive
integer power.  In the case of immediate 
interest we have
\beq
\transint_1 =
{{2\pi}\over{\Qprobe^2}}
\Biggl[
\ln\biggl({{\Qprobe^2}\over{4}}\biggr) + 2\gamma_E
\Biggr]^2.
\label{trans1d}
\eeq

We deal with $\transint_2$ by first writing
\beq
\transint_2 =
2 \int d^2\xt \ts
{{J_1(\Qprobe x)}\over{\Qprobe x}}  \ts
\Bigl[ \ln(\Qprobe x) - \ln\Qprobe \Bigr]
\ln(\xt-\deltat)^2.
\label{trans2a}
\eeq
Next, we rescale $\xt \equiv \yt/\Qprobe$ in the first term
to arrive
at a form which possesses a transparent $\Qprobe\rightarrow\infty$
limit:
\beqa
\transint_2 &=&
2 \int d^2\yt \ts
{{J_1(y)}\over{y}}  \ts \ln y \ts
\ln\biggl( {{\yt}\over{\Qprobe}} - \deltat \biggr)^2
\cr &-&
\ln\Qprobe^2 \int d^2\xt \ts
{{J_1(\Qprobe x)}\over{\Qprobe x}}  \ts
\ln(\xt-\deltat)^2.
\label{trans2b}
\eeqa
In the first term of~(\ref{trans2b}), we may neglect $\yt/\Qprobe$
in the logarithm and do the remaining integral 
with the aid of Eq.~(6.772.2) of Ref.~\cite{PhysicistsFriend}.
In the second term we simply apply~(\ref{KeyLimit}).  The result reads
\beq
\transint_2 =
-{{2\pi}\over{\Qprobe^2}} 
\Biggl[
\ln\biggl({{\Qprobe^2}\over{4}}\biggr) + 2\gamma_E
\Biggr]
\ln\deltat^2 .
\eeq

The third term of Eq.~(\ref{derivative}) contains among
its contributions only one new non-trivial transverse integral:
\beq
\transint_3 \equiv 
\int d^2\xt \ts
{{J_1(\Qprobe x)}\over{\Qprobe x}} \ts
{{\xt\cdot\deltat}\over{\xt^2}} \ts
\ln(\xt-\deltat)^2.
\label{trans3}
\eeq
We obtain the large-$\Qprobe$ limit of~(\ref{trans3}) by
rescaling $\xt \equiv \yt/\Qprobe$ and writing
\beq
\ln\biggl(\deltat - {{\yt}\over{\Qprobe}} \biggr)^2
= \ln\deltat^2 - {{2\yt\cdot\deltat}\over{\Qprobe\deltat^2}}
+ {\cal O}\biggl({{1}\over{\Qprobe^2}}\biggr).
\eeq
Then,~(\ref{trans3}) becomes
\beqa
\transint_3 &=&
{{1}\over{\Qprobe}} \ln\deltat^2
\int d^2\yt \ts
{{J_1(y)}\over{y}} \ts
{{\yt\cdot\deltat}\over{\yt^2}} 
\cr \phantom{\Biggl[} &-&
{{2}\over{\Qprobe^2}} 
\int d^2\yt \ts
{{J_1(y)}\over{y}} \ts
{{(\yt\cdot\deltat)^2}\over{\yt^2}} .
\label{trans3a}
\eeqa
The first term of~(\ref{trans3a}) posseses a vanishing angular
integral, while the second term  is straightforward to evaluate.
Thus, 
\beqa
\transint_3 &=&
-{{2\pi}\over{\Qprobe^2}} .
\label{trans3b}
\eeqa

Finally, we remark that the fourth term of Eq.~(\ref{derivative}),
which consists of contributions which are cubic and higher
in the charge density squared vanishes in the large-$\Qprobe$
limit.  Essentially, this comes about because the small-$\xt$
behavior of $\partial_i L(\xt)$ is $x_i [ 1+\ln\xt^2]$.
Thus, the factor $[\partial_i L(\xt)]^2$ contains sufficient
powers of $\xt$ in this region to kill off the diverging
logarithms.

After working through all of the necessary algebra, we arrive
at the surprisingly simple result
\beqa
\langle \qt^2 \rangle {{dN}\over{d\xf}} &=& 
{{1}\over{\xf}}
{{\Qtotal^2}\over{4\pi^2}}
\Bigl[ \Qprobe^2 - 4\pi C(0) \Bigr]
\cr &+& 
{{1}\over{\xf}}
{{\Qtotal^2}\over{4\pi^2}}
{{N_c\CHI_\infty} \over{8\pi}}
\left\{
\Biggl[
\ln\biggl({{\Qprobe^2}\over{4}}\biggr) + 2\gamma_E
+ \int d^2\deltat \ts C(\Delta)\ln\Delta^2 
\Biggr]^2   \right.
\cr && \left. \phantom{\Biggr]^2} \qquad\qquad\qquad 
-2 \Biggl[
\ln\biggl({{\Qprobe^2}\over{4}}\biggr) + 2\gamma_E
+ \int d^2\deltat \ts C(\Delta)\ln\Delta^2 
\Biggr]
- 2 \right\},
\label{numerFINAL}
\eeqa
where we have denoted the total charge-squared of the nucleus
$(N_c^2-1)\CHI_\infty \pi R^2$  by $\Qtotal^2$.
Each occurence of $\ln\Qprobe^2$ in Eq.~(\ref{numerFINAL})
comes with the same combination of $\gamma_E$ and integral
over $C(\Delta)$, leading to the definition of $\Qprobe_0$
presented in Eq.~(\ref{LogMom}).

%%%%%%%%%%%%%%%%%%%%%%%%%%%%%%%%%%%%%%%%%%%%%%%%%%%%%%%%%%%%%%%%
%%
%%            DENOMINATOR
%%
%%%%%%%%%%%%%%%%%%%%%%%%%%%%%%%%%%%%%%%%%%%%%%%%%%%%%%%%%%%%%%%%

\subsection{Calculation of $dN/d\xf$ to Order $1/Q^2$}

We now turn to the denominator of~(\ref{Q2def}).
Previously, we presented Eq.~(\ref{preFactorized}),
which writes this quantity in terms of the
momentum space representation of the charge density correlation
function $\widetilde\curlyD(\qt)$.
However, since the numerator has been expressed in
terms of the position space function $\Ihat(\Delta)$,
it will be useful to do the same for the denominator as well.
Besides, Eq.~(\ref{preFactorized}) is
correct only to leading order, and we require the $1/\Qprobe^2$
terms as well.

To obtain such an expression, we again employ Eq.~(\ref{Glue3}) for the
integrand:
\beq
{{dN}\over{d\xf}} =
{{N_c^2-1}\over{2\pi^3}} \ts
\pi R^2 \ts
{{1}\over{\xf}}
\int_{-\infty}^{\infty} dx^{-} \mu^2(x^{-})
\int d^2\xt
\int_{\vert\qt\vert\le\Qprobe}\nts\nts\nts d^2\qt\ts
e^{i\qt\cdot\xt} \curlyL(\xt) e^{N_c \CHI(x^{-})L(\xt)}.
\label{denom1}
\eeq
The momentum integration appearing in~(\ref{denom1}) 
may be done by applying Eq.~(\ref{Bessel}),
producing
\beq
{{dN}\over{d\xf}} =
{{N_c^2-1}\over{\pi^2}} \ts
\pi R^2 \ts
{{1}\over{\xf}} \ts
\Qprobe^2
\int_{-\infty}^{\infty} dx^{-} \mu^2(x^{-})
\int d^2\xt \ts
{{J_1(\Qprobe x)}\over{\Qprobe x}}  \ts
\curlyL(\xt) e^{N_c \CHI(x^{-})L(\xt)}.
\label{denom2}
\eeq
Since we are working to order $1/\Qprobe^2$, it is not
enough to simply apply Eq.~(\ref{KeyLimit}) to replace
the Bessel function by a delta function when it is safe to
do so, since this would make order $1/\Qprobe^2$ errors.
Instead we will have to expand the exponential and apply
an assortment of scaling arguments to determine the order
of each contribution.  We will denote the contribution to
$dN/d\xf$ which contains $m$ powers of $N_c\CHI(x^{-})L(\xt)$
by $dN^{(m)}/d\xf$.

At zeroth order in this expansion we have
\beq
{{dN^{(0)}}\over{d\xf}} = -
{{\Qtotal^2}\over{8\pi^3}} \ts
{{1}\over{\xf}} \ts
\Qprobe^2 
\Biggl\{
\int d^2\xt \ts
{{J_1(\Qprobe x)}\over{\Qprobe x}}  \ts
\ln\xt^2
-\int d^2\deltat \ts \Ihat(\Delta)
\int d^2\xt \ts
{{J_1(\Qprobe x)}\over{\Qprobe x}}  \ts
\ln(\xt-\deltat)^2
\Biggr\}.
\label{denom3}
\eeq
where we have used Eqs.~(\ref{curlyLgen}) and~(\ref{TwoTermsTwo})
to write out $\curlyL(\xt)$.
The $\xt$-integration in the first term of~(\ref{denom3})
may be performed exactly
with the
aid of Eq.~(6.772.2) of Ref.~\cite{PhysicistsFriend}:
\beq
\int d^2\xt \ts
{{J_1(\Qprobe x)}\over{\Qprobe x}}  \ts
\ln \xt^2
= -{{2\pi}\over{\Qprobe^2}} \ts
\Biggl[ 
\ln\biggl({{\Qprobe^2}\over{4}}\biggr) + 2\gamma_E
\Biggr].
\label{BesselLog}
\eeq
The second term, which will combine with the logarithm 
in~(\ref{BesselLog}) to make its argument dimensionless, 
requires a bit more work.
To this end we define
\beq
\transint_4 \equiv
\Qprobe^2
\int d^2\deltat \ts \Ihat(\Delta)
\int d^2\xt \ts
{{J_1(\Qprobe x)}\over{\Qprobe x}}  \ts
\ln(\xt-\deltat)^2.
\label{trans4}
\eeq
The rescaling $\xt \equiv \wt \Delta$ produces
\beqa
\transint_4 &=&
\Qprobe
\int d^2\deltat \ts \Ihat(\Delta)
\ts \Delta \ln\Delta^2
\int d^2\wt \ts
{{J_1(w\Qprobe\Delta)}\over{w}}  \phantom{\Biggl[}
\cr &+&
\Qprobe
\int d^2\deltat \ts \Ihat(\Delta)
\ts \Delta 
\int d^2\wt \ts
{{J_1(w\Qprobe\Delta)}\over{w}} \ts \ln(\wt-\deltahat)^2
\label{trans4a}
\eeqa
where $\deltahat$ is a unit vector in the $\deltat$ direction.
We may immediately perform the $\wt$ integration exactly
in the first term of~(\ref{trans4a}).  In the second term,
we rescale, $\deltat \equiv \deltat'/(w\Qprobe)$:
\beqa
\transint_4 &=&
\int d^2\deltat \ts \Ihat(\Delta)
\ts \ln\Delta^2
\cr &+&
{{1}\over{\Qprobe^2}} 
\int d^2\wt 
\int d^2\deltat' \ts 
\Ihat\biggl( {{\Delta'}\over{w\Qprobe}} \biggr)
\ts \Delta'  J_1(\Delta')
{{1}\over{w^4}} \ts \ln(\wt-\deltahat)^2.
\label{trans4b}
\eeqa
The only angular dependence in the second term is contained
in the logarithm.  Thus, Eq.~(\ref{amusing1}) may be applied
to yield
\beqa
\transint_4 &=&
\int d^2\deltat \ts \Ihat(\Delta)
\ts \ln\Delta^2
\cr &+&
{{2\pi}\over{\Qprobe^2}} 
\int_{1}^{\infty} {{dw}\over{w^3}} \ln w^2
\int d^2\deltat' \ts 
\Ihat\biggl( {{\Delta'}\over{w\Qprobe}} \biggr)
\ts \Delta'  J_1(\Delta').
\label{trans4c}
\eeqa
Thanks to the theta-function from Eq.~(\ref{amusing1}), the
radial part of the $w$ integration begins at $w=1$, not $w=0$.
If we now expand the correlation function in powers of 
$\Delta'/(w\Qprobe)$, the $w$ integration will always converge.
Likewise, we may always do the $\Delta'$ integration via the
analytic continuation of Eq.~(6.561.14) of Ref.~\cite{PhysicistsFriend}.
Thus, we obtain a well-defined expansion of the second term
of~(\ref{trans4c}) in powers of $1/\Qprobe$.  In particular,
the $1/\Qprobe^2$ term vanishes, since
\beq
\int_0^{\infty} d\Delta \Delta^2 J_1(\Delta) = 0.
\eeq
Thus, only the first term of Eq.~(\ref{trans4c}) contributes
through order $1/\Qprobe^2$.  Combining this result with 
Eqs.~(\ref{denom3}) and~(\ref{BesselLog}) we arrive at
\beq
{{dN^{(0)}}\over{d\xf}} = 
{{1}\over{\xf}}
{{\Qtotal^2}\over{4\pi^2}}
\Biggl[
\ln\biggl({{\Qprobe^2}\over{4}}\biggr) + 2\gamma_E
+ \int d^2\deltat \ts C(\Delta)\ln\Delta^2
\Biggr].
\label{denom5}
\eeq

The next term in the expansion of the exponential of Eq.~(\ref{denom2})
will produce contributions of order $1/\Qprobe^2$.  Our overview
of this part of the calculation begins with the result of
doing the expansion, which reads
\beq
{{dN^{(1)}}\over{d\xf}} \equiv
{{\Qtotal^2}\over{2\pi^2}} \ts N_c\CHI_\infty \ts
{{1}\over{q^{+}}}\ts \Qprobe^2
\int d^2\xt \ts
{{J_1(\Qprobe x)}\over{\Qprobe x}} \ts
\curlyL(\xt) L(\xt).
\label{denom6}
\eeq
Eqs.~(\ref{Lgen}) and~(\ref{curlyLgen}) when combined with
Eq.~(\ref{TwoTermsTwo}) tell us that there are three kinds
of contributions to Eq.~(\ref{denom6}).  The first is generated
by multiplying the $\delta$ function containing terms of $\curlyL(\xt)$
and $L(\xt)$.  It may be evaluated exactly with the help of
\beq
\int_0^{\infty} dx \ts x^2 \ln^2 x \ts J_1(Qx)
= {{2}\over{Q^3}}
\Biggl[
\ln\biggl( {{Q^2}\over{4}} \biggr)
+ 2\gamma_E - 1
\Biggr],
\label{YetAnotherIntegral}
\eeq
which was obtained via the same methods used to 
integrate Eq.~(\ref{trans1a}).

The second type of contribution to Eq.~(\ref{denom6}) consists
of terms produced by combining a $\delta$ function from one
factor [$\curlyL(\xt)$ or $L(\xt)$] with the smooth part
of the other.  Generally, these contributions are handled in
the same fashion as $\transint_4$ 
[Eqs.~(\ref{trans4})--(\ref{trans4c})].  In particular, we
extract the $1/\Qprobe^2$ terms via the 
rescaling $\xt = \wt \Delta$ followed by $\deltat = \deltat'/(wQ)$.
The only complication that arises is the appearance of the
integral
\beq
\transint_6 \equiv
-{{\Qprobe}\over{4\pi}}
\int d^2\deltat \ts \Ihat(\Delta) \ts \Delta^3
\int d^2\wt \ts {{\wt\cdot\deltahat}\over{w}} \ts
J_1(w\Qprobe\Delta) 
\ln(w\Delta)^2 \ts \ln(\wt-\deltahat)^2.
\eeq
The angular part of the $\wt$ integration is governed by 
Eq.~(\ref{amusing2}), leading to
\beq
\transint_6 =
{{\Qprobe}\over{2}}
\int d^2\deltat \ts \Ihat(\Delta) \ts \Delta^3
\int_0^{\infty} dw 
J_1(w\Qprobe\Delta) 
\ln(w\Delta)^2 
\Bigl[ \Theta(w-1) - w^2\Theta(1-w) \Bigr].
\label{trans6a}
\eeq
We may apply our usual rescaling of $\deltat$ to the first
term of~(\ref{trans6a}) to deduce that it is of order
$(\ln Q^2)/Q^4$, and so may be neglected.  However, in the
second term, we cannot rescale $\deltat$ in this manner
since the $w$ integration would diverge.  Instead, we write
$w\equiv v/(\Qprobe\Delta)$, yielding
\beq
\transint_6 =
{{1}\over{2\Qprobe^2}}
\int d^2\deltat \ts \Ihat(\Delta) 
\int_0^{\Qprobe\Delta} dv
\ts v^2 (\ln v^2 - \ln\Qprobe^2) \ts
J_1(v).
\label{trans6b}
\eeq
If we let the upper limit of the $v$ integration go to infinity,
we see that the integral remains finite.  Therefore, we conclude
that
\beq
\transint_6 = -{{2}\over{\Qprobe^2}},
\eeq
correct through order $1/\Qprobe^2$.

The final type of contribution to~(\ref{denom6}) which
we encounter is generated by
multiplying the smooth parts of both factors, leading to
the formidible-looking integral
\beq
\int d^2\xt \ts
{{J_1(\Qprobe x)}\over{\Qprobe x}}
\int d^2\deltat \ts
\Ihat(\Delta) \Bigl[ (\xt-\deltat)^2 \ln(\xt-\deltat^2)
                     - \Delta^2 \ln\Delta^2 \Bigr]
\int d^2\deltat' \ts
\Ihat(\Delta') \ln(\xt-\deltat')^2.
\label{Ugh}
\eeq
By employing the rescalings $\deltat \equiv \overline{\deltat} x$
and $\deltat' \equiv \overline{\deltat}' x$ followed by
$\xt \equiv \yt/\Qprobe$, it is straightforward to show that
the leading contribution to~(\ref{Ugh}) is of 
order $(\ln^2 \Qprobe^2)/Q^6$.

At the end of this lengthy procedure we arrive at 
\beq
{{dN^{(1)}}\over{d\xf}} = 
-{{1}\over{\xf}}
{{\Qtotal^2}\over{4\pi^2}}
{{N_c \CHI_\infty}\over{4\pi\Qprobe^2}}
\Biggl[
\ln\biggl({{\Qprobe^2}\over{4}}\biggr) + 2\gamma_E
+ \int d^2\deltat \ts C(\Delta)\ln\Delta^2
\Biggr].
\label{denomSubLead}
\eeq
When combined with Eq.~(\ref{denom5}), we obtain the 
result~(\ref{zeroth}) presented in Sec.~\ref{MOMENTTEXT}.

%%%%%%%%%%%%%%%%%%%%%%%%%%%%%%%%%%%%%%%%%%%%%%%%%%%%%%%%%%%%%%%%
%%%%%%%%%%%%%%%%%%%%%%%%%%%%%%%%%%%%%%%%%%%%%%%%%%%%%%%%%%%%%%%%
%%%%%%%%%%%%%%%%%%%%%%%%%%%%%%%%%%%%%%%%%%%%%%%%%%%%%%%%%%%%%%%%
%%%%%%
%%%%%%      APPENDIX:  INTEGRALS FOR THE UNIFORM DISTRIBUTION
%%%%%%
%%%%%%%%%%%%%%%%%%%%%%%%%%%%%%%%%%%%%%%%%%%%%%%%%%%%%%%%%%%%%%%%
%%%%%%%%%%%%%%%%%%%%%%%%%%%%%%%%%%%%%%%%%%%%%%%%%%%%%%%%%%%%%%%%
%%%%%%%%%%%%%%%%%%%%%%%%%%%%%%%%%%%%%%%%%%%%%%%%%%%%%%%%%%%%%%%%

\section{The Integrals Appearing in Kovchegov's Model}\label{INTEG1}

%%%%%%%%%%%%%%%%%%%%%%%%%%%%%%%%%%%%%%%%%%%%%%%%%%%%%%%%%%%%%%%%
%%
%%                         IHAT
%%
%%%%%%%%%%%%%%%%%%%%%%%%%%%%%%%%%%%%%%%%%%%%%%%%%%%%%%%%%%%%%%%%

\subsection{The Charge Density Correlation Function} 

In Sec.~\ref{EXAMPLE}, we wrote
the charge density correlation function derived from 
Kovchegov's model~\cite{paper12} in terms of the two
integrals $\Isng$ and $\Ismth$, given by Eqs.~(\ref{Isng})
and~(\ref{Ismth}) respectively.
In this Appendix, we will explicitly evaluate these expressions
assuming that the nuclear radius $R$ is much larger than the
nucleon radius $a$.

The easier of the two integrals to evaluate is $\Isng$.
Because the nucleus and nucleons are spherical and we assume a sharp
cut-off in the allowed positions of the nucleons and the quarks,
we have
\beq
\Isng(\xt,\xprimet) =
{{9}\over{4}}\ts
{{\delta^2(\xt-\xprimet)}\over{\pi^2 R^3 a^3}}
\int_{\vert\xit\vert \le a} 
\negthinspace\negthinspace\negthinspace\negthinspace 
d^2\xit
\int_{\vert\rt\vert \le R} 
\negthinspace\negthinspace\negthinspace\negthinspace 
d^2\rt
\thinspace
\sqrt{R^2-\rt^2} \sqrt{a^2-\smash\xit^2}
\ts\delta^2(\xt-\rt-\xit).
\label{sng1}
\eeq
Now an exact evaluation of~(\ref{sng1}) would require that
we satisfy the $\delta$-function by  setting $\rt \equiv \xt-\xit$
and integrating over only those values of $\xit$ which are compatible
with this.  However, since we take $R \gg a$, we may approximate
by setting $\rt^2 \approx \xt^2$ in the integrand and performing
the $\xit$ integration over the entire disk of radius $a$.
By doing this, we are allowing nucleons which are located near
the edge of the nucleus to ``pop out'' by a (small) amount of
order $a$.  In this approximation, a straightforward evaluation
yields
\beq
\Isng(\sigmat,\deltat) =
{{3}\over{2\pi R^2}}\ts
\sqrt{ 1 - {{\sigmat^2}\over{R^2}}  }
\ts\Theta\Bigl(R^2-\sigmat^2\Bigr) \ts \delta^2(\deltat).
\label{sng2}
\eeq
where we have switched to the sum and difference 
variables~(\ref{SumDiff}).
Now Eq.~(\ref{sng2}) depends upon $\sigmat$ whereas the expressions
employed in Sec.~\ref{EXAMPLE} do not.
The $\sigmat$-dependence of Eq.~(\ref{sng2}) is not
really surprising, since the pancake obtained by Lorentz-contracting 
a spherical nucleus
should be thinner near the edges than near the center.
What we call $\CHI_\infty$ in Eq.~(\ref{mu-sqr}) is obtained
from~(\ref{sng2}) by first integrating 
over $\sigmat$:
{\it i.e.}\ instead of simply writing $\pi R^2$ for the
$\sigmat$ integral, as was done in the discussion
following Eq.~(\ref{GlueStart}),
we perform that integral including the geometric factors
indicated in~(\ref{sng2}):
\beq
\int d^2\sigmat \ts\ts \Isng(\sigmat,\deltat) =
\delta^2(\deltat).
\label{sng3}
\eeq
Comparing this result with Eqs.~(\ref{2to3}) and (\ref{KovForm}),
we conclude that
\beq
\pi R^2 \CHI_\infty = {{g^2 N}\over{N_c}}.
\eeq

The other integral is somewhat more difficult to evaluate.
Doing the longitudinal integrations and making the $R \gg a$
approximation described above yields
\beqa
\Ismth(\sigmat,\deltat) &=& 
{{27}\over{\pi^3R^2a^6}}\ts
\sqrt{ 1 - {{\sigmat^2}\over{R^2}}  } \ts \Theta(R^2-\sigmat^2)
\cr\nonumber\\[0.1cm] &\quad& \quad\times
\int_{\vert\xit\vert \le a} 
\negthinspace\negthinspace\negthinspace\negthinspace 
d^2\xit
\int_{\vert\xiprimet\vert \le a} 
\negthinspace\negthinspace\negthinspace\negthinspace 
d^2\xiprimet \thinspace\thinspace
\sqrt{a^2-\smash\xit^2} \sqrt{a^2-{\smash\xiprimet}^2}
\thinspace
\delta^2(\deltat-\xit+\xiprimet)
\cr\nonumber\\[0.2cm] &\equiv&
{{3}\over{2\pi R^2}}\ts
\sqrt{ 1 - {{\sigmat^2}\over{R^2}}  } \ts \Theta(R^2-\sigmat^2)
\ts\Ihat(\deltat)
\label{smoooth}
\eeqa
where $\Ihat$ is the function introduced in Eq.~(\ref{IHATu}).
Since we have defined $\Ihat(\deltat)$ by extracting the same
$\sigmat$-dependent prefactor which appeared in~(\ref{sng2}),
at the end of the day we find that 
$\curlyD(\deltat) = \delta^2(\deltat) - \Ihat(\deltat)$,
as written in Eq.~(\ref{TwoTerms}).

\begin{figure}[h]

\vspace*{15cm}
\includegraphics{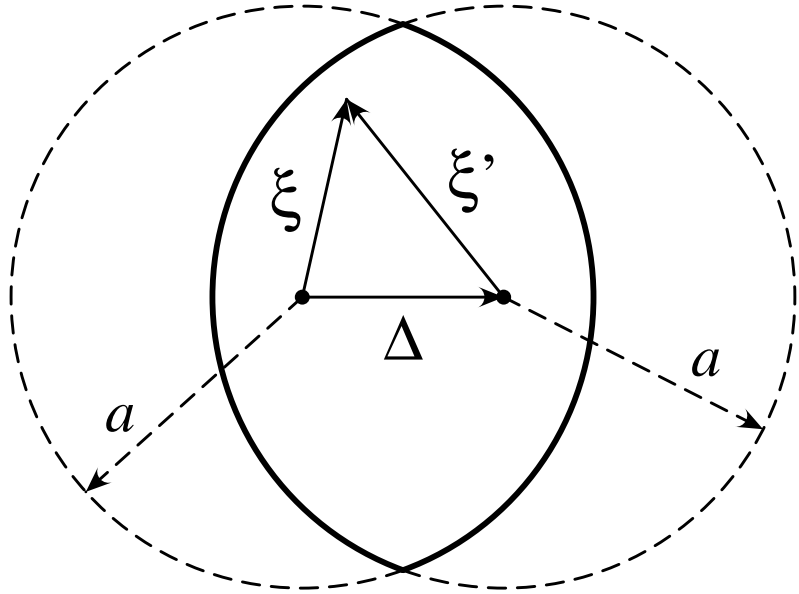}

\caption[]{The integration region encountered
in Eq.~(\protect\ref{smoooth}) for the
smooth part of the correlation function in the uniform
version of Kovchegov's model~\protect\cite{paper12}.
}
\label{IhatIntegral}
\end{figure}

A direct evaluation of the integral defining $\Ihat$ is
difficult because of the $\delta$-function constraint:
we are instructed to consider the area of overlap between
two disks of radius $a$ separated by a distance $\Delta$.
This area must be weighted by the product of the distances
from the two centers to the integration point 
(see Fig.~\ref{IhatIntegral}).  Although this sounds simple,
the only immediate conclusion which may be drawn directly from
the integral is that when
$\vert\deltat\vert \ge 2a$ the function vanishes:  the disks
do not overlap.

To deal with the situation when 
$\vert\deltat\vert\le 2a$,
we begin by computing the 
Fourier transform of $\Ihat(\deltat)$:
\beqa
\widetilde\Ihat(\qt) &\equiv&
\int d^2\deltat  \ts\ts
\hbox{\boldmath ${\it e}^{{\it i}q\cdot\Delta}$ \unboldmath}
C(\deltat)
\cr &=&
{ {9}\over{4\pi^2 a^6} } \thinspace\thinspace
\int_{\vert\xit\vert \le a} 
\negthinspace\negthinspace\negthinspace\negthinspace 
d^2\xit
\int_{\vert\xiprimet\vert \le a} 
\negthinspace\negthinspace\negthinspace\negthinspace 
d^2\xiprimet 
\int d^2\deltat  \ts\ts
\hbox{\boldmath ${\it e}^{{\it i}q\cdot\Delta}$ \unboldmath}
\sqrt{a^2-\smash\xit^2} \sqrt{a^2-{\smash\xiprimet}^2}
\thinspace
\delta^2(\deltat-\xit+\xiprimet)
\eeqa
Since the $\deltat$ integration is over all space, the $\delta$-function
constraint can be satisfied for all values of $\xit$ and $\xiprimet$:
no messy $\Theta$-functions are introduced.  In fact, the
entire expression factorizes:
\beq
\widetilde\Ihat(\qt) =
\Biggl\{
{ {3}\over{2\pi a^3} } \thinspace\thinspace
\int_{\vert\bbox{\xi}\vert \le a} 
\negthinspace\negthinspace\negthinspace\negthinspace 
d^2\xit \ts\ts
e^{i\bbox{q\cdot\xi}}
\sqrt{a^2-\smash\xit^2} 
\Biggr\}^2.
\eeq
Performing the angular integration and rescaling the
radial integral to unit range we obtain
\beq
\widetilde\Ihat(\qt) =
\Biggl\{
3 
\int_{0}^{1}
dv
\ts v \sqrt{1-v^2}
\ts J_0(aqv)
\Biggr\}^2.
\eeq
The radial integral may be performed with the help of
Ref.~\cite{PhysicistsFriend} ({\it c.f.}\ Eq.~(6.567.1)):
\beq
\tilde\Ihat(\qt) =
\Biggl\{
3 \sqrt{{\pi}\over{2}} \ts
{ {J_{3/2}(aq)} \over {(aq)^{3/2}} }
\Biggr\}^2.
\eeq
Happily, the half-integer Bessel functions are expressible as
trigonometric polynomials.  Thus, we arrive finally at
\beq
\widetilde\Ihat(\qt) =
{{9}\over{(aq)^6}} 
\Bigl[
\sin(aq) - (aq)\cos(aq)
\Bigr]^2.
\label{Ctilde}
\eeq
In spite of the apparent high-order singularity at $q=0$,
$\tilde\Ihat(0)$ is nevertheless finite:  
\beq
\lim_{q\rightarrow 0} \ts\ts\widetilde\Ihat(\qt) = 1.
\label{finite0}
\eeq

It is amusing to note that the function implied by~(\ref{finite0})
for the Fourier transform of $\curlyD$, namely
\beq
\widetilde\curlyD(\qt) = 1 -
{{9}\over{(aq)^6}} 
\Bigl[
\sin(aq) - (aq)\cos(aq)
\Bigr]^2.
\label{Dtilde}
\eeq
is exactly the same function that arises in position space for
the parallel spin pair correlation function in a noninteracting
gas of spin-$1\over2$ fermions\cite{Baym}.  In the Fermi gas,
the role of the sphere of radius $a$ containing a uniform 
distribution of quarks is played by the momentum-space filling
of energy levels up to the Fermi surface.  The condition that
the nucleons be color neutral on large scales leads to the
vanishing of $\curlyD$ at zero momentum.  The Pauli exclusion
principle leads to the vanishing of the parallel spin pair
correlation function at zero separation.\footnote{The opposite
spin pair correlation function  for the Fermi gas is unity
(the exclusion principle does not affect particles of opposite
spin).  This case is analagous to the non-charge conserving
correlator of Eq.~(\protect\ref{MVtrans}).} 

Returning to the issue at hand,
to obtain $\Ihat(\deltat)$, we simply invert the Fourier transform.
The angular integration is straightforward, and leaves us with
\beq
\Ihat(\deltat) =
{{9}\over{4\pi a^2}} 
\int_0^{\infty} dq \ts
J_0\biggl({{q\Delta}\over{a}}\biggr)
\biggl[
{{1}\over{q^3}}
+ {{1}\over{q^5}}
+ {{\cos 2q}\over{q^3}}
- {{2\sin 2q}\over{q^4}}
- {{\cos 2q}\over{q^5}}
\biggr].
\label{smth1}
\eeq
Each of the five terms on the right-hand-side of~(\ref{smth1})
is individually divergent.   In order to perform the $q$ integration
term-by-term, we must insert a regulator.
Since the final combination of terms in~(\ref{smth1}) is 
guaranteed to be finite by~(\ref{finite0}), the result
will be insensitive to the details of the regulator employed.
A convenient means of regulation is to make the replacement
\beq
J_0\biggl({{q\Delta}\over{a}}\biggr)
\rightarrow
J_{2\eps}\biggl({{q\Delta}\over{a}}\biggr)
\eeq
in~(\ref{smth1}).  We may then apply Eqs.~(6.561.14), (6.699.1),
and~(6.699.2) of Ref.~\cite{PhysicistsFriend} to obtain 
\beqa
\Ihat(\deltat) =
{{9}\over{4\pi a^2}} 
\Biggl\{ &&
-{{X^2}\over{2}} \ts {{1}\over{\eps}}
\ts {{1}\over{1-\eps^2}}
+{{X^4}\over{8}} \ts {{1}\over{\eps}}
\ts {{1}\over{(1-\eps^2)(1-\eps^2/4)}}
\cr\nonumber\\[0.1cm] &&
- 2 \biggl( {{X^2}\over{4}} \biggr)^\eps \ts
{{\cos\pi\eps}\over{\eps(1-2\eps)(2-2\eps)(3-2\eps)(4-2\eps)}}
\cr\nonumber\\[0.1cm] && 
\quad\qquad\times
\biggl[ (3-2\eps)(4-2\eps)
{\ts}_2F_1(-1+\eps,-\hbox{$1\over2$}+\eps;1+2\eps;X^2)
\phantom{\Biggl[}
\cr &&
\qquad\quad\quad
-4(4-2\eps) {\ts}_2F_1(-1+\eps,-\hbox{$3\over2$}+\eps;1+2\eps;X^2)
\phantom{\Biggl[}
\cr &&
\quad\qquad\quad
+4 {\ts}_2F_1(-2+\eps,-\hbox{$3\over2$}+\eps;1+2\eps;X^2)
\biggr]
\Biggr\},
\eeqa
which is valid for $\Delta \le 2a$.  In this expression we have
defined the dimensionless distance $X\equiv \Delta/(2a)$.
The hypergeometric functions may be expanded, and all of the 
poles in $\eps$ cancelled.  The resulting series may then
be converted back into a different (generalized) hypergeometric
function.  The result of this tedious algebra is
\beq
\Ihat(\deltat) = 
{ {9}\over{64\pi a^2} } \thinspace\thinspace
\Biggl[ 2X^2(4-X^2) \ln\biggl({X^2\over4}\biggr) 
+ 8 + X^4  
 - {{X^6}\over{2}} \thinspace {}_3F_2(1,1,\hbox{$5\over2$};3,4;X^2)
\Biggr].
\label{smth2}
\eeq

Finally, we notice that all but one of the
parameters in the hypergeometric function appearing in~(\ref{smth2})
are integers.  Hence, we may eliminate this hypergeometric function 
by combining
Eqs.~(7.512.12) and~(9.111) of Ref.~\cite{PhysicistsFriend}
to obtain
\beqa
{}_3F_2(a,b,c;p,q;z) &=&
{
{\Gamma(p)\Gamma(q)}
\over
{\Gamma(b)\Gamma(p-b)\Gamma(c)\Gamma(q-c)}
}
\cr && \quad\times
\int_0^1 dt \int_0^1 du \enspace
t^{b-1} (1-t)^{p-b-1} 
u^{c-1} (1-u)^{q-c-1}
(1-tuz)^{-a}.
\label{3F2}
\eeqa
The required integrations are all straightforward, yielding
\beqa
{}_3F_2(1,1,\hbox{$5\over2$};3,4;z) &=&
-{{4}\over{z}} \biggl( 1 - {{4}\over{z}} \biggr)
\biggl[ \ln\biggl({{z}\over{4}}\biggr) 
       +2\tanh^{-1}\sqrt{1-z} \biggr] 
\phantom{\Biggl[} \cr && 
+{{2}\over{z}} + {{16}\over{z^3}}
-{{8}\over{z^2}}\biggl( 1+{{2}\over{z}} \biggr) \sqrt{1-z}
\phantom{\Biggl[}
\label{hyper32}
\eeqa
Inserting~(\ref{hyper32}) into~(\ref{smth2}) at last gives us
\beq
\Ihat(\deltat) = 
{ {9} \over {16\pi a^2} }
\thinspace
\Bigl[
(2+X^2)\sqrt{1-X^2}
-X^2(4-X^2)\tanh^{-1}\sqrt{1-X^2}
\Bigr],
\eeq
which is true for $X\le1$.

%%%%%%%%%%%%%%%%%%%%%%%%%%%%%%%%%%%%%%%%%%%%%%%%%%%%%%%%%%%%%%%%
%%
%%                         curlyL
%%
%%%%%%%%%%%%%%%%%%%%%%%%%%%%%%%%%%%%%%%%%%%%%%%%%%%%%%%%%%%%%%%%

\subsection{Computation of $\curlyL(\xt-\xprimet)$}

We now turn to the evaluation of $\curlyL(\xt-\xprimet)$.
This function may be determined by the same procedure
used to determine $\Ihat(\deltat)$.  
Since Eq.~(\ref{curlyLgen}) expresses $\curlyL(\xt-\xprimet)$
as the convolution of $\curlyD$ with a logarithm, we may 
use~(\ref{TwoTerms}) and~(\ref{Ctilde}) to immediately write
\beqa
\widetilde\curlyL(\qt) &=&
{{1}\over{2q^2}}
\biggl\{ 1 -
{{9}\over{(aq)^6}} 
\Bigl[
\sin(aq) - (aq)\cos(aq)
\Bigr]^2
\biggr\}. 
\label{curlyL3}
\eeqa
Once again we encounter an expression with a fake pole at
the origin:  the $\qt\rightarrow 0$ limit is actually
\beq
\lim_{q\rightarrow 0} \ts\ts\tilde\curlyL(\qt) = {{a^2}\over{10}}.
\label{curlyL0}
\eeq
So, 
to invert the Fourier transform, we may employ the same trick as
was used on Eq.~(\ref{smth1}):  after the angular integration
we replace the Bessel function of order zero by one of order
$2\eps$ and integrate term-by-term.  After verifying that
all of the poles in $\eps$ do indeed cancel, we are left with
\beqa
\curlyL(\xt-\xprimet) = 
{ {1}\over{64\pi} } \thinspace &&
\biggl[
-(1-X^2)(8+8X^2-X^4)\ln\biggl({X^2\over4}\biggr)
\phantom{\Biggl[} \cr &&
- {1\over6}(1-X^2)(169-47X^2+7X^4) 
\phantom{\Biggl[} \cr &&
+ {1\over6} - {{9X^8}\over{64}} \thinspace 
{}_4F_3(1,1,\hbox{$5\over2$},4;3,5,5;X^2)
\biggr]. \phantom{\Biggl[}
\label{curlyL4}
\eeqa
This expression is valid only for $X\le 1$.
Eq.~(\ref{curlyL4}) displays the precise form predicted
for $\curlyL(\xt-\xprimet)$ at the end of Appendix~\ref{SIGMA}:
it is a power series in $X^2$ plus $\ln X^2$ times another
series in $X^2$.  No odd powers of $X$ appear in this series
since we employed a rotationally invariant correlation function
depending upon the {\it vector}\ $\xt-\xprimet$, as opposed
to an arbitrary function of $\vert\xt-\xprimet\vert$.

The hypergeometric function may be eliminated from~(\ref{curlyL4}) by
combining~(\ref{3F2}) with Eq.~(7.512.12) of
Ref.~\cite{PhysicistsFriend}:
\beqa
{}_4F_3(a,b,c,d;p,q,r;z) &=& 
{
{\Gamma(p)\Gamma(q)\Gamma(r)}
\over
{\Gamma(b)\Gamma(p-b)\Gamma(c)\Gamma(q-c)\Gamma(d)\Gamma(r-d)}
}
\cr \phantom{\Biggl[} && \times
\int_0^1 ds \int_0^1 dt \int_0^1 du \enspace
s^{b-1} (1-s)^{p-b-1} 
t^{c-1} (1-t)^{q-c-1}
\cr \phantom{\Bigl[} && \qquad\qquad\qquad\qquad\times
u^{d-1} (1-u)^{r-d-1}
(1-stuz)^{-a},
\label{4F3}
\eeqa
from which we learn that
\beqa
{}_4F_3(1,1,\hbox{$5\over2$},4;3,5,5;z) &=&
-{{64}\over{9z}} \biggl( 1 - {{9}\over{z}} + {{8}\over{z^3}} \biggr)
\biggl[ \ln\biggl({{z}\over{4}}\biggr) 
       +2\tanh^{-1}\sqrt{1-z} \biggr] 
\phantom{\Biggl[} \cr && 
-{{128}\over{9z^2}}\biggl( 1+{{13}\over{z}} 
                           - {{14}\over{z^2}} \biggr) \sqrt{1-z}
\phantom{\Biggl[} \cr && 
+{{224}\over{27z}} - {{64}\over{z^2}}
+ {{256}\over{z^3}} - {{1792}\over{9z^4}}
\phantom{\Biggl[}
\label{page1054G}.
\eeqa
The result of inserting~(\ref{page1054G}) into~(\ref{curlyL4})
is precisely~(\ref{curlyLu}) except for the step-function.
For $X>1$, it is easy to combine the integral representation
for $\curlyL$ given in Eq.~(\ref{curlyLgen}) with the integral
form of $\Ihat$ from Eq.~(\ref{IHATu}) to show that
that $\curlyL(\xt-\xprimet)$ vanishes in this region.

%%%%%%%%%%%%%%%%%%%%%%%%%%%%%%%%%%%%%%%%%%%%%%%%%%%%%%%%%%%%%%%%
%%
%%                         plainL
%%
%%%%%%%%%%%%%%%%%%%%%%%%%%%%%%%%%%%%%%%%%%%%%%%%%%%%%%%%%%%%%%%%

\subsection{Computation of $L(\xt-\xprimet)$}

Finally, we come to $L(\xt-\xprimet)$.
We should like to begin with the Fourier transform function
$\tilde{L}(\qt)$.  Note that~(\ref{decomp})
implies 
\beq
\qt^2 \ts \tilde{L}(\qt) = 2 \widetilde\curlyL(\qt).
\label{ambiguous}
\eeq
Unfortunately, this relation is not sufficient to
determine $\tilde{L}(\qt)$ uniquely:  we may add an
arbitrary multiple of $\delta^2(\qt^2)$ to $\tilde{L}(\qt)$
without contradicting Eq.~(\ref{ambiguous}).
According to the integral representation of $L(\xt-\xprimet)$,
$L({\bf 0})$ vanishes.  Therefore, we could adjust this
arbitrary term so that this is true when we invert the Fourier
transform.  Operationally, we may simply invert the expression
implied by~(\ref{ambiguous}) using the same method we have
used on the other integrals in this Appendix
and drop the pole and constant terms. 
Although this procedure lacks the satisfying feeling that comes
when one sees the poles cancel automatically, it does produce
the correct result, as has been verified by a direct numerical
evaluation of the integral in~(\ref{Lgen}).  The bottom
line is that for $X\le1$ we find
\beqa
L(\xt-\xprimet) = 
{ {a^2}\over{4\pi} } \thinspace &&
\Biggl[
\biggl(
X^2 - {{X^6}\over{8}} + {{X^8}\over{128}} 
\biggr) \ln\biggl({X^2\over4}\biggr) 
\phantom{\Biggl[} \cr &&
+ {{3X^2}\over{2}}
- {{9X^4}\over{8}}
+ {{5X^6}\over{24}}
- {{5X^8}\over{384}} 
\phantom{\Biggl[} \cr &&
+ {{9X^{10}}\over{12800}} \thinspace 
  {}_4F_3(1,1,\hbox{$5\over2$},4;3,6,6;X^2)
\Biggr] \phantom{\Biggl[}
\label{plainL5}
\eeqa
whereas for $X\ge1$
\beqa
L(\xt-\xprimet) = 
-{ {a^2}\over{20\pi} } \thinspace 
\Biggl[ {{93}\over{20}} 
+ \ln\biggl({{X^2}\over{4}}\biggr) 
\Biggr].
\label{plainL6}
\eeqa
This latter result may be obtained inserting the integral
representation~(\ref{IHATu}) of $\Ihat$ into the integral 
representation~(\ref{Lgen}) of $L$.
Using the identity~(\ref{4F3}) to evaluate the hypergeometric
function appearing in~(\ref{plainL5}) gives:
\beqa
{}_4F_3(1,1,\hbox{$5\over2$},4;3,6,6;z) &=&
-{{20}\over{9z}} \biggl( 5 - {{80}\over{z}} 
                            + {{640}\over{z^3}} 
                            + {{128}\over{z^4}} \biggr)
\biggl[ \ln\biggl({{z}\over{4}}\biggr) 
       +2\tanh^{-1}\sqrt{1-z} \biggr] 
\phantom{\Biggl[} \cr && 
-{{8}\over{9z^2}}\biggl( 25+{{982}\over{z}} 
                           -{{2984}\over{z^2}} 
                           -{{1488}\over{z^3}} \biggr) \sqrt{1-z}
\phantom{\Biggl[} \cr && 
+{{500}\over{27z}} - {{8000}\over{27z^2}}
+ {{1600}\over{z^3}} - {{6400}\over{3z^4}}
- {{3968}\over{3z^5}}
\phantom{\Biggl[}
\label{page1060I}
\eeqa
Eqs.~(\ref{plainL5})--(\ref{page1060I}) combine to reproduce
the expression presented in Eq.~(\ref{plainLu}).

%%%%%%%%%%%%%%%%%%%%%%%%%%%%%%%%%%%%%%%%%%%%%%%%%%%%%%%%%%%%%%%%
%%%%%%%%%%%%%%%%%%%%%%%%%%%%%%%%%%%%%%%%%%%%%%%%%%%%%%%%%%%%%%%%
%%%%%%%%%%%%%%%%%%%%%%%%%%%%%%%%%%%%%%%%%%%%%%%%%%%%%%%%%%%%%%%%
%%%%%%
%%%%%%      REFERENCES
%%%%%%
%%%%%%%%%%%%%%%%%%%%%%%%%%%%%%%%%%%%%%%%%%%%%%%%%%%%%%%%%%%%%%%%
%%%%%%%%%%%%%%%%%%%%%%%%%%%%%%%%%%%%%%%%%%%%%%%%%%%%%%%%%%%%%%%%
%%%%%%%%%%%%%%%%%%%%%%%%%%%%%%%%%%%%%%%%%%%%%%%%%%%%%%%%%%%%%%%%

%%%%%%%%%%%%%%%%%%%%%%%%%%%%%%%%%%%%%%%%%%%%%%%%%%%%%%%%%%%%%%%%
%%%%%%%%%%%%%%%%%%%%%%%%%%%%%%%%%%%%%%%%%%%%%%%%%%%%%%%%%%%%%%%%
%%%%%%%%%%%%%%%%%%%%%%%%%%%%%%%%%%%%%%%%%%%%%%%%%%%%%%%%%%%%%%%%
%%%%%%
%%%%%%          FIGURES and FIGURE CAPTIONS
%%%%%%
%%%%%%%%%%%%%%%%%%%%%%%%%%%%%%%%%%%%%%%%%%%%%%%%%%%%%%%%%%%%%%%%
%%%%%%%%%%%%%%%%%%%%%%%%%%%%%%%%%%%%%%%%%%%%%%%%%%%%%%%%%%%%%%%%
%%%%%%%%%%%%%%%%%%%%%%%%%%%%%%%%%%%%%%%%%%%%%%%%%%%%%%%%%%%%%%%%

%%%%%%%%%%%%%%%%%%%%%%%%%%%%%%%%%%%%%%%%%%%%%%%%%%%%%%%%%%%%%%%%
%%
%%      FIGURE 1              (Plot of Ihat)
%%
%%%%%%%%%%%%%%%%%%%%%%%%%%%%%%%%%%%%%%%%%%%%%%%%%%%%%%%%%%%%%%%%

%%%%%%%%%%%%%%%%%%%%%%%%%%%%%%%%%%%%%%%%%%%%%%%%%%%%%%%%%%%%%%%%
%%
%%      FIGURE 2              ( Plot of Tr<AA>(x) )
%%
%%%%%%%%%%%%%%%%%%%%%%%%%%%%%%%%%%%%%%%%%%%%%%%%%%%%%%%%%%%%%%%%

%%%%%%%%%%%%%%%%%%%%%%%%%%%%%%%%%%%%%%%%%%%%%%%%%%%%%%%%%%%%%%%%
%%
%%      FIGURE 3     ( Plot of F(q) -- i.e. dN/d3q at fixed q+ )
%%
%%%%%%%%%%%%%%%%%%%%%%%%%%%%%%%%%%%%%%%%%%%%%%%%%%%%%%%%%%%%%%%%

%%%%%%%%%%%%%%%%%%%%%%%%%%%%%%%%%%%%%%%%%%%%%%%%%%%%%%%%%%%%%%%%
%%
%%      FIGURE 4                       ( Plot of q^2*F(q) )
%%
%%%%%%%%%%%%%%%%%%%%%%%%%%%%%%%%%%%%%%%%%%%%%%%%%%%%%%%%%%%%%%%%

%%%%%%%%%%%%%%%%%%%%%%%%%%%%%%%%%%%%%%%%%%%%%%%%%%%%%%%%%%%%%%%%
%%
%%      FIGURE 5             ( Plot of all-orders/lowest-order )
%%
%%%%%%%%%%%%%%%%%%%%%%%%%%%%%%%%%%%%%%%%%%%%%%%%%%%%%%%%%%%%%%%%

%%%%%%%%%%%%%%%%%%%%%%%%%%%%%%%%%%%%%%%%%%%%%%%%%%%%%%%%%%%%%%%%
%%
%%      FIGURE 6     ( Plot of F(q) -- i.e. dN/d3q at fixed q+ )
%%
%%%%%%%%%%%%%%%%%%%%%%%%%%%%%%%%%%%%%%%%%%%%%%%%%%%%%%%%%%%%%%%%

%%%%%%%%%%%%%%%%%%%%%%%%%%%%%%%%%%%%%%%%%%%%%%%%%%%%%%%%%%%%%%%%
%%
%%      FIGURE 7      (Plot of dN/dq+
%%
%%%%%%%%%%%%%%%%%%%%%%%%%%%%%%%%%%%%%%%%%%%%%%%%%%%%%%%%%%%%%%%%

%%%%%%%%%%%%%%%%%%%%%%%%%%%%%%%%%%%%%%%%%%%%%%%%%%%%%%%%%%%%%%%%
%%
%%      FIGURE 8      (Plot of <q^2>
%%
%%%%%%%%%%%%%%%%%%%%%%%%%%%%%%%%%%%%%%%%%%%%%%%%%%%%%%%%%%%%%%%%

%%%%%%%%%%%%%%%%%%%%%%%%%%%%%%%%%%%%%%%%%%%%%%%%%%%%%%%%%%%%%%%%
%%
%%      FIGURE 9             (IhatIntegral)
%%
%%%%%%%%%%%%%%%%%%%%%%%%%%%%%%%%%%%%%%%%%%%%%%%%%%%%%%%%%%%%%%%%

\end{document}